\documentclass[aps,prx,reprint,groupedaddress]{revtex4-2}

\usepackage{amssymb,amsmath,bm}
\usepackage{graphicx}
\usepackage{enumerate}
\usepackage[utf8]{inputenc}
\usepackage{cancel}
\usepackage{color}
\usepackage{hyperref}
\usepackage{float}

\newcommand{\mean}[1]{\left\langle #1 \right\rangle}
\newcommand{\st}{\text{s}}
\newcommand{\hcs}{\text{HCS}}
\newcommand{\ini}{\text{i}}
\newcommand{\fin}{\text{f}}
\newcommand{\ther}{\text{th}}
\newcommand{\rel}{\text{rel}}
\newcommand{\calL}{\mathcal{L}}
\newcommand{\calC}{\mathcal{C}}

\newcommand{\tI}{\text{(I)}}
\newcommand{\tII}{\text{(II)}}
\newcommand{\tIII}{\text{(III)}}

\begin{document}


\title{Optimising the relaxation route with optimal control}


\author{A. Prados}
\email[]{prados@us.es}
\affiliation{F\'{\i}sica Te\'orica, Universidad de Sevilla, Apartado de Correos 1065, 41080
Sevilla, Spain}


\date{\today}

\begin{abstract}
  We look into the minimisation of the connection time between
  non-equilibrium steady states. As a prototypical example of an
  intrinsically non-equilibrium system, a driven granular gas is
  considered. For time-independent driving, its natural time scale for
  relaxation is characterised from an empirical---the relaxation
  function---and a theoretical---the recently derived classical speed
  limits---point of view. Using control theory, we find that bang-bang
  protocols---comprising two steps, heating with the largest possible
  value of the driving and cooling with zero driving---minimise the
  connecting time.  The bang-bang time is shorter than both the
  empirical relaxation time and the classical speed limit: in this
  sense, the natural time scale for relaxation is beaten. Information
  theory quantities stemming from the Fisher information are also
  analysed over these optimal protocols. The implementation of the
  bang-bang processes in numerical simulations of the dynamics of the
  granular gas show an excellent agreement with the theoretical
  predictions. Moreover, general implications of our results are
  discussed for a wide class of driven non-equilibrium systems.
  Specifically, we show that analogous bang-bang protocols, with a
  number of bangs equal to the number of relevant physical variables,
  give the minimum connecting time under quite general conditions.
\end{abstract}


\maketitle

\section{Introduction}\label{sec:intro}

Very recent developments make it possible to define the natural time
scale for the dynamical evolution---or, in other words, a speed
limit---in classical systems from a fundamental point of
view~\cite{shanahan_quantum_2018,okuyama_quantum_2018,
  ito_stochastic_2018,
  shiraishi_speed_2018,ito_stochastic_2020,
  nicholson_timeinformation_2020}. In
the quantum realm, speed limits have been known for a long time: the
so-called Mandelstam-Tamm~\cite{tamm_uncertainty_1991} and
Margolus-Levitin~\cite{margolus_maximum_1998} bounds. A recent review
on the matter is provided by Ref.~\cite{deffner_quantum_2017}. Roughly
speaking, the quantum speed limit entails a trade-off between
operation time and uncertainty in energy, i.e. the time--energy
uncertainty relation. This idea has been extended to classical systems
with Markovian dynamics: taking advantage of the similarities of the
mathematical structure of the respective Hilbert spaces, the different
versions of speed limits in
Refs.~\cite{shanahan_quantum_2018,okuyama_quantum_2018,
  ito_stochastic_2018, shiraishi_speed_2018,ito_stochastic_2020,
  nicholson_timeinformation_2020}
have been derived.

Very recently, a speed limit that is the classical analog of the
Mandelstam-Tamm bound has been
derived~\cite{nicholson_timeinformation_2020}. It is
valid for a completely general dynamics, not necessarily Markovian and
includes, as a particular case, the one derived in
Ref.~\cite{ito_stochastic_2020} starting from the Cramér-Rao
inequality. These speed limits in
Refs.~\cite{nicholson_timeinformation_2020,ito_stochastic_2020} can be
understood as a trade-off between time and cost in the considered
process. It must be noted, however, that their being the most
restrictive bounds on operation time has not been yet
proved. Currently, this is an open question for the classical speed
limits, whereas for their quantum counterparts it has been rigorously
established that the unification of the Mandelstam-Tamm and
Margolus-Levitin bounds is tight~\cite{levitin_fundamental_2009}.

The possibility of accelerating the dynamical evolution of a given
physical system has been recently analysed in different contexts, both
for
classical~\cite{martinez_engineered_2016,muratore-ginanneschi_application_2017,li_shortcuts_2017,chupeau_engineered_2018,albay_thermodynamic_2019,martikyan_comparison_2020,funo_shortcuts_2020,albay_realization_2020,plata_finite-time_2020,baldassarri_engineered_2020}
and quantum
systems~\cite{chen_fast_2010,chen_shortcut_2010,deffner_quantum_2013,campbell_trade-off_2017,xu_effects_2020,ding_smooth_2020}---for
a recent review, see Ref.~\cite{guery-odelin_shortcuts_2019}. In the
classical case, the focus has been put on engineering the connection
between equilibrium states for Markovian systems, the dynamics of
which is described by a Fokker-Planck or a master equation. This has
especially been done in the simple harmonic potential
case~\cite{martinez_engineered_2016,muratore-ginanneschi_application_2017,li_shortcuts_2017,chupeau_engineered_2018,albay_thermodynamic_2019,albay_realization_2020,plata_finite-time_2020,baldassarri_engineered_2020},
for which the fact that the probability distribution remains Gaussian
for all times strongly simplifies the mathematical
treatment~\footnote{Very recently, the connection between two
  non-equilibrium steady states of a Brownian gyrator has been
  analysed~\cite{baldassarri_engineered_2020}, but still the
  probability distribution remains exactly Gaussian in that case.}.

Here, not only do we show how to speed up the connection between
non-equilibrium steady states (NESS) but also how to optimise this
connection. This is done in a system that is a benchmark for
out-of-equilibrium systems, a driven granular gas.  In the kinetic
description, neither the dynamics is Markovian---the
Boltzmann-Fokker-Planck equation is non-linear---nor the velocity
distribution function is Gaussian---even in the long-time limit, when
the granular gas reaches a non-equilibrium steady state
(NESS).

It must be stressed that there is no ``thermodynamic'' description for
granular fluids. Extending thermodynamics concepts to them is far from
trivial: inelastic collisions break time reversal invariance and make
the system intrinsically out-of-equilibrium, which has many, some of
them unexpected, implications. For example, Shannon's entropy no
longer increases monotonically in the non-driven, freely cooling, case
and there is no clear formulation of the second principle for granular
fluids~\cite{bena_stationary_2006,marconi_about_2013,garcia_de_soria_towards_2015}.

Moreover, results derived under the assumption of Markovian
dynamics~\cite{okuyama_quantum_2018,shanahan_quantum_2018,ito_stochastic_2018,
  shiraishi_speed_2018}
are in principle not valid in the framework of kinetic
theory. Nevertheless, the very recent results based on information
geometry apply because the underlying dynamics is very
general~\cite{nicholson_timeinformation_2020,ito_stochastic_2020}. Central
to the latter approach is the concept of Fisher information $I(t)$,
which is the curvature of the Kullback-Leibler divergence and is
related to entropy production for Markovian
dynamics~\cite{nicholson_nonequilibrium_2018,
  ito_stochastic_2018,hasegawa_uncertainty_2019,
  ito_stochastic_2020}.

The concept of a thermodynamic length was first introduced in the
context of finite-time thermodynamics about forty years
ago~\cite{salamon_thermodynamic_1983,salamon_length_1985,feldmann_thermodynamic_1985},
employing an approach similar to that used for defining a statistical
distance in Hilbert space for quantum mechanical
systems~\cite{wootters_statistical_1981}. More recently, the relation
between the thermodynamic length and Fisher information was
unveiled~\cite{crooks_measuring_2007}. Later works showed how to
employ this formalism to find optimal protocols, in the sense that the
relevant physical quantities attain a minimal
value~\cite{sivak_thermodynamic_2012,kim_geometric_2016}. Over the
last few years, further work has linked the Fisher information $I(t)$
with the so-called thermodynamic uncertainty
relations~\cite{nicholson_nonequilibrium_2018,hasegawa_uncertainty_2019,
  dechant_multidimensional_2019},
also showing that $I(t)$ is related to the entropic acceleration,
i.e. to the second time derivative of Shannon's entropy~\cite{nicholson_nonequilibrium_2018}.

The natural time scale for connecting two NESS corresponding to
different values of the driving can be characterised both empirically
and theoretically. Let us consider relaxation at constant driving: at
$t=0$, the driving is instantaneously changed from its initial to its
final value.  From an empirical standpoint, the relaxation time in
such a process can be measured by looking for the point over the
relaxation curve at which the granular temperature equals its steady
value, up to a certain small precision. From a theoretical standpoint,
the relaxation time is bounded from below by the classical speed limit
$\Delta t\geq \calL^{2}/(2\calC)$~\cite{ito_stochastic_2020}, where
$\calL$ and $\calC$ are the integrals over time of $\sqrt{I(t)}$ and
$I(t)$, respectively.

One of the main objectives of this paper is to engineer a protocol to
minimise the connection time between the two NESS. Note that the existence of
non-holonomic constraints impinges on the connecting time: it is not
possible to have an arbitrarily short connecting time since this
leads, in general, to the violation of the constraints~\footnote{This
  is a practical shortcoming of the usual ``shortcut to adiabaticity''
  or ``engineered swift relaxation'' processes. The emergence of
  negative values of the stiffness of the harmonic trap for too fast
  protocols is a well-known issue of such
  transformations~\cite{chupeau_thermal_2018,
    plata_optimal_2019,albay_realization_2020}.}. Therefore, a
non-vanishing minimum connection time emerges associated to a suitable
time-dependent $\chi(t)$ protocol for the driving. To work out the
optimal connection, we leverage Pontryagin's maximum principle, a key
result in control
theory~\cite{pontryagin_mathematical_1987,liberzon_calculus_2012}.

Our work shows the feasibility of beating the---constant
driving---relaxation times, both the empirical one and the theoretical
speed limits, with optimal control.  Specifically, the optimal process
comprises two time windows: one with the largest possible value of the
driving, $\chi=\chi_{\max}$, and the other with no driving at all,
$\chi=0$. In the context of control theory, the kind of processes in
which the control function changes abruptly between its limiting
values are known as bang-bang. Here, we have two bangs---because the
description of our system involves two variables, see below. The order
of the bangs depends on the value of the target granular temperature
$T_{\fin}$ being larger or smaller than the initial one.

In addition, we argue that similar bang-bang processes also minimise
the connecting time for a quite general class of systems. Despite the
non-linear dependence on the relevant physical variables of the
evolution equations, the latter are often linear in
the ``control function(s)''. A few illustrative examples are a
colloidal particle trapped in a harmonic potential (controls:
stiffness of the trap and (or) temperature of the
bath~\cite{martinez_engineered_2016,chupeau_thermal_2018}), active
lattice gases (diffusion
coefficient~\cite{kourbane-houssene_exact_2018}, noise strength and
(or) density~\cite{manacorda_lattice_2017}), and a particle in an
electric field (intensity of electric
field~\cite{martikyan_comparison_2020}). Moreover, in most of these
situations the controls (stiffness, temperature, diffusion
coefficient, noise strength) are non-negative and a non-holonomic
constraint arises. Bang-bang protocols thus emerge as the optimal
ones, because of the linearity of Pontryagin's Hamiltonian in the
control function, with the number of bangs depending on the number of
independent variables.


This manuscript is organised as follows. In Sec.~\ref{sec:model}, we
introduce our model system and write down the evolution equations for
the granular temperature and the excess kurtosis. The characteristic
relaxation times for relaxation at constant driving are analysed in
Sec.~\ref{sec:charac-time}, including the classical speed
limits. Section~\ref{sec:ESR} is devoted to the possibility of
accelerating the connection between two NESS corresponding to
different values of the driving. Therein, we put forward the control
problem for the minimisation of the connection time and show that the
optimal processes are of bang-bang type. The bang-bang processes are
explicitly built in Sec.~\ref{sec:analysis-bang-bang}, and the
associated physical properties over them---minimum connecting time,
length and cost---are derived in
Sec.~\ref{sec:physical-props-bang-bang}. Numerical simulations of the
dynamics are presented and compared with our analytical predictions in
Sec.~\ref{sec:numerics}. The generality of the bang-bang protocols is
investigated in Sec.~\ref{sec:generality-bang-bang}. We illustrate the general situation by briefly
analysing the optimal connection for a colloidal particle trapped in a
three-dimensional harmonic well. Finally,
Sec.~\ref{sec:discussion} discusses the main results of our work,
their implications for a wide class of driven systems, and possible
future developments. The Appendices deal with some technicalities that
complement the main text.

\section{Evolution equations}\label{sec:model}

We consider a uniformly heated granular gas of $d$-dimensional hard
spheres of mass $m$ and diameter $\sigma$, with number density $n$. In
addition to inelastic collisions, with restitution coefficient
$\alpha$, the gas particles are submitted to a white noise force of
variance $m^{2}\xi^{2}$---the so-called stochastic thermostat. In the
low density limit, the dynamics of the system is accurately described
by the Boltzmann-Fokker-Planck
equation~\cite{van_noije_velocity_1998}.

Our analysis is mainly done in the so-called first Sonine
approximation for the kinetic equation. This approach characterises
the gas in terms of the granular temperature $T$ and the excess
kurtosis $a_{2}$. The latter incorporates non-Gaussianities in the
velocity distribution function in the simplest possible way, it is the
first non-trivial cumulant. The Sonine approximation accurately
describes the granular gas in many different
situations~\cite{van_noije_velocity_1998,montanero_computer_2000,garcia_de_soria_universal_2012,trizac_memory_2014,prados_kovacs-like_2014,lasanta_when_2017},
and we employ it here to investigate the classical speed
limits. Nevertheless, at some points of the paper we will make use of
the harsher Gaussian approximation, which---as a rule of thumb---works
when the property being analysed does not vanish.  Non-Gaussianities,
i.e. the excess kurtosis in the Sonine approximation, only introduce
corrections to the predicted behaviour in such a case.

We start by defining the granular temperature $T$ and the excess
kurtosis $a_{2}$,
\begin{equation}
  \label{eq:T-a2-def}
  T\equiv \frac{m \mean{v^{2}}}{d}, \qquad a_{2}\equiv\frac{d}{d+2}\frac{\mean{v^{4}}}{\mean{v^{2}}^{2}}-1.
\end{equation}
Higher-order cumulants are neglected, which makes it possible to get a
closed set of equations for $T$ and $a_{2}$. In addition,
non-linearities in $a_{2}$ are dropped, because the typical values of
the excess kurtosis are quite
small.

In the long time limit, the granular gas reaches a NESS. Therein,
energy loss from collisions is compensated, in average, by the energy
input from the stochastic thermostat. The stationary value of the
temperature and the excess kurtosis are given
by~\cite{van_noije_velocity_1998,montanero_computer_2000}
\begin{subequations}
\begin{equation}\label{eq:Ts} T_\st^{3/2}=\frac{m\xi^2}
{\zeta_0(1+\frac{3}{16}a_2^\st)}\equiv\chi, \quad \zeta_0=\frac{2 n
\sigma^{d-1} \left(1-\alpha^2\right) \pi^{\frac{d-1}{2}}}{\sqrt{m}
d\,\Gamma(d/2)},
\end{equation}
\begin{equation}\label{eq:a2s}
  a_{2}^{\st}=\frac{16(1-\alpha)(1-2\alpha^2)}
{73+56d-24d\alpha-105\alpha+30(1-\alpha)\alpha^2}.
\end{equation}
\end{subequations}

The temperature and the excess kurtosis obey the evolution equations
\begin{subequations}\label{eq:evol-with-dim}
\begin{align}
  \dot{T}&=\zeta_{0}\left[\chi\left(1+\frac{3}{16}a_{2}^{\st}\right)-T^{3/2}
  \left(1+\frac{3}{16}a_{2}\right)\right], \\
  \dot{a_{2}}&=\frac{2\zeta_{0}}{T}\left[\left(T^{3/2}-\chi\right)a_{2}+B
  \left(a_{2}^{\st}-a_{2}\right)\right],
\end{align}
\end{subequations}
which are non-linear in the temperature but linear in $a_{2}$, as a
consequence of the Sonine approximation. The parameter $B$ is only a
function of $\alpha$ and $d$, namely
\begin{align}\label{eq:B}
  B=\frac{a_{2}^{\hcs}}{a_{2}^{\hcs}-a_{2}^{\st}},
\end{align}
where
\begin{align}
  a_{2}^{\hcs}&=\frac{16(1-\alpha)(1-2\alpha^{2})} {25
+2\alpha^{2}(\alpha-1)+ 24d+\alpha(8d-57)},
\end{align}
is the value of the excess kurtosis in the homogeneous cooling
state---the long-time time-dependent state that the system tends to
approach when cools freely, i.e. with no driving.

Before proceeding
further, we introduce dimensionless variables by taking adequate units
for $T$,
$\chi$, and $t$:
\begin{equation}
  \label{eq:nondim}
  T^{*}=T/T_{\ini}, \quad \chi^{*}=\chi/T_{\ini}^{3/2}, \quad
  t^{*}=\zeta_{0}T_{\ini}^{1/2}t,
\end{equation}
where $T_{\ini}$ is the initial value of the
temperature. Consistently, velocities are made dimensionless with
$\sqrt{T_{\ini}/m}$, $\bm{v}^{*}=\sqrt{m/T_{\ini}}\bm{v}$. The excess
kurtosis is already dimensionless, but for our purposes it is convenient to define the
scaled variable
\begin{equation}\label{eq:A2-def}
  A_{2}\equiv a_{2}/a_{2}^{\st}.
\end{equation}
We see in what follows that $A_{2}$ is
basically non-negative, whereas $a_{2}$ changes sign with the
inelasticity (specifically, $a_{2}^{\st}=0$ for
$\alpha=1/\sqrt{2}$).

In the remainder of the paper, we always work with dimensionless
variables---therefore, we drop the asterisks not to clutter our
formulae. The corresponding evolution equations can be written as
\begin{subequations}\label{eq:evol-non-dim}
  \begin{align}
    \dot{T}=&f_{1}(T,A_{2};\chi),  \label{eq:evol-non-dim-T}\\ f_{1}(T,A_{2};\chi)\equiv&
\chi\left(1+\frac{3}{16}a_{2}^{\st}\right)- T^{3/2}
                                                                                          \left(1+\frac{3}{16}a_{2}^{\st}A_{2}\right), \label{eq:f1-def}\\
\dot{A_{2}}=&f_{2}(T,A_{2};\chi), \label{eq:evol-non-dim-a2}\\
f_{2}(T,A_{2};\chi)\equiv
&\frac{2}{T}\left[\left(T^{3/2}-\chi\right)A_{2}+B\, T^{3/2}
\left(1-A_{2}\right)\right]. \label{eq:f2-def}
\end{align}
\end{subequations}
Apart from a factor $d$, $T$ is basically the (dimensionless) energy
per particle. Therefore, the first term in $f_{1}$,
$\chi(1+\frac{3}{16}a_{2}^{\st})$ is the rate of energy input from the
stochastic thermostat, while the second term,
$-T^{3/2}(1+\frac{3}{16}a_{2}^{\st}A_{2})$, is the rate of energy
dissipation in collisions. Equations~\eqref{eq:evol-non-dim} must be
supplemented with suitable initial conditions. With our choice of
units, the initial temperature equals unity. Since we are interested
in processes that start from the NESS corresponding to the initial
temperature,
\begin{equation}\label{eq:initial-conditions}
  T(t=0)=A_{2}(t=0)=1.
\end{equation}

\section{Characteristic relaxation time}\label{sec:charac-time}

Initially, our granular fluid is in the NESS corresponding to
$\chi_{\ini}=1$. A typical relaxation process is constructed by
suddenly changing the noise intensity from $\chi_{\ini}=1$ to a
different value $\chi_{\fin}$ at $t=0$. Then, the system relaxes to a
new NESS with granular temperature $T_{\fin}$ corresponding to the
noise intensity $\chi_{\fin}\equiv T_{\fin}^{3/2}$. Note that the stationary
value of the excess kurtosis $a_{2}^{\st}$ is independent of the noise
intensity and so is $A_{2}^{\st}$, namely $A_{2}^{\st}=1$. Relaxation
in this process has a certain characteristic time $t_{R}$, at which
the temperature has almost completely reached---complete relaxation
only happens for infinite time---its steady state value.

To characterise the relaxation time from an empirical point of view,
we define the relaxation function of the temperature as
$\phi(t)=(T(t)-T_{\fin})/(1-T_{\fin})$, such that $\phi(t=0)=1$ and
$\phi(t\to\infty)=0$. The granular temperature has almost relaxed to
$T_{\fin}$ when $\phi(t_{R})=\epsilon\ll 1$, i.e. for a temperature
$T_{R}(T_{\fin},\epsilon)=T_{\fin}+\epsilon(1-T_{\fin})$. We consider
$\epsilon=10^{-4}$ for the sake of concreteness. This relaxation time
$t_{R}$ can be estimated by numerically solving the system of
equations~\eqref{eq:evol-non-dim}.

\begin{figure}
  \centering
  \includegraphics[width=3.375in]{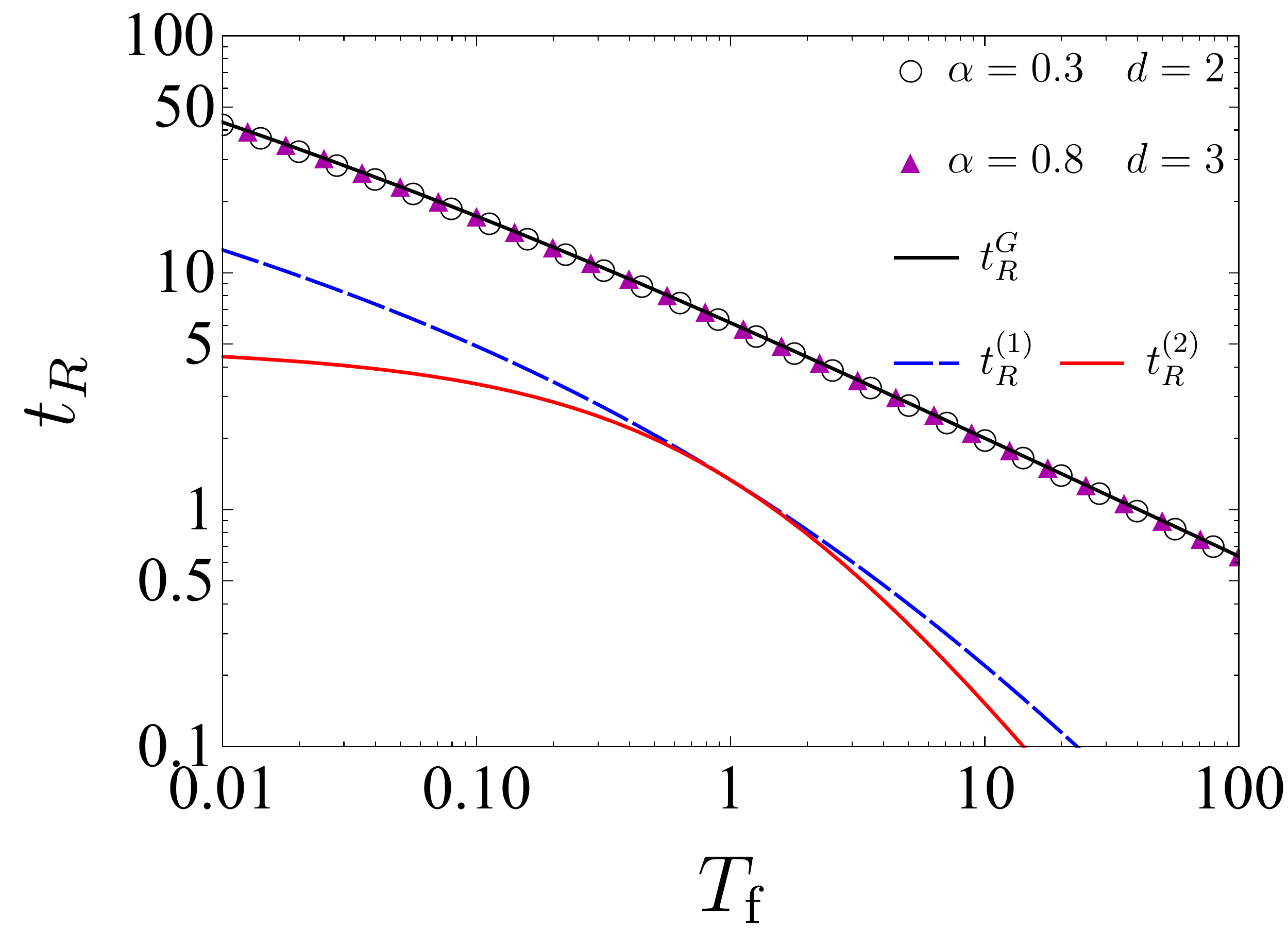}
  \caption{Characteristic relaxation time as a function of the target
    temperature. The numerical value of $t_{R}$ (symbols) is obtained
    by integrating the system of equations~\eqref{eq:evol-non-dim}
    numerically for the considered pair of parameters
    $(\alpha,d)$. Note that $t_{R}$ depends very weakly on
    $(\alpha,d)$ and is very well predicted by the Gaussian
    approximation $t_{R}^{G}$ (black solid line), as given by
    Eq.~\eqref{eq:tRG}. Also plotted are the speed limits
    $t_{R}^{(1)}$ (blue broken) and $t_{R}^{(2)}$ (red solid) for the
    relaxation process, as given by Eq.~\eqref{eq:tR1-tR2}, for
    $d=2$. }
  \label{fig:tR}
\end{figure}

Figure~\ref{fig:tR} shows $t_{R}$ as a function of the final
temperature $T_{\fin}$ for a couple of values of $(\alpha,d)$, namely
$(0.3,2)$ (circles) and $(0.8,3)$ (open triangles). Other $(\alpha,d)$
pairs are not shown because all the curves would be basically
superimposed. Therefore, the ``natural'' time scale for the relaxation
of the granular temperature to its final value $T_{\fin}$ is basically
independent of $\alpha$ and $d$ in our dimensionless time scale
defined in Eq.~\eqref{eq:nondim}~\footnote{We have made time
  dimensionless with $\zeta_{0}$, which depends on $d$ and is
  proportional to $(1-\alpha)^{2}$.}. It is also observed that $t_{R}$
is a decreasing function of the final temperature and vanishes in the
limit as $T_{\fin}\to\infty$.

The weak dependence of $t_{R}$ on $(\alpha,d)$ suggests that it can be
quite accurately predicted by the Gaussian approximation, in which the
excess kurtosis is set to zero in Eq.~\eqref{eq:evol-non-dim}. This yields
\begin{align}\label{eq:tRG}
  t_{R}^{G}(T_{\fin},\epsilon)=\!\!\int_{1}^{T_{R}}\!\!
  \frac{dT}{T_{\fin}^{\frac 3 2}-T^{\frac 3 2}}=\!
  \frac{\Omega\left(\!\sqrt{\frac{T_R}{T_{\fin}}}\right)-
  \Omega\left(\!\sqrt{\frac{1}{T_{\fin}}}\right)}{3\sqrt{T_{\fin}}},
\end{align}
where $\Omega$ is given by~\cite{van_noije_randomly_1999}
\begin{align}
  \Omega(x)=\ln\frac{1+x+x^{2}}{\left|1-x \right|^{2}} -
  2\sqrt{3}\arctan\left( \frac{1+2x}{\sqrt{3}}\right).
\end{align}
Figure~\ref{fig:tR} also shows $t_{R}^{G}$ as a function of the final
temperature $T_{\fin}$ (solid line). The agreement with the numerical
estimates for $t_{R}$ is excellent over the whole range of
temperatures considered, which covers four orders of magnitude,
$0.01\leq T_{\fin}\leq 100$.  Eq.~\eqref{eq:tRG} entails that
$t_{R}^{G}$ vanishes algebraically in the high temperature limit
$T_{\fin}\gg 1$, specifically
\begin{equation}\label{eq:tRG-highT}
  t_{R}^{G}\sim \frac{2|\ln\epsilon|}{3}T_{\fin}^{-1/2}, \quad
  T_{\fin}\gg 1.
\end{equation}

So far, we have characterised the relaxation time from an empirical
standpoint. Henceforth, we consider the classical speed limits  that have been recently proposed in the
literature~\cite{shanahan_quantum_2018,okuyama_quantum_2018,
  ito_stochastic_2018,
  shiraishi_speed_2018,ito_stochastic_2020,
  nicholson_timeinformation_2020}. Specifically,
we analyse those in Ref.~\cite{ito_stochastic_2020} within the
framework of information geometry, which are valid for a general
dynamics---not necessarily Markovian.

We denote the one-particle PDF for the velocity by $P(\bm{v},t)$. The
Fisher information is defined as
\begin{equation}
  \label{eq:Fisher-inf-def}
  I(t)\equiv \int
  d\bm{v}\frac{\left(\partial_{t}P(\bm{v},t)\right)^{2}}{P(\bm{v,t})}
  =\langle \left(\partial_{t}\ln P(\bm{v},t)\right)^{2}\rangle\geq 0
\end{equation}
and plays a central role in information
geometry~\cite{amari_information_2016}.  Therefrom, the statistical
length is introduced as~\cite{wootters_statistical_1981,crooks_measuring_2007}
\begin{equation}
  \label{eq:stat-length}
  \calL=\int_{0}^{t_{\fin}}dt\, \sqrt{I(t)},
\end{equation}
which represents the distance swept by the probability distribution in
the time interval $(0,t_{\fin})$.  Since the probability distribution
is normalised for all times, $P(\bm{v},t)$ moves on the unit
sphere. As a result, the statistical length $\calL$ is bounded from
below by the arc-length between $P_{\ini}(\bm{v})\equiv P(\bm{v},0)$
and $P_{\fin}(\bm{v})\equiv P(\bm{v},t_{\fin})$, i.e. the so-called
Bhattacharyya
angle~\cite{wootters_statistical_1981,ito_stochastic_2020}
\begin{equation}
  \label{eq:bhatta-angle}
  \Lambda=2\arccos\left(\int
    d\bm{v}\sqrt{P_{\ini}(\bm{v})P_{\fin}(\bm{v})}\right), \quad
  \calL\geq\Lambda.
\end{equation}
The equality $\calL=\Lambda$ is attained only over the geodesic in
probability space, along which the Fisher information $I(t)$ remains
constant---e.g. see Appendix E of Ref.~\cite{ito_stochastic_2020} for
details.

It has recently been proved that the
Cauchy-Schwartz inequality leads to the classical speed limits
\begin{equation}
  \label{eq:CSL-general}
  t_{\fin} \geq \frac{\calL^{2}}{2\calC}\geq \frac{\Lambda^{2}}{2\calC},
\end{equation}
where $t_{\fin}$ and 
\begin{equation}
  \label{eq:deltat-cost-def}
  \calC\equiv \frac{1}{2}\int_{0}^{t_{\fin}}dt\, I(t)
\end{equation}
are the operation time and the cost of the process,
respectively~\cite{ito_stochastic_2020}. Eq.~\eqref{eq:CSL-general}
expresses a trade-off between time and cost operation,
$2 t_{\fin} \,\calC\geq \calL^{2}\geq \Lambda^{2}$. The bound provided
by $\calL$ is tighter but, in general, depends on the whole dynamical
evolution, whereas $\Lambda$ only depends on the initial and final
distributions.

For our system, the speed limits above can be exactly calculated
within the Gaussian approximation. Therein,
$I(t)=I_{G}(t)=d/2(\dot{T}/T)^{2}$, $T$ is a monotonic function of
time, and both bounds are completely determined by $T_{\fin}$---as
detailed in Appendix~\ref{sec:inform-geom}. With the definitions
\begin{equation}
  \label{eq:phi-def}
  \gamma(T)\equiv\left(\frac{2 \sqrt{T}}{1+T}\right)^{d/2},
  \qquad \varphi(T)\equiv T^{3/2}-3T^{1/2}+2, 
\end{equation}
we have that~\footnote{Not only does $\calL$ but also $\calC$ depend
  on the specific protocol followed to connect the initial and final
  states. This is explicitly taken into account in our notation by
  writing $\calL^{\rel}$ and $\calC^{\rel}$ for the relaxation
  process.}
\begin{equation}\label{eq:L-and-C-Gaussian}
  \Lambda_{G}=2 \arccos\gamma(T_{\fin}), \;
  \calL_{G}^{\rel}=\sqrt{\frac{d}{2}}\left| \ln T_{\fin}\right|, \; 
  \calC_{G}^{\rel}=\frac{d}{4}\varphi(T_{\fin}). 
\end{equation}
Making use of these expressions, the connecting time $t_{\fin}^{\rel}$
in a relaxation process verifies the inequality
\begin{equation}
  \label{eq:CSL-relax}
  t_{\fin}^{\rel}\geq t_{R}^{(1)}\geq t_{R}^{(2)},
\end{equation}
where
\begin{equation}
  \label{eq:tR1-tR2}
  t_{R}^{(1)}=\frac{|\ln T_{\fin}|^{2}}{\varphi(T_{\fin})},
  \qquad
  t_{R}^{(2)}=
  \frac{8\left[\arccos\gamma(T_{\fin})\right]^{2}}{d\,\varphi(T_{\fin})}.
\end{equation}
It is the geodesic in probability space that the smallest bound
$t_{R}^{(2)}$ corresponds to. A relevant question is thus the
attainability of the geodesic for the granular fluid: we
discuss this issue in Appendix~\ref{sec:inform-geom}.

Both $t_{R}^{(1)}$ and $t_{R}^{(2)}$ are shown in Fig.~\ref{fig:tR}
for the two-dimensional case. Consistently, we have that the Gaussian
estimate $t_{R}^{G}$ for the relaxation time lies above both of them,
specifically $t_{R}^{G}/t_{R}^{(1)}$ changes
from---approximately---$4$ to $30$ across the range
$0.01\leq T_{\fin}\leq 100$~\footnote{It should be remarked that the
  value of the empirical relaxation time depends on the specific value
  chosen for $\epsilon$.  For instance, taking $\epsilon=10^{-2}$
  instead of $10^{-4}$ makes the empirical relaxation time roughly
  one-half of the one plotted in
  Fig.~\ref{fig:tR}.}. Non-Gaussianities in the velocity distribution
function will affect the speed limits $t_{R}^{(1)}$ and
$t_{R}^{(2)}$. However, we expect the smallness of $a_{2}$ to
introduce only slight changes to the results above, as was the case of
the empirical relaxation time.

\section{Engineered swift relaxation}\label{sec:ESR}

Our idea is engineering a protocol, by controlling the noise intensity
$\chi(t)$, that connects the initial and final NESS---the ones
corresponding to $\chi_{\ini}=1$ and $\chi_{\fin}=T_{\fin}^{3/2}$---in
a given time $t_{\fin}$, as short as possible. A relevant question
thus arises: whether or not it is possible to beat the characteristic
relaxation time of the system---not only $t_{R}^{G}$ but also the
classical speed limits $t_{R}^{(1)}$ and $t_{R}^{(2)}$ for the
relaxation process. Note that the latter is possible only for time
dependent driving.

In order to connect the two
NESS, the solution to Eq.~\eqref{eq:evol-non-dim} must verify the
initial conditions~\eqref{eq:initial-conditions} and also
\begin{equation}
  \label{eq:boundary-conditions-tf}
  T(t=t_{\fin})=T_{\fin}, \quad A_{2}(t=t_{\fin })=1.
\end{equation}
Therefore, Eqs.~\eqref{eq:initial-conditions} and
\eqref{eq:boundary-conditions-tf} constitute the boundary conditions
for our \textit{Engineered Swift Relaxation} (ESR) protocol. If a
solution to Eq.~\eqref{eq:evol-non-dim} satisfies these boundary
conditions and the control function $\chi(t)$ is such that
$\chi(t=0)=1$, $\chi(t=t_{\fin})=T_{\fin}^{3/2}$, the
system is really stationary at both the initial and final time, i.e.
$\dot{T}(t=0)=\dot{T}(t=t_{\fin})=0$ and
$\dot{A_{2}}(t=0)=\dot{A_{2}}(t=t_{\fin})=0$.

First, we show that it is indeed possible to connect the two NESS in a
finite time, by a reverse-engineering procedure~\footnote{The idea is
  similar to that employed in
  Refs.~\cite{martinez_engineered_2016,plata_finite-time_2020} for
  connecting two equilibrium states.}. We start from a certain
function (protocol) $T_{p}(t)$ that connects the initial and final
values of the temperature and, in addition, is stationary at both
$t=0$ and $t=t_{\fin}$, i.e.
\begin{equation}\label{eq:Tp-conditions}
  T_{p}(0)=1, \quad T_{p}(t_{\fin})=T_{\fin}, \quad
  \dot{T}_{p}(t=0)=\dot{T}_{p}(t=t_{\fin})=0.
\end{equation}
We aim at finding a driving $\chi_{p}(t)$ and a time evolution for the
scaled kurtosis $A_{2p}(t)$, such that (i) $(T_{p}(t),A_{2p}(t))$ is a
solution to Eq.~\eqref{eq:evol-non-dim} for the driving $\chi_{p}(t)$,
(ii) the boundary conditions for $A_{2}(t)$ are verified,
$A_{2p}(0)=A_{2p}(t_{\fin})=1$, and (iii) the driving
verifies the boundary conditions $\chi(0)=1$,
$\chi(t_{\fin})=T_{\fin}^{3/2}$---which ensure stationarity at both
$t=0$ and $t=t_{\fin}$.

Now, we employ Eqs.~\eqref{eq:evol-non-dim-T} and
\eqref{eq:f1-def} to write the driving in terms of
$(T_{p}(t),A_{2p}(t))$,
\begin{equation}
  \label{eq:chi-p}
  \chi_{p}(t)=\frac{\dot{T_{p}}(t)+
    \left[T_{p}(t)\right]^{3/2}
    \left[1+\frac{3}{16}a_{2}^{\st}A_{2p}(t)\right]}{1+\frac{3}{16}a_{2}^{\st}}.
\end{equation}
Since we do not know $A_{2p}(t)$---yet, $\chi_{p}(t)$ is not
completely determined at this point. However, insertion of
Eq.~\eqref{eq:chi-p} into \eqref{eq:evol-non-dim-a2} and
\eqref{eq:f1-def} gives us a closed equation for $A_{2p}(t)$,
which we can solve with the initial condition
$A_{2p}(0)=1$. Therefore, we need one free parameter---to be included
in our choice for $T_{p}(t)$---to ``tune'' $A_{2p}(t)$ to verify
$A_{2p}(t_{\fin})=1$. Eqs.~\eqref{eq:Tp-conditions} and
\eqref{eq:chi-p} ensure that $\chi_{p}(0)=1$,
$\chi_{p}(t_{\fin})=T_{\fin}^{3/2}$ in such a case: the solution found
in this way is indeed stationary at the initial and final times and an
ESR protocol has been successfully constructed. We show how to build a
simple polynomial connection in Appendix~\ref{sec:simple-ESR}.

\subsection{The control problem}\label{sec:control}

Let us consider the ESR connection problem from the following point of
view. For a given---in general time-dependent---choice of the driving
intensity $\chi(t)$, the system of ODEs
\eqref{eq:evol-non-dim} predicts the corresponding time
evolutions for the granular temperature $T$ and the excess kurtosis
$A_{2}$. Therefore, $\chi(t)$ plays the role of a control function.

We restrict ourselves to a certain set of admissible control
functions, specifically those that make it possible to connect the
two NESS in a certain time $t_{\fin}$,
\begin{equation}
  \label{eq:bc-T-a2}
  T(0)=1, \; T(t_{\fin})=T_{\fin}, \quad A_{2}(0)=A_{2}(t_{\fin})=1,
\end{equation}
and ensure stationarity at the initial and final times, i.e.
\begin{equation}
  \label{eq:bc-chi}
  \chi(0)=1, \quad \chi(t_{\fin})=T_{\fin}^{3/2}.
\end{equation}
The control function $\chi(t)$ is assumed to be piecewise continuous
in the time interval $[0,t_{\fin}]$. The presence of finite
jumps in $\chi(t)$ is not problematic from a physical point of view:
already in the ``basic'' relaxation process $\chi$ jumps from $1$ to
$\chi_{\fin}=T_{\fin}^{3/2}$ at $t=0$, and $T$ and $A_{2}$
are always continuous functions of time.

Above, we have shown that there exist control functions $\chi(t)$ that
do the job, at least for not too short connecting times---see also
Appendices~\ref{sec:geodesic-Gaussian} and \ref{sec:simple-ESR}. Here,
we would like to consider the problem in the light of optimal control
theory: our control verifies the inequality $\chi(t)\geq 0$ and thus
the possible optimisation problems, such as minimising the connection
time, have a non-holonomic constraint. Therefore, we leverage
Pontryagin's maximum
principle~\cite{pontryagin_mathematical_1987,liberzon_calculus_2012}
to solve the optimisation problem and find the optimal control
$\chi(t)$ for the corresponding physical situation. For the sake of
mathematical rigour, we also consider that the noise intensity is
bounded from above, $\chi(t)\leq \chi_{\max}$; afterwards we will take
the limit $\chi_{\max}\to\infty$.

\subsection{Optimising the connection
}

Let us consider the following optimisation problem: we want to obtain
the minimum time for making the connection between the two NESS,
i.e. we want to minimise $t_{\fin}=\int_{0}^{t_{\fin}}dt$.  In order
to apply Pontryagin's procedure, we define a variable $y_{0}(t)$ such
that the optimisation problem is equivalent to the minimisation of
$y_{0}(t_{\fin})$, i.e.
\begin{equation}\label{eq:y0-def}
  \dot{y}_{0}=f_{0}(T,A_{2},\chi), \quad f_{0}(T,A_{2},\chi)\equiv 1,
  \quad y_{0}(t_{\fin})=t_{\fin}.
\end{equation}

To proceed, we introduce Pontryagin's Hamiltonian
\begin{equation}\label{eq:Pontry-Hamilt}
  \Pi(\bm{y},\bm{\psi},\chi)\equiv
  \psi_{0}f_{0}(\bm{y},\chi)+\psi_{1}f_{1}(\bm{y},\chi)+
  \psi_{2}f_{2}(\bm{y},\chi).
\end{equation}
In this context, we employ the notation $y_{1}\equiv T$,
$y_{2}\equiv A_{2}$, $\bm{y}\equiv(y_{0},y_{1},y_{2})$,
$\bm{\psi}\equiv(\psi_{0},\psi_{1},\psi_{2})$ to simplify some
formulas. The variables $y_{i}$ and their conjugate momenta
$\psi_{i}$, $i=0,1,2$, evolve following the Hamiltonian system
\begin{equation}
  \label{eq:Hamilt-sys}
  \dot{y_{i}}=\frac{\partial\Pi}{\partial\psi_{i}}, \quad
  \dot{\psi_{i}}=-\frac{\partial\Pi}{\partial
    y_{i}}=-\sum_{j=0}^{2}\psi_{j}\frac{\partial f_{j}}{\partial y_{i}}.
\end{equation}
From the construction above, the functions $f_{j}$ do not depend on
$y_{0}$ and thus $\dot{\psi_{0}}=0$, $\psi_{0}=\text{const}$.

Pontryagin's maximum principle states necessary conditions for optimal
connection: in order that $(\chi^{*}(t),T^{*}(t),A_{2}^{*}(t))$ be
optimal, it is necessary that there exists a non-zero continuous
vector function
$\bm{\psi}^{*}(t)=(\psi_{0}^{*}(t),\psi_{1}^{*}(t),\psi_{2}^{*}(t))$
corresponding to  $(\chi^{*}(t),T^{*}(t),A_{2}^{*}(t))$ such that for all $t$, $0\leq t\leq t_{\fin}$, (i)
the canonical system \eqref{eq:Hamilt-sys} holds, (ii) if we define
the supremum of $\Pi$ as function of the control,
$H(\bm{y},\bm{\psi})=\sup_{\chi}\Pi(\bm{y},\bm{\psi},\chi)$, we have that
\begin{equation}
  \label{eq:Pontry-condit-sup}
  H(\bm{y}^{*}(t),\bm{\psi}^{*}(t))=\Pi(\bm{y}^{*}(t),
  \bm{\psi}^{*}(t),\chi^{*}(t)),
\end{equation}
and (iii) the two constants of motion $\psi_{0}^{*}$ and
$H^{*}\equiv H(\bm{y}^{*}(t),\bm{\psi}^{*}(t)) $ satisfy
$\psi_{0}^{*}\leq 0$ and $H^{*}=0$.

To find the supremum of $\Pi$ with respect to $\chi$, we
calculate $\partial\Pi/\partial\chi$: either $\chi^{*}$  follows from the
condition $\partial\Pi/\partial\chi|_{\chi^{*}}=0$ or lies at the
boundaries of the interval $[0,\chi_{\max}]$. Making use of
Eqs.~\eqref{eq:f1-def}, \eqref{eq:f2-def}, \eqref{eq:y0-def}, and
\eqref{eq:Pontry-Hamilt}, we obtain
\begin{equation}\label{eq:der-Pi-resp-chi}
  \frac{\partial\Pi}{\partial\chi}=\psi_{1}\left(1+
    \frac{3}{16}a_{2}^{\st}\right)-2\psi_{2}\frac{A_{2}}{T},
\end{equation}
which does not depend on $\chi$ and thus does not allow for finding
$\chi^{*}$. This is a consequence of $\Pi$ being a linear function of
$\chi$ and therefore either $\chi^{*}=0$ or $\chi^{*}=\chi_{\max}$,
depending on the sign of $\partial\Pi/\partial\chi$. The optimal
control jumps from $\chi^{*}=0$ to $\chi^{*}=\chi_{\max}$ at those
times for which $\partial\Pi/\partial\chi$ changes from negative to
positive, and vice versa.  This kind of discontinuous optimal controls
are commonly known as
bang-bang~\cite{liberzon_calculus_2012,chen_fast_2010,martikyan_comparison_2020,ding_smooth_2020}.

The simplest situation is thus a two-step bang-bang process, with two
possibilities: (i) high driving window $\chi^{*}(t)=\chi_{\max}$,
$0\leq t\leq t_{J}$, followed by free cooling $\chi^{*}(t)=0$,
$t_{J}\leq t\leq t_{\fin}$, and (ii) first free cooling
$\chi^{*}(t)=0$, $0\leq t\leq t_{J}$, followed by high driving
$\chi^{*}(t)=\chi_{\max}$, $t_{J}\leq t\leq t_{\fin}$. From our study
of the polynomial connection, we may guess that (i) is the optimal
protocol for $T_{\fin}>1$, but this ansatz has to be checked.

\section{Bang-bang optimal controls}\label{sec:analysis-bang-bang}

In this section we carry out an in-depth study of the bang-bang
controls we have just described above. For the sake of simplicity, we
explicitly build such protocols for the case
$\chi_{\max}\gg 1$~\footnote{The existence of the restriction
  $\chi\leq\chi_{\max}$ is of practical nature, we assume that the
  intensity of the heat bath cannot be arbitrarily large, whereas the
  constraint $\chi>0$ is of fundamental nature: in average, the
  stochastic forcing always increases the kinetic energy of the
  particles.}. 

In general, we focus on the motion of point
describing the state of the system in the \textit{phase space} plane
$(A_{2},T)$: Eq.~\eqref{eq:evol-non-dim} is a system of first-order
ODEs and trajectories in the phase space plane cannot
intersect. Making use of them, we arrive at
\begin{equation}
  \label{T-A2-evol}
  \frac{2}{T}\frac{dT}{dA_{2}}=
  \frac{\chi\left(1+\frac{3}{16}a_{2}^{\st}\right)-
    T^{3/2}\left(1+\frac{3}{16}a_{2}^{\st}A_{2}\right)}
 {\left(T^{3/2}-\chi\right)A_{2}+B T^{3/2} \left(1-A_{2}\right)}
\end{equation}

\subsection{Heating-cooling bang-bang}\label{sec:h-c-bangbang}

Here, we analyse the bang-bang process in which the granular fluid is
first heated, $\chi(t)=\chi_{\max}\gg 1$, $0\leq t\leq t_{J}$, and
afterwards freely cools, $\chi(t)=0$, $t_{J}\leq t\leq t_{\fin}$. 
Taking the limit $\chi_{\max}\gg 1$ in Eq.~\eqref{T-A2-evol} and
solving the resulting separable ODE with initial condition $(A_{2,\ini},T_{\ini})=(1,1)$ in
the $(A_{2},T)$ plane, we get
\begin{equation}\label{eq:T-A2-evol-chimax-hc}
T^{2}A_{2}^{1+\frac{3}{16}a_{2}^{\st}}=T_{\ini}^{2}A_{2,\ini}^{1+\frac{3}{16}a_{2}^{\st}}=1, \quad 0\leq
t\leq t_{J}.
\end{equation}

Now we investigate the behaviour of the system in the second time
window, $t_{J}\leq t\leq t_{\fin}$. Putting $\chi=0$ in
Eq.~\eqref{T-A2-evol} and taking into account Eq.~\eqref{eq:B}, we
arrive again at a separable first order ODE, the solution of which is
given by
\begin{align}
2\ln&\left(\frac{T}{T_{\fin}}\right)=\frac{3}{16}a_{2}^{\st}
(A_{2}^{\hcs}-1)(A_{2}-1) \nonumber \\ &+
\left(1+\frac{3}{16}a_{2}^{\st}A_{2}^{\hcs}\right)
(A_{2}^{\hcs}-1)\ln\left(\frac{A_{2}^{\hcs}-A_{2}}{A_{2}^{\hcs}-1}\right)
, \nonumber \\ & \qquad t_{J}\leq t\leq t_{\fin}.
                 \label{eq:T-A2-sol-chi0-hc}
\end{align}
For the final time, $t=t_{\fin}$, we have that $T=T_{\fin}$ and
$A_{2}=A_{2\fin}=1$. We obtain a relation between $T_{J}$ and
$T_{\fin}$ by particularising Eq.~\eqref{eq:T-A2-sol-chi0-hc} for the
joining time $t=t_{J}$, specifically
\begin{align}
  2\ln&\left(\frac{T_{\fin}}{T_{J}}\right)=\frac{3}{16}a_{2}^{\st}
(A_{2}^{\hcs}-1)(1-A_{2J}) \nonumber \\ &+
\left(1+\frac{3}{16}a_{2}^{\st}A_{2}^{\hcs}\right)
(A_{2}^{\hcs}-1)\ln\left(\frac{A_{2}^{\hcs}-1}{A_{2}^{\hcs}-A_{2J}}\right)
                                          .
                                          \label{eq:Tf-TJ-relation}
\end{align}
In turn, $T_{J}$ and $A_{2J}$ are related by
\begin{equation}\label{eq:TJ-A2J}
  T_{J}^{2}A_{2J}^{1+\frac{3}{16}a_{2}^{\st}}=1,
\end{equation}
as implied by Eq.~\eqref{eq:T-A2-evol-chimax-hc}. As a consequence,
Eq.~\eqref{eq:Tf-TJ-relation} gives a one to one relation between
$T_{\fin}$ and $T_{J}$---or $T_{\fin}$ and $A_{2J}$~\footnote{In the
  first part of the bang-bang process, the system heats with
  $\chi_{\max}$ and thus $\dot{T}\geq 0$ and $T_{J}\geq 1$, which
  entails that $A_{2J}\leq1$. In the limit as $T_{J}\to\infty$, the
  velocity distribution becomes Gaussian at the joining time,
  $A_{2J}\to 0$.}.

\begin{figure}
  \centering
   \includegraphics[width=3.375in]{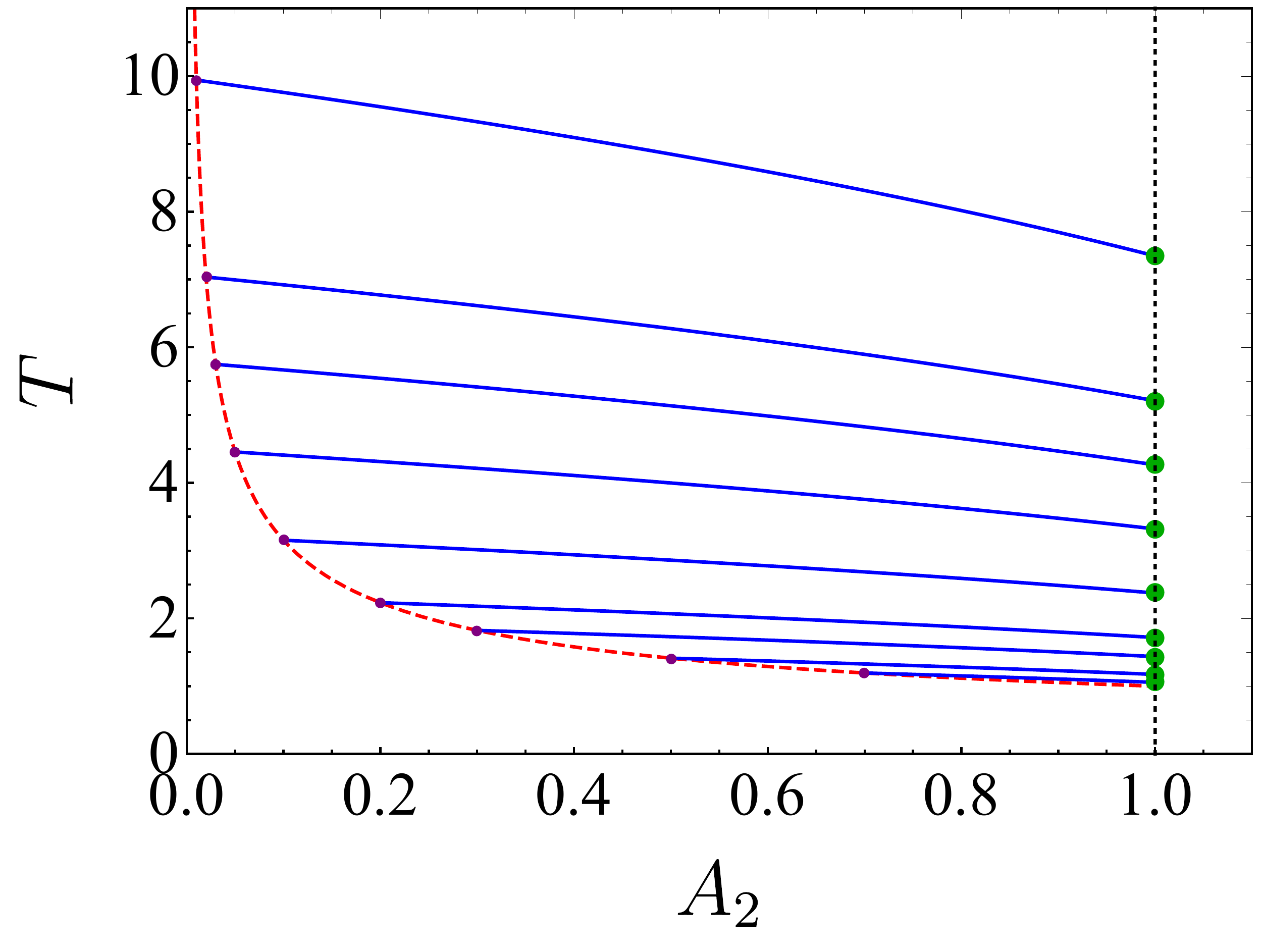}
   \caption{Bang-bang protocol for $T_{\fin}>1$. A representative
     example of the motion of the granular gas in the $(A_{2},T)$
     plane is shown: specifically, we have considered $\alpha=0.8$ and
     $d=2$. Other values of $(\alpha,d)$ lead to a completely
     analogous picture. The bang-bang process connects the initial
     NESS with $(A_{2\ini}=1,T_{\ini}=1)$ with the final state
     $(A_{2\fin}=1,T_{\fin}>1)$ and comprises two parts: first heating
     (red dashed line) followed by cooling (solid blue
     lines). Different target points $(A_{2\fin}=1,T_{\fin}>1)$ over
     the vertical line $A_{2}=1$ (dotted) are reached by starting
     the cooling part from different points $(A_{2J},T_{J})$ over the
     heating curve.}
  \label{fig:hc-bangbang}
\end{figure}

A qualitative plot of the motion of the system in the $(A_{2},T)$
plane is shown in Fig.~\ref{fig:hc-bangbang}. In the first part of the
protocol, $0\leq t\leq t_{J}$, the system is heated with
$\chi(t)=\chi_{\max}$ and follows
Eq.~\eqref{eq:T-A2-evol-chimax-hc}. The second part of the bang-bang
process starts at a given point $(A_{2J},T_{J})$ over this
line. Therefrom, for $t_{J}<t<t_{\fin}$, the system freely cools with
$\chi(t)=0$ and thus follows Eq.~\eqref{eq:T-A2-sol-chi0-hc}. This
part of the bang-bang finishes when the system hits the vertical line
$A_{2}=1$ at the corresponding target point
$(A_{2\fin}=1,T_{\fin})$. In order to keep the system stationary for
$t\leq 0$ and $t\geq t_{\fin}$, the control function has sudden jumps
at these points, $\chi(t\leq 0)=1$, $\chi (t)=\chi_{\max}\gg 1$ for
$0<t\leq t_{J}$, $\chi(t)=0$ for $t_{J}<t<t_{\fin}$,
$\chi(t\geq t_{\fin})=T_{\fin}^{3/2}$.

Note that with this order of the bangs, the bang-bang protocol always
leads the system to a final NESS with $T_{\fin}>1$. The impossibility
of reaching $T_{\fin}<1$ can be physically understood in the following
way: in the first part of the bang-bang process, the system always
heats, $T_{J}>1$, and the corresponding excess kurtosis decreases in
absolute value---the velocity distribution function becomes closer to
a Gaussian, $A_{2J}<1$. Therefore, the initial slope, i.e. at the point
$(A_{2J},T_{J})$, of the curve for the second part of the bang-bang
process (blue solid in Fig.~\ref{fig:hc-bangbang}) is always
larger than the slope of the curve for the first part (red dashed) at the same
point. This can be shown by inspecting the corresponding expressions
for $dT/dA_{2}$ and taking into account that $A_{2J}<1$. Since
evolution curves corresponding to differential initial points cannot
intersect in the $(A_{2},T)$ plane, it must be concluded that
$T_{\fin}>1$.

To reach NESS with $T_{\fin}<1$, one intuitively thinks that inverting
the bangs, i.e. first cooling and afterwards heating should be
necessary. We prove this is indeed the case in the next section.

\subsection{Cooling-heating bang-bang}\label{sec:c-h-bangbang}

Next, we look into the bang-bang protocol in which the granular fluid
freely cools first, $\chi(t)=0$, $0\leq t\leq t_{J}$, and afterwards
is strongly heated, $\chi(t)=\chi_{\max}$, $t_{J}\leq t\leq
t_{\fin}$. The same separable first-order ODEs in the
$(A_{2},T)$ plane have to be solved, but with different initial
conditions. In the cooling stage, the resulting evolution is
\begin{align}
  2\ln&T=\frac{3}{16}a_{2}^{\st}
(A_{2}^{\hcs}-1)(A_{2}-1) \nonumber \\ &+
\left(1+\frac{3}{16}a_{2}^{\st}A_{2}^{\hcs}\right)
(A_{2}^{\hcs}-1)\ln\left(\frac{A_{2}^{\hcs}-A_{2}}{A_{2}^{\hcs}-1}\right)
, \nonumber \\ & \qquad 0\leq t\leq t_{J}.\label{eq:T-A2-sol-chi0-ch}
\end{align}
For $t_J\leq t\leq t_{\fin}$, the system evolves with
$\chi(t)=\chi_{\max}\gg 1$, and we have that
\begin{equation}\label{eq:T-A2-evol-chimax-ch}
T^{2}A_{2}^{1+\frac{3}{16}a_{2}^{\st}}=T_{\fin}^{2},
\quad
t_{J}\leq t\leq t_{\fin}.
\end{equation}
This equation is similar to~\eqref{eq:T-A2-evol-chimax-hc}, but here
the second part of the protocol ends at the point
$(A_{2\fin}=1,T_{\fin})$.  Since it starts from at $t=t_{J}$ from the
point $(A_{2J},T_{J})$, we get the relation
\begin{equation}\label{Tf-TJ_A2J-ch}
  T_{\fin}^{2}=T_{J}^{2}A_{2J}^{1+\frac{3}{16}a_{2}^{\st}}.
\end{equation}
In turn, $T_{J}$ and $A_{2J}$ are related by the particularisation of
Eq.~\eqref{eq:T-A2-sol-chi0-ch} for $t=t_{J}$,
\begin{align}
  2\ln&T_{J}=\frac{3}{16}a_{2}^{\st} (A_{2}^{\hcs}-1)(A_{2J}-1)
\nonumber \\ &+ \left(1+\frac{3}{16}a_{2}^{\st}A_{2}^{\hcs}\right)
(A_{2}^{\hcs}-1)
\ln\left(\frac{A_{2}^{\hcs}-A_{2J}}{A_{2}^{\hcs}-1}\right).
                            \label{eq:TJ-A2J-ch}
\end{align}

\begin{figure}
  \centering
  \includegraphics[width=3.375in]{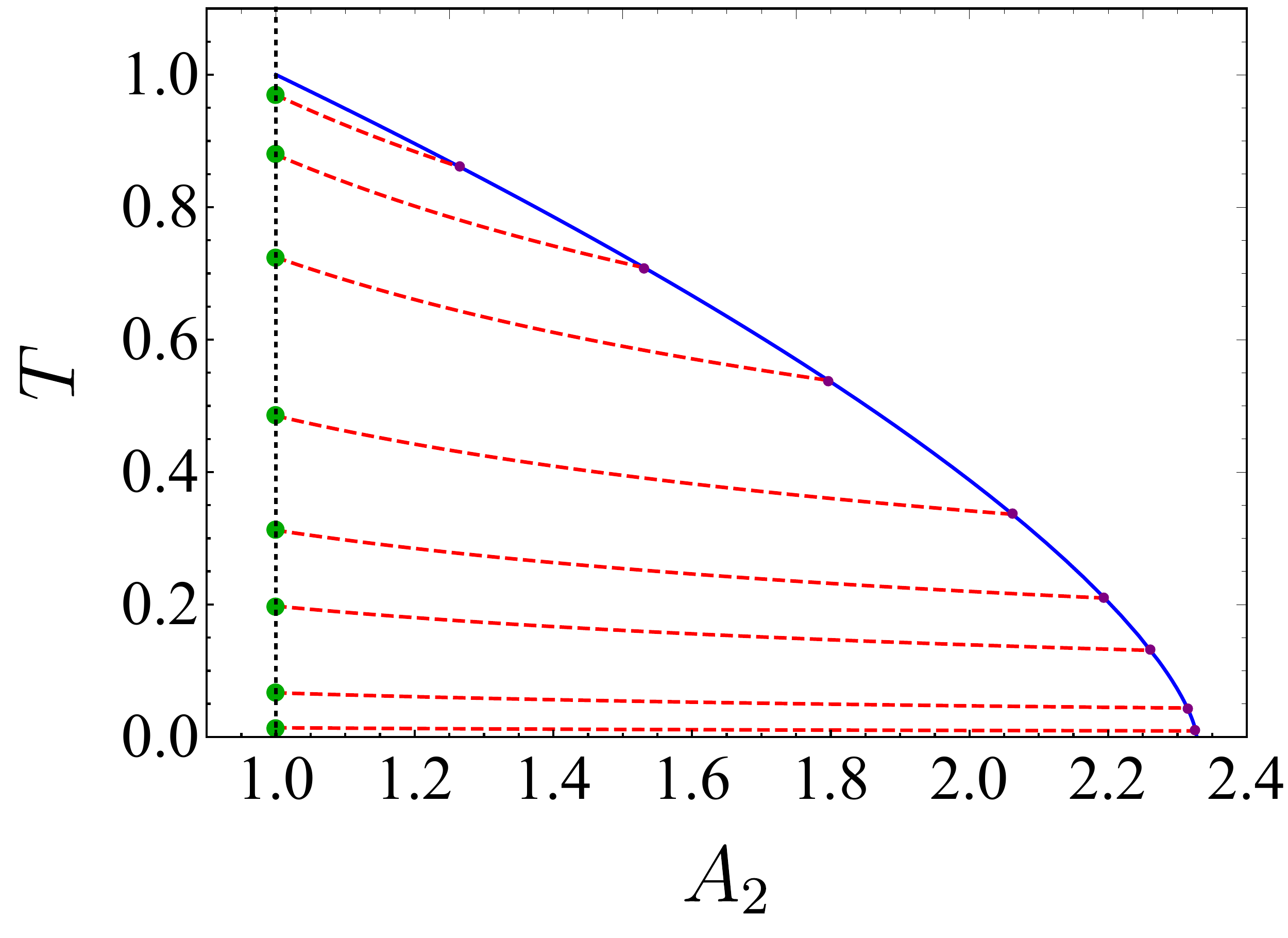}
  \caption{Bang-bang protocol for $T_{\fin}<1$. As a representative
    example we show the case $(\alpha=0.3,d=2)$---the qualitative
    picture of the motion of the possible in the $(A_{2},T)$ plane is the
    same for other values of $(\alpha,d)$. The bang-bang process
    connects the initial NESS $(A_{2\ini}=1,T_{\ini}=1)$ with the
    target NESS $(A_{2\fin}=1,T_{\fin}<1)$. Again, it comprises two
    parts, but the order of the bangs is reversed, as compared with
    Fig.~\ref{fig:hc-bangbang}: first the system is cooled (blue solid
    line) and afterwards is heated (red dashed lines). Different
    target NESS over the vertical line
    $A_{2}=1$ (dotted) are reached by starting the heating part from
    different points $(A_{2J},T_{J})$ over the cooling curve.}
  \label{fig:ch-bangbang}
\end{figure}

Figure~\ref{fig:ch-bangbang} shows the motion of the system in the
$(A_{2},T)$ plane for this bang-bang process. Therefore, it is
analogous to Fig.~\ref{fig:hc-bangbang}, but with the order of the
bangs reversed. In the first part of the bang-bang process, the system
follows the curve given by Eq.~\eqref{eq:T-A2-sol-chi0-ch}, for
$0\leq t\leq t_{J}$. In its second part, starting from a given point
$(A_{2J},T_{J})$ over this line, the system evolves according to
Eq.~\eqref{eq:T-A2-evol-chimax-ch}. This bang-bang process connects
the initial NESS $(A_{2\ini}=1,T_{\ini}=1)$ with the final NESS
$(A_{2\fin}=1,T_{\fin})$, but now we have that $T_{\fin}\leq 1$. In
order to keep the system stationary for $t=0$ and $t=t_{\fin}$, the
control function has again sudden jumps at the initial and final
times: at $t=0^{+}$, it changes from $1$ to $0$; at $t=t_{\fin}^{-}$,
it changes from $\chi_{\max}$ to $\chi_{\fin}=T_{\fin}^{3/2}$.

\section{Physical properties for the bang-bang optimal controls}\label{sec:physical-props-bang-bang}

Let us analyse in more detail the just described bang-bang protocols,
which drive the system from the initial NESS
$(A_{2\ini}=1,T_{\ini}=1)$ to the final NESS
$(A_{2\fin}=1,T_{\fin}\neq 1)$.  The two-step bang-bang processes
provide us with the minimum connecting time, and we obtain it as a
function of $T_{\fin}$ both for $T_{\fin}>1$ and for
$T_{\fin}<1$~\footnote{In linear response, when $|T_{\fin}-1|\ll 1$,
  the sub-optimality of bang-bang protocols with more than two steps
  is a consequence of a theorem in the number of switchings, see for
  instance theorem 10 in Sec.~III.17 of Pontryagin's
  book~\cite{pontryagin_mathematical_1987}. The formal proof for this
  specific non-linear case is quite lengthy and will be published
  elsewhere. A physical argument for the number of steps of the
  optimal bang-bang for a general non-linear case with $n$ variables
  is provided in Sec.~\ref{sec:generality-bang-bang}.}. In addition,
we calculate the statistical length and the cost for them.

In the following, we investigate the cases $T_{\fin}>1$ and
$T_{\fin}<1$ separately.

\subsection{Heating-cooling bang-bang:
  $T_{\fin}>1$}\label{sec:heating}

We start by analysing the heating-cooling bang-bang process described
in Sec.~\ref{sec:h-c-bangbang}, which makes it possible to reach
temperatures that are larger than the initial one, $T_{\fin}>1$. It
comprises two steps: (i) $\chi=\chi_{\max}$ for $0\leq t\leq t_{J}$,
and (ii) $\chi(t)=0$ for $t_{J}\leq t\leq t_{\fin}$.

\subsubsection{Minimum connecting time}\label{sec:min-time-heat}

Along the first part of the heating-cooling bang-bang, i.e. in the
time window $0\leq t\leq t_{J}$, we have
$\dot{T}\sim\chi_{\max}\left(1+\frac{3}{16}a_{2}^{\st}\right)$. Therefore,
we get
\begin{equation}\label{eq:tJ-heat}
  t_{J}=\frac{T_{J}-1}{\chi_{\max}\left(1+\frac{3}{16}a_{2}^{\st}\right)}\to
  0, \quad \chi_{\max}\to\infty.
\end{equation}
Note that $t_{J}\to 0$, but $\chi_{\max}t_{J}$ remains finite.

In the second part of the process, $t_{J}\leq t\leq t_{\fin}$, the
system freely cools with $\chi=0$. Therefore, making use of
Eq.~\eqref{eq:evol-non-dim} and taking into account that $t_{J}\to
0$, 
\begin{equation}\label{eq:tf-tJ-hc}
  t_{\fin}=\int_{T_{\fin}}^{T_{J}}
  \frac{dT}{T^{3/2}\left[1+\frac{3}{16}a_{2}^{\st}A_{2}(T)\right]},
\end{equation}
in which $A_{2}(T)$ is implicitly given by
Eq.~\eqref{eq:T-A2-sol-chi0-hc}: it is thus impossible to carry out
this integral analytically, at least in an exact manner.

We can obtain an approximate analytical expression for the connecting
time if we bring to bear that $3a_{2}^{\st}/16$ is quite small over
the whole range of restitution coefficient, $0\leq\alpha\leq 1$, and
$A_{2}$ is expected to be of the order of unity. Accordingly, denoting
by $t_{\fin}^{(0)}$ the connecting time obtained by putting
$a_{2}^{\st}=0$ in Eq.~\eqref{eq:tf-tJ-hc}~\footnote{Note that we are
  keeping $A_{2}$ and thus this is not equivalent to the Gaussian
  approximation, in which $a_{2}$ is completely disregarded from the
  very beginning. Therein, the optimal connecting time vanishes
  because only one bang with $\chi_{\max}$ suffices to reach the final
  temperature and $t_{f}^{G}=(T_{\fin}-1)/\chi_{\max}\to 0$.}, we get
\begin{equation}\label{eq:tf0-tJ-hc}
  t_{\fin}^{(0)}=\int_{T_{\fin}}^{T_{J}^{(0)}}
  \frac{dT}{T^{3/2}}=2\left[T_{\fin}^{-1/2}-\left(T_{J}^{(0)}
    \right)^{-1/2}\right].
\end{equation}
Above, $T_{J}^{(0)}$ means that $T_{J}$ must be
consistently put in terms of $T_{\fin}$ by considering
Eqs.~\eqref{eq:Tf-TJ-relation} and \eqref{eq:TJ-A2J} for
$a_{2}^{\st}=0$ but $A_{2}=O(1)$, which yields
\begin{equation}\label{eq:Tf-TJ0}
  T_{\fin}=T_{J}^{(0)}\left(\frac{A_{2}^{\hcs}-1}
    {A_{2}^{\hcs}-A_{2J}^{(0)}}\right)^{\frac{A_{2}^{\hcs}-1}{2}},
  \quad T_{J}^{(0)}=\left(A_{2J}^{(0)}\right)^{-1/2}.
\end{equation}
Eqs.~\eqref{eq:tf0-tJ-hc} and \eqref{eq:Tf-TJ0} provide us with
the connecting time $t_{\fin}^{(0)}$ as a function of the final
temperature $T_{\fin}$---both of them are given in terms of
$A_{2J}^{(0)}$, $0<A_{2J}^{(0)}<1$.

\begin{figure}
  \centering
  \includegraphics[width=3.375in]{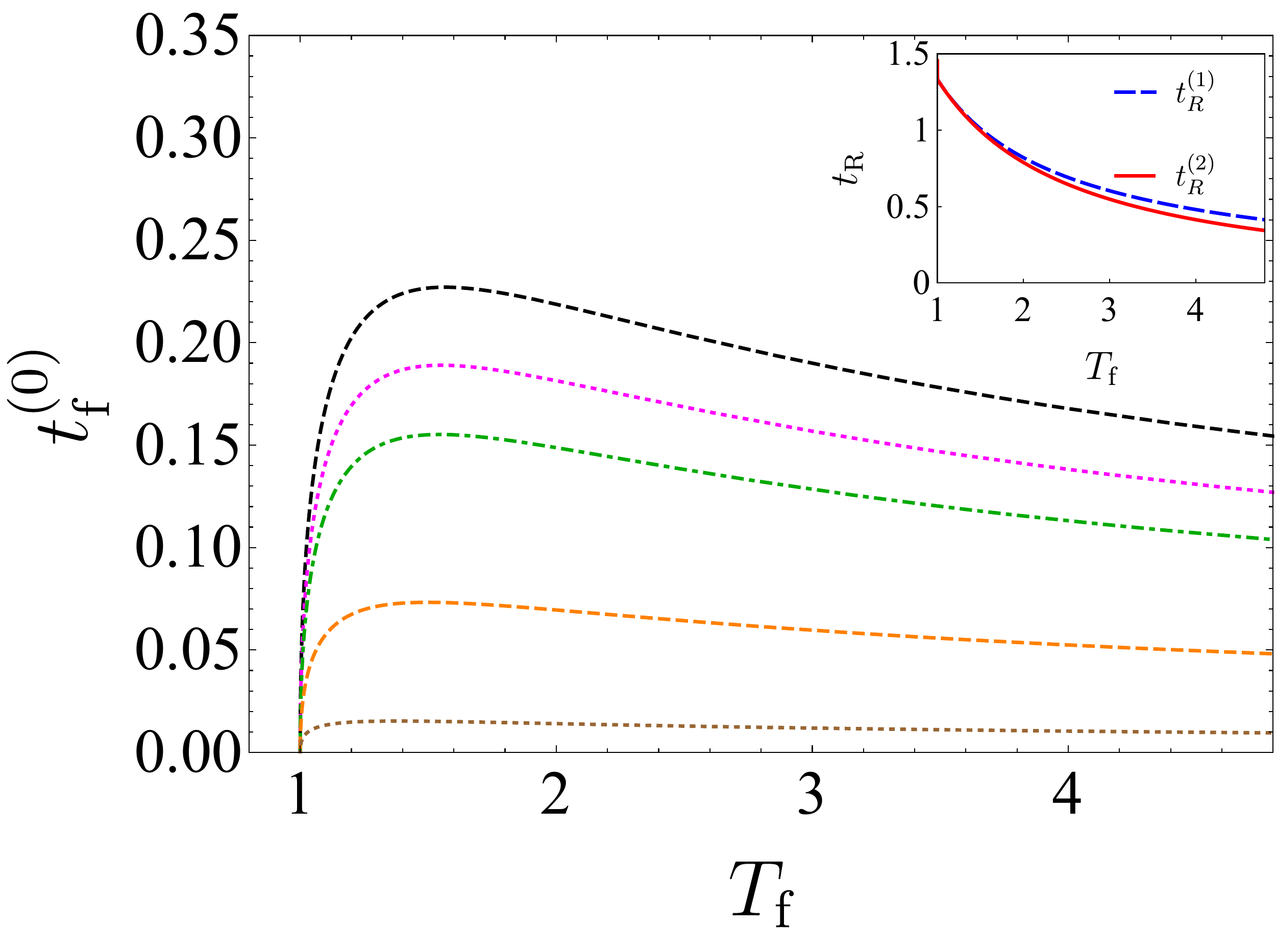}
  \caption{Minimum connection time as a function of the target
    temperature, for $T_{\fin}>1$. All lines correspond to $d=2$, and
    different values of the restitution coefficient are
    considered. From top to bottom, $\alpha=0.3$ (black dashed),
    $\alpha=0.8$ (magenta dotted), $\alpha=0.9$ (green dot-dashed),
    $\alpha=0.98$ (orange dashed), and $\alpha=0.998$ (brown
    dotted). Note that $t_{\fin}^{(0)}$ vanish in the limit as
    $T_{\fin}\to 1$ in all cases, whereas its high temperature
    behaviour depends on the inelasticity. For reference, the
    speed limits for relaxation, $t_{R}^{(1)}$ and $t_{R}^{(2)}$, are
    plotted in the inset, which lie well above $t_{\fin}^{(0)}$.}
  \label{fig:tf-heating}
\end{figure}

Figure~\ref{fig:tf-heating} shows $t_{\fin}^{(0)}$, as given by
Eq.~\eqref{eq:tf0-tJ-hc} and \eqref{eq:Tf-TJ0}, as a function of the
target temperature $T_{\fin}$. Over the scale of the figure,
$t_{\fin}^{(0)}$ is indistinguishable from the numerical integration
of Eq.~\eqref{eq:tf-tJ-hc}. Specifically, we have plotted the curves
for the two-dimensional case and several values of the
inelasticity. The minimum connection time given by control theory
clearly beats the speed limits for relaxation $t_{R}^{(1,2)}$, given
by Eq.~\eqref{eq:tR1-tR2}, which are shown in the inset. It is
observed that $t_{\fin}^{(0)}$ decreases as the restitution
coefficient $\alpha$ increases, vanishing in the elastic limit as
$\alpha\to 1$. Physically, this can be understood as follows: the
system does not cool in the second part of the process for
$\alpha\to 1$. Thus, $T_{J}^{(0)}\to T_{\fin}$ and
$t_{\fin}^{(0)}\to 0$. Mathematically,
$A_{2}^{\hcs}\to 1$ in the elastic limit, which ensures that
$T_{J}^{(0)}\to T_{\fin}$.

Asymptotic expressions for $t_{\fin}^{(0)}$ can be derived in some
limits. First, in the high temperature limit, $T_{J}^{(0)}$ becomes
large and $A_{2J}^{(0)}$ small; therefore we have that
\begin{equation}\label{eq:tf0-hc-highT}
  t_{\fin}^{(0)}\sim 2T_{\fin}^{-1/2}\left[1-\left(\frac{A_{2}^{\hcs}-1}
    {A_{2}^{\hcs}}\right)^{\frac{A_{2}^{\hcs}-1}{4}}\right], \quad
T_{\fin}\gg 1.
\end{equation}
Note that the rhs vanishes in the elastic limit, in which
$A_{2}^{\hcs}\to 1$. Second, we consider the linear response limit,
$T_{\fin}-1\ll 1$. Therein, Eq.~\eqref{eq:Tf-TJ0} implies that
$T_{J}^{(0)}-1\sim(T_{\fin}-1)^{1/2}$ and then $t_{\fin}^{(0)}$
vanishes as
\begin{equation}\label{eq:tf0-hc-lin-resp}
  t_{\fin}^{(0)}\sim \left(\frac{A_{2}^{\hcs}-1}{A_{2}^{\hcs}}\right)^{1/2} (T_{\fin}-1)^{1/2}, \quad
T_{\fin}-1\ll 1. 
\end{equation}
Again, the factor $A_{2}^{\hcs}-1$ makes the rhs vanish in the elastic
limit.

\begin{figure}
  \centering
  \includegraphics[width=3.375in]{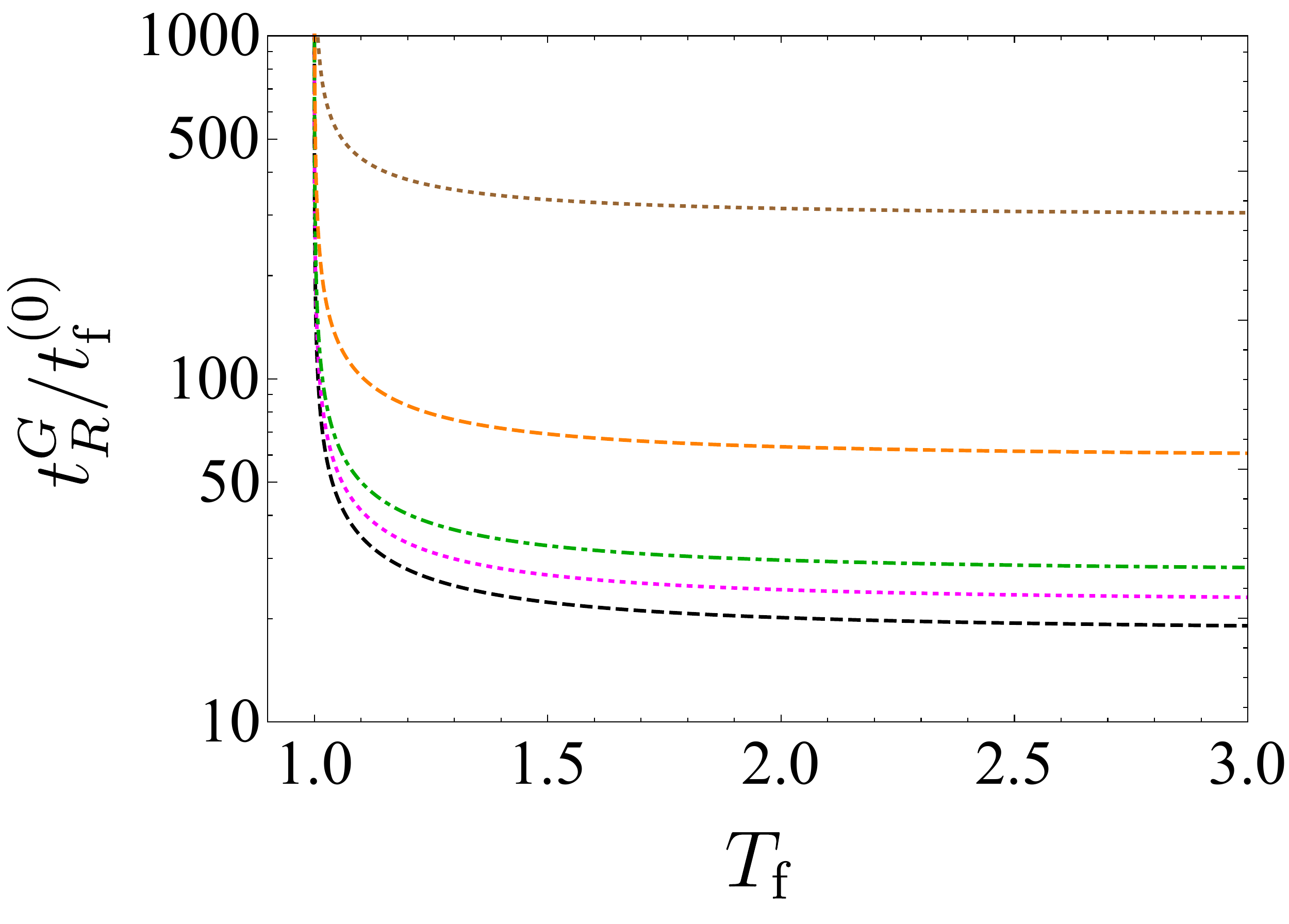}
  \caption{Acceleration factor as a function of the final temperature,
    for $T_{\fin}>1$. Different lines correspond to different values
    of the restitution coefficient $\alpha$, for $d=2$, with the same
    code as in Fig.~\ref{fig:tf-heating}. Note the logarithmic scale
    in the vertical axis.}
  \label{fig:accel-heating}
\end{figure}

As already commented above, the minimum value of the connecting time
$t_{\fin}^{(0)}$ beats the speed limits in
Eq.~\eqref{eq:tR1-tR2}. Therefore, it entails a really large
acceleration of the relaxation, as compared with the characteristic
relaxation time $t_{R}^{G}$ given by Eq.~\eqref{eq:tRG}.  We can
measure the acceleration factor in the bang-bang process by the ratio
$t_{R}^{G}/t_{\fin}^{(0)}$. In Fig.~\ref{fig:accel-heating}, we plot
this ratio as a function of the target temperature for $d=2$ and the
same values of the restitution coefficient as in
Fig.~\ref{fig:tf-heating}. Specifically, relaxation is speeded up by
more than one of order of magnitude for high temperatures and by a
diverging amount as the final temperature approaches unity, i.e. in
the linear response limit. For high target temperatures,
$t_{R}^{G}/t_{\fin}^{(0)}$ goes to a constant value that depends on
the inelasticity: both times vanish as $T_{\fin}^{-1/2}$, see
Eqs.~\eqref{eq:tRG-highT} and \eqref{eq:tf0-hc-highT}.

\subsubsection{Associated length and cost}\label{sec:length-cost-heatingt}

It is worth investigating the length $\calL$ traversed by the system
in probability space and the cost $\calC$ of the bang-bang
process. There is a trade-off between operation and cost, as expressed
by the ``thermodynamic uncertainty relation''
\eqref{eq:CSL-general}. As a result, it is expected that minimising
the operation time, as we have done, should entail a neat separation
from the geodesic in probability space, for which $\calL=\Lambda$, and
an increase in the cost $\calC$.

In Appendix \ref{sec:Fisher-info-Sonine}, we show that the Fisher
information is given by
\begin{equation}
  I(t)=I^{(0)}(t)\left[1+O(a_{2}^{\st})\right], \quad
  I^{(0)}(t)=\frac{d}{2}\!\left(\frac{\dot{T}(t)}{T(t)}\right)^{\!\!2}. 
\end{equation}
In the following we calculate the lowest order contribution to the
length and cost, i.e.
\begin{equation}\label{eq:L0}
  \calL^{(0)}=\int_{0}^{t_{\fin}}dt\,
  \sqrt{I^{(0)}(t)}=\sqrt{\frac{d}{2}} \int_{0}^{t_{\fin}} dt\,
  \left|\frac{\dot{T}(t)}{T(t)}\right| ,
\end{equation}
and
\begin{equation}\label{eq:C0}
  \calC^{(0)}=\int_{0}^{t_{\fin}}dt\,
  I^{(0)}(t)=\frac{d}{2}\int_{0}^{t_{\fin}} dt
  \left(\frac{\dot{T}(t)}{T(t)}\right)^{\!\!2}. 
\end{equation}

First, we look into the length swept by the probability
distribution. Taking into account that $T_{J}>T_{\fin}>1$, we have to
split the integral in Eq.~\eqref{eq:L0} into two summands: the first
one for the time interval $[0,t_{J}]$, in which the temperature
monotonically increases from $T_{\ini}=1$ to $T_{J}$, and the second
one for $[t_{J},t_{\fin}]$, in which the temperature monotonically
decreases from $T_{J}$ to $T_{\fin}$. Therefore, we have that
\begin{align}
  \calL^{(0)}=&\sqrt{\frac{d}{2}}\left[ \int_{0}^{t_{J}}dt\,
                   \frac{\dot{T}}{T}-
                   \int_{t_{J}}^{t_{\fin}}dt\,
               \frac{\dot{T}}{T} \right]
         \nonumber \\
  =&
  \sqrt{\frac{d}{2}}\left[\int_{1}^{T_{J}}
                   \frac{dT}{T}+
                   \int_{T_{\fin}}^{T_{J}}
     \frac{\dot{T}}{T}\right]=\sqrt{\frac{d}{2}}\,
     \ln\left(\frac{T_{J}^{2}}{T_{\fin}}\right).
         \label{eq:length-granular}
\end{align}

\begin{figure}
  \centering
  \includegraphics[width=3.375in]{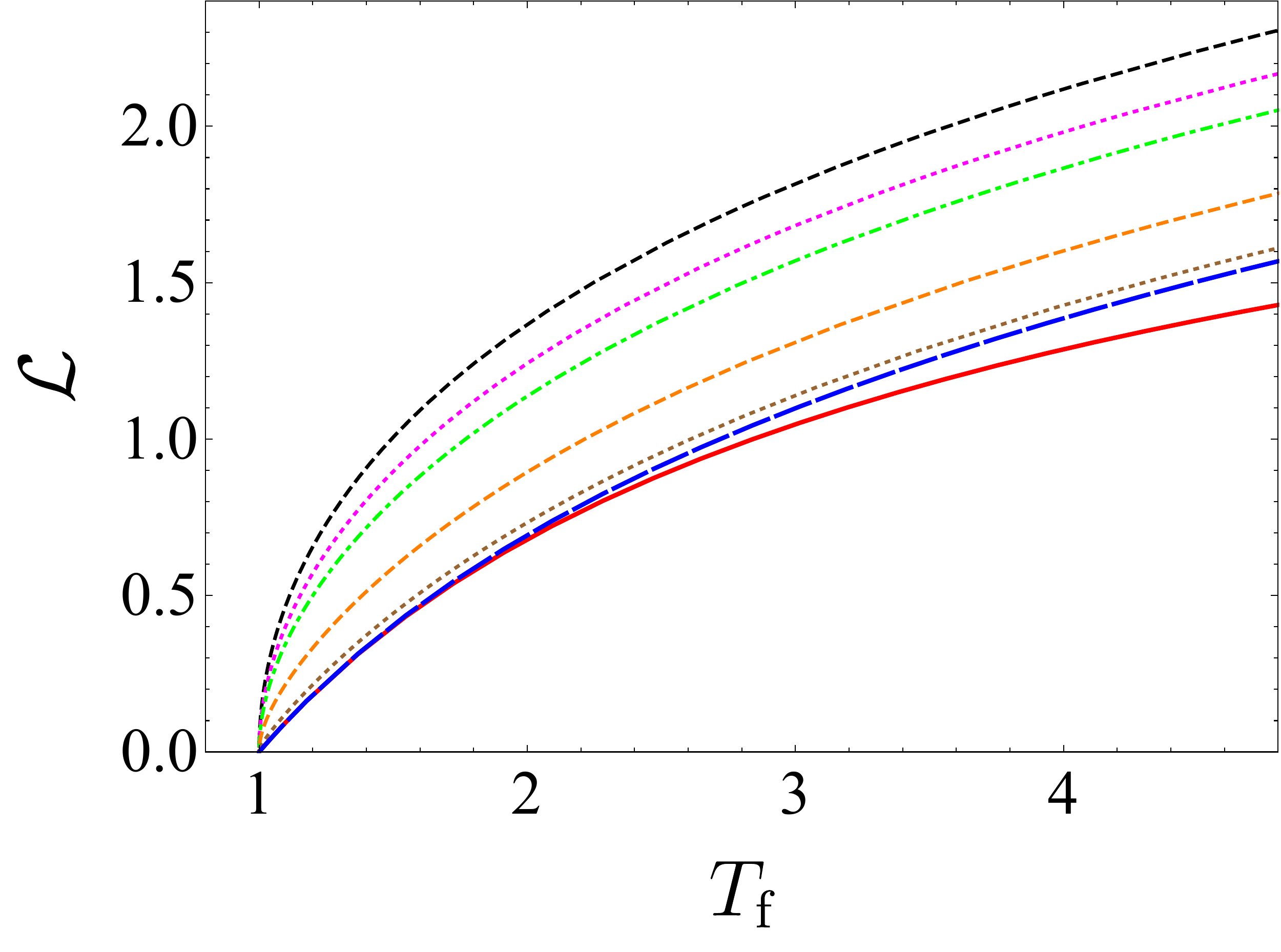}
  \caption{Length in probability space for as a function of the final
    temperature, for $T_{\fin}>1$. The two lowest lines correspond to
    the length over the geodesic $\Lambda_{G}$ (red solid), that for
    the relaxation process $\calL_{G}^{\rel}$ (blue broken)---both
    defined in Eq.~\eqref{eq:L-and-C-Gaussian}. The remainder above
    are those corresponding to the optimal heating-cooling bang-bang
    connections for different values of the restitution
    coefficient $\alpha$ ($d=2$), with the same code as in
    Fig.~\ref{fig:tf-heating}.}
  \label{fig:L-heating}
\end{figure}

We plot our estimate $\calL^{(0)}$ for the
length over the optimal bang-bang connection as a function of the
target temperature in Fig.~\ref{fig:L-heating}. Also plotted are the
length for the relaxation process $\calL_{G}^{\rel}$ (blue broken) and
the length over the geodesic $\Lambda_{G}$ (solid red), given in
Eq.~\eqref{eq:L-and-C-Gaussian}. Despite the heating-cooling bang-bang
minimises the connection time, which is much shorter than the
relaxation time $t_{R}^{G}$, the corresponding length is always larger
than $\calL_{G}^{\rel}$, since
\begin{equation}
  \calL^{(0)}-\calL_{G}^{\rel}=\sqrt{2d}\,
     \ln\left(\frac{T_{J}}{T_{\fin}}\right)\geq 0,
\end{equation}
with the equality holding when $T_{J}=T_{\fin}$.  It is inelasticity
that makes $T_{J}$ different from $T_{\fin}$ and thus increases the
length swept by the system in probability space. Only in the elastic
limit $\alpha\to 1$ we have that $\calL^{(0)}\to\calL_{G}^{\rel}$, as
neatly observed in the plot.

Second, we consider the cost of this heating-cooling bang-bang
process.  Since the bang-bang process minimises the connection time, a
large value of the cost associated with the speed limit is
expected. Similarly to what we have just done for the length, the
smallness of non-Gaussianities allows us to estimate the cost with
$\calC^{(0)}$. Again, by splitting the time integral into the
subintervals $[0,t_{J}]$ and $[t_{J},t_{\fin}]$ and integrating over
temperature in each subinterval, one gets
\begin{align}
  \calC^{(0)}=& \frac{d}{4}\int_{0}^{t_{\fin}}dt\,
\frac{\dot{T}^{2}}{T^{2}} =\frac{d}{4}\left(\int_{1}^{T_{J}}dT\,
\frac{\dot{T}}{T^{2}}+ \int_{T_{J}}^{T_{\fin}}dt\,
                   \frac{\dot{T}}{T^{2}} \right) \nonumber \\
  =& \frac{d}{4}\left[\int_{1}^{T_{J}}dT\,
\frac{\chi_{\max}-T^{3/2}}{T^{2}}-\int_{T_{J}}^{T_{\fin}}dT\,
     T^{-1/2}\right] \nonumber \\
  = & \frac{d}{4}\left[\chi_{\max}\frac{T_{J}-1}{T_{J}}-
  2\left(T_{\fin}^{1/2}-1\right)\right].
         \label{eq:cost-granular}
\end{align}
We have taken into account that $\chi(t)=\chi_{\max}$ ($\chi(t)=0$) in
the time interval $[0,t_{J}]$ ($[t_{J},t_{\fin}]$). Also, and
consistently, we have put $a_{2}^{\st}=0$ in the evolution equation
for the temperature of Eq.~\eqref{eq:evol-non-dim}, since we are
evaluating $\calC^{(0)}$ to the lowest order.

Our main conclusion is thus that $\calC^{(0)}$ diverges for the
optimal bang-bang connection,
\begin{align}
  \calC^{(0)}\sim \frac{d}{4}\chi_{\max}\frac{T_{J}-1}{T_{J}}\to\infty, &&
\chi_{\max}\to\infty.
   \label{eq:cost-heating-G}
\end{align}
If $\chi_{\max}\gg 1$ but not infinite, the above equation gives
the leading behaviour of the cost. In that case,
$t_{J}=O(\chi_{\max}^{-1})$ is small and the cooling part still rules 
the operation time, $t_{\fin}^{(0)}-t_{J}\gg t_{J}$, whereas the
heating pulse still prevails for the cost.

Despite the divergence of $\calC^{(0)}$, the energy input from the
stochastic thermostat remains finite. Since the energy of the granular
fluid is proportional to the temperature, we identify energy with
temperature in the following discussion. The stochastic forcing is
switched off in the cooling step of the bang-bang process, so the energy
input comes from the heating step and equals
\begin{equation}
  \int_{0}^{t_{J}}\!\!dt\, \chi_{\max}\left(
    1+\frac{3}{16}a_{2}^{\st}\right)\!=\chi_{\max}t_{J}\left(
    1+\frac{3}{16}a_{2}^{\st}\right)\!\to T_{J}-1.   
\end{equation}
In fact, for a fixed connection time $t_{\fin}$, the bang-bang process
minimises the energy input by the stochastic thermostat---and thus
also the energy dissipated in collisions, since the total energy
increment is given by $T_{\fin}-1$ for the initial and final NESS. The
reason is simple: the only change is Pontryagin's scheme is that of
the function $f_{0}$, in which $f_{0}=1$ for time minimisation---see
\eqref{eq:y0-def}---is substituted with
$f_{0}=\chi\left( 1+\frac{3}{16}a_{2}^{\st}\right)$ for energy input
minimisation. Still, Pontryagin's Hamiltonian $\Pi$ is linear in the
control $\chi$ and therefore the bang-bang protocol also emerges as
the optimal solution in this case.

\subsection{Cooling-heating bang-bang:
  $T_{\fin}<1$}\label{sec:cooling}

Now we turn our attention to the  case in which the target
temperature is smaller than the initial one, $T_{\fin}<1$. Similarly
to what we have done in the previous section, we consider the two-step
bang-bang process but with the order of the bangs reversed: $\chi=0$
for $0\leq t\leq t_{J}$, and $\chi(t)=\chi_{\max}$ for $t_{J}\leq
t\leq t_{\fin}$, as described in Sec.~\ref{sec:c-h-bangbang}.

\subsubsection{Minimum connecting time}\label{sec:min-time-cool}

In the second part of the process, a line of reasoning similar to the
one leading to Eq.~\eqref{eq:tJ-heat} gives us that
\begin{equation}
  t_{\fin}-t_{J}=\frac{T_{\fin}-T_{J}}{\chi_{\max}\left(1+\frac{3}{16}a_{2}^{\st}
    \right)}\to 0, \quad \chi_{\max}\to\infty.
\end{equation}
This means that $t_{\fin}\to t_{J}$, the second part of the process is
instantaneous in the limit as $\chi_{\max}\to\infty$. On the other
hand,  the system freely cools in the first part of the process, and then
\begin{equation}
  t_{\fin}=\int_{T_{J}}^{1}\frac{dT}{T^{3/2}\left[1+
      \frac{3}{16}a_{2}^{\st}A_{2}(T)\right]},
\end{equation}
where $A_{2}(T)$ in now given by
Eq.~\eqref{eq:T-A2-sol-chi0-ch}.

Again, the integral cannot be carried out analytically but it is
possible to derive an approximate expression for $t_{\fin}$ by
recalling that $a_{2}^{\st}$ is small and $A_{2}=O(1)$. In this way,
we obtain
\begin{equation}\label{eq:tf0-tJ-ch}
  t_{\fin}^{(0)}=\int_{T_{J}^{(0)}}^{1}\frac{dT}{T^{3/2}}=2\left[\left(
      T_{J}^{(0)}\right)^{-1/2}-1\right],
\end{equation}
where
\begin{equation}
  T_{\fin}=T_{J}^{(0)} \left(A_{2J}^{(0)}\right)^{1/2}, \quad 
    T_{J}^{(0)}=\left(\frac{A_{2}^{\hcs}-A_{2J}^{(0)}}{A_{2}^{\hcs}-1}
  \right)^{\frac{A_{2}^{\hcs}-1}{2}}.
\end{equation}
Once more, the last two equations give the connecting time
$t_{\fin}^{(0)}$ as a function of the final temperature $T_{\fin}$,
since both of them are given in terms of $A_{2J}^{(0)}$---here,
$1<A_{2J}^{(0)}<A_{2}^{\hcs}$.

\begin{figure}
  \centering
  \includegraphics[width=3.375in]{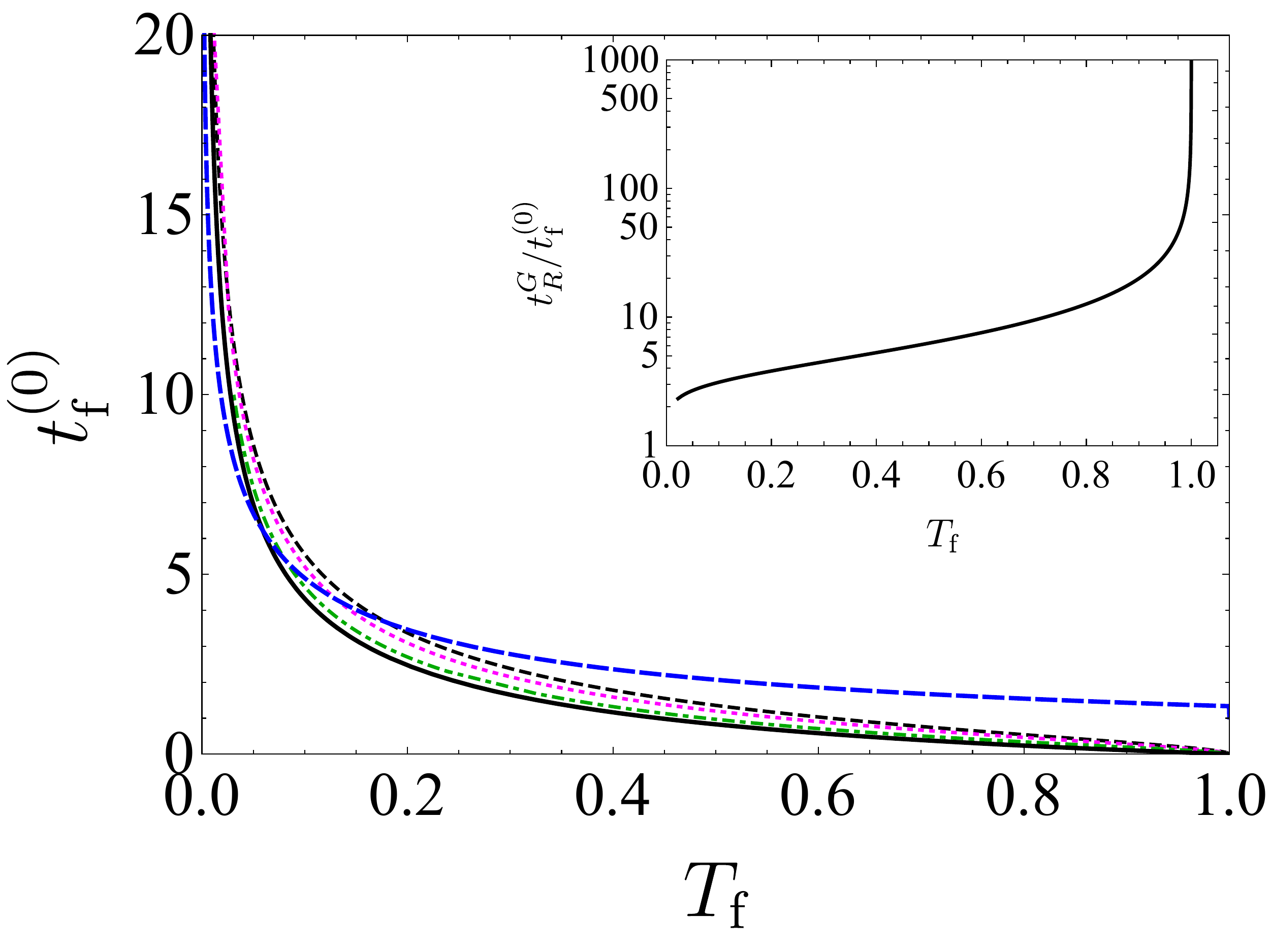}
  \caption{Minimum connection time $t_{\fin}^{(0)}$ as a function of
    the target temperature, for $T_{\fin}<1$. We consider the
    two-dimensional case and different values of the restitution
    coefficient: from top to bottom, $\alpha=0.3$ (black dashed line),
    $\alpha=0.8$ (magenta dotted), and $\alpha=0.95$ (green
    dot-dashed). Note that $t_{\fin}^{(0)}$ depends very weakly on the
    inelasticity, thus it is quite close to the Gaussian estimate for
    the connecting time $t_{\fin}^{G}$ (black solid). Also plotted is
    the speed limit for the relaxation process $t_{R}^{(1)}$ (blue
    broken), defined in Eq.~\eqref{eq:tR1-tR2}. In the inset, the
    acceleration factor $t_{R}^{G}/t_{\fin}^{(0)}$ is shown for
    $\alpha=0.3$, other values of $\alpha$ are basically
    superimposed. Similarly to the case $T_{\fin}>1$, the acceleration
    factor diverges in the limit as $T_{\fin}\to 1$, for which the
    bang-bang connecting time $t_{\fin}^{(0)}$ vanishes.}
  \label{fig:tf-cooling}
\end{figure}

We show the behaviour of $t_{\fin}^{(0)}$ as a function of the target
temperature in Fig.~\ref{fig:tf-cooling}, for $T_{\fin}<1$. All curves
correspond to $d=2$ but different values of the restitution
coefficient $\alpha$. In this case, $t_{\fin}^{(0)}$ beats the speed
limit $t_{R}^{(1)}$ for relaxation for high enough $T_{\fin}$---in the
limit as $T_{\fin}\to 1^{-}$, we have that $t_{\fin}^{(0)}\to 0$ but
$t_{R}^{(1)}$ remains finite--- but lies above it for
$T_{\fin}\lesssim 0.15$, approximately. This contrasts with the
situation for $T_{\fin}>1$, shown in Fig.~\ref{fig:tf-heating}. Physically, this asymmetry between the cases
$T_{\fin}>1$ and $T_{\fin}<1$ can be understood as stemming from the
non-holonomic constraint $\chi\geq 0$, which limits the rate at which
the system can be cooled down---whereas no such limit exists for
$T_{\fin}>1$ because we have considered that $\chi_{\max}\to\infty$.

At difference with the case $T_{\fin}>1$, $t_{\fin}^{(0)}$ depends
very weakly on $\alpha$~\footnote{We have made time dimensionless with
  $\zeta _{0}$, which depends on $d$ and is proportional to
  $(1-\alpha )^{2}$.}. Since the excess kurtosis is small, we can
obtain a rough estimate of the behaviour of the system by completely
neglecting it---the so-called Gaussian approximation, which we have
already employed in Sec.~\ref{sec:ESR}. Therein, it is clear that only
one bang with $\chi=0$ suffices: the fastest way of reaching a
temperature $T_{\fin}$ below the initial one is to turn off the
stochastic thermostat. Putting $\chi$ and $a_{2}^{\st}$ to zero in
Eq.~\eqref{eq:evol-non-dim}, we obtain the Gaussian estimate for the
connecting time $t_{\fin}^{G}=2\left(T_{\fin}^{-1/2}-1\right)$, which
is also plotted in Fig.~\ref{fig:tf-cooling}.

We can obtain asymptotic expressions for $t_{\fin}^{(0)}$ in two
relevant limits. In the low target temperature limit, $T_{\fin}\ll 1$,
$T_{J}^{(0)}$ is also small and $A_{2J}^{(0)}\to A_{2}^{\hcs}$, which
leads to 
\begin{equation}\label{eq:tf0-ch-highT}
  t_{\fin}^{(0)}\sim 2\left(A_{2}^{\hcs}\right)^{1/4}T_{\fin}^{-1/2}, \quad
T_{\fin}\ll 1.
\end{equation}
In the elastic limit, $A_{2}^{\hcs}\to 1^{+}$, and thus
$t_{\fin}^{(0)}\sim t_{\fin}^{G}$, $T_{\fin}\ll 1$, $\alpha\to
1^{-}$. For larger inelasticity $t_{\fin}^{(0)}$ lies above
$t_{\fin}^{G}$ but it is of the same order of magnitude---for example,
$t_{\fin}^{(0)}/t_{\fin}^{G}\sim (A_{2}^{\hcs})^{1/4}$ for $T_{\fin}\ll
1$. This is in accordance with the behaviour observed in
Fig.~\ref{fig:tf-cooling}.  In the linear response limit,
$1-T_{\fin}\ll 1$, the behaviour is completely similar to that of
$T_{\fin}>1$: Eq.~\eqref{eq:tf0-hc-lin-resp} still holds replacing
$T_{\fin}-1$ with its absolute value.

\subsubsection{Associated length and
  cost}\label{sec:length-cost-cooling}

Let us evaluate the length and cost of the
cooling-heating bang-bang protocol. We start by calculating the length
$\calL^{(0)}$: here, $T_{J}<T_{\fin}<1$ and once more the integral
Eq.~\eqref{eq:L0} has to be split into the two time subintervals inside
which the temperature is monotonic, which yields
\begin{align}
  \calL^{(0)}=&\sqrt{\frac{d}{2}}\left[ -\int_{0}^{t_{J}}dt\,
                   \frac{\dot{T}}{T}+
                   \int_{t_{J}}^{t_{\fin}}dt\,
               \frac{\dot{T}}{T} \right]
         \nonumber \\
  =&
  \sqrt{\frac{d}{2}}\left[\int_{T_{J}}^{1}
                   \frac{dT}{T}+
                   \int_{T_{J}}^{T_{\fin}}
     \frac{dT}{T}\right]=\sqrt{\frac{d}{2}}\,
     \ln\left(\frac{T_{\fin}}{T_{J}^{2}}\right).
         \label{eq:length-granular}
\end{align}

\begin{figure}
  \centering
  \includegraphics[width=3.375in]{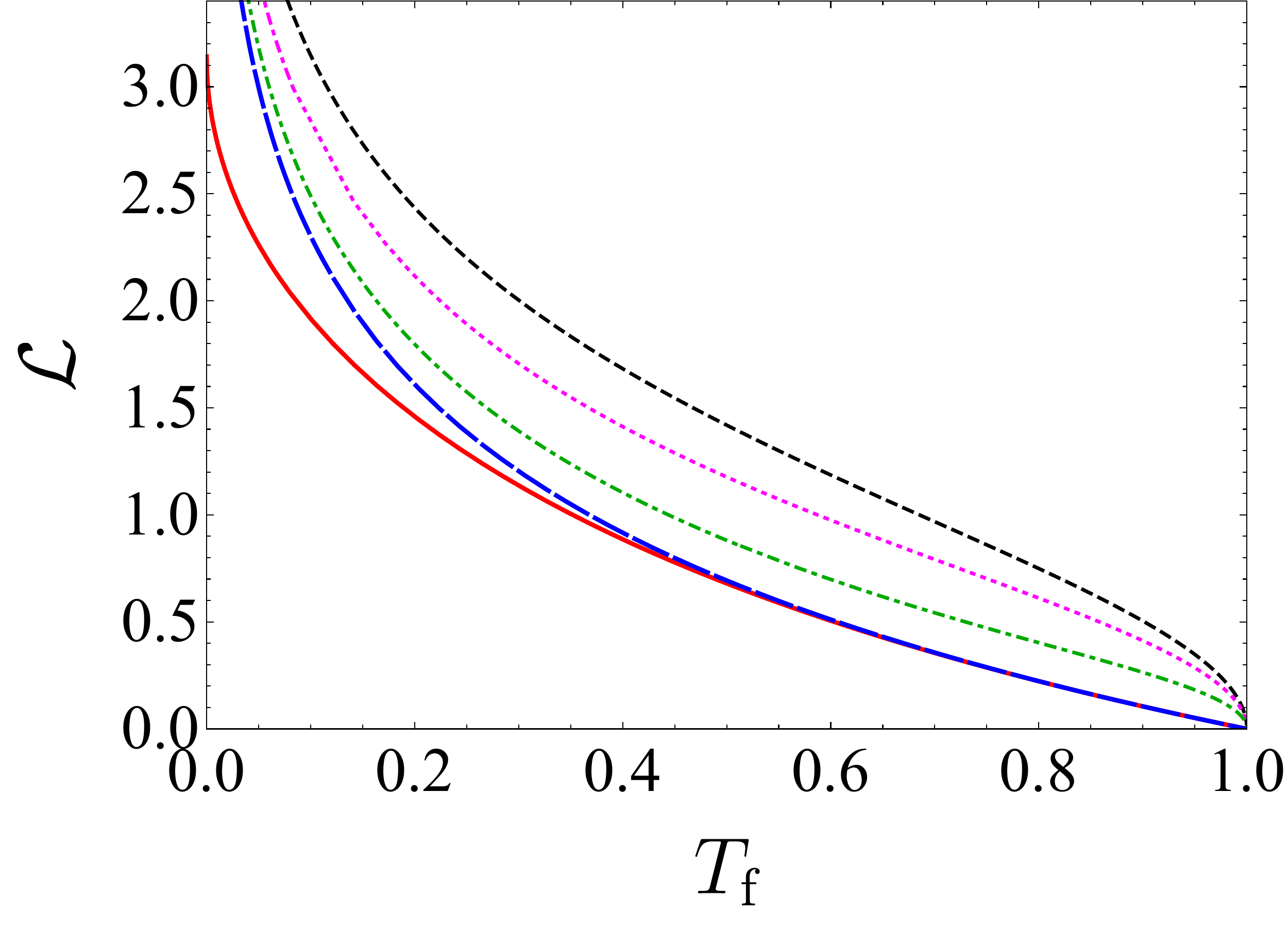}
  \caption{Length in probability space for as a function of the final
    temperature, for $T_{\fin}<1$. From top to bottom, the following
    lines are shown. The first three
    ones correspond to the optimal cooling-heating
    bang-bang protocols for different values of the restitution
    coefficient in the two-dimensional case: $\alpha=0.3$,
    $\alpha=0.8$ and $\alpha=0.95$; the two lowest ones are the
    lengths for the relaxation process $\calL_{G}^{\rel}$ (blue
    broken) and the geodesic length $\Lambda_{G}$ (red solid).}
  \label{fig:L-cooling}
\end{figure}

Figure~\ref{fig:L-cooling} is devoted to the comparison of the length
$\calL^{(0)}$ over the optimal bang-bang connection with the lengths
for the relaxation process $\calL_{G}^{\rel}$ and over the geodesic
$\Lambda_{G}$, which are given by
Eq.~\eqref{eq:L-and-C-Gaussian}. Similarly to the case $T_{\fin}>1$,
$\calL^{(0)}$ is always larger than its value for relaxation
$\calL_{G}^{\rel}$
\begin{equation}
  \calL^{(0)}-\calL_{G}^{\rel}=\sqrt{2d}\,
     \ln\left(\frac{T_{\fin}}{T_{J}}\right)\geq 0.
\end{equation}
Again, the equality only holds when $T_{J}=T_{\fin}$, which
corresponds to the elastic limit.

Now we move on to the cost of the cooling-heating bang-bang. One more
time, we split the time integral and change to integrate over
temperature in each part of the bang-bang process. For the case $T_{\fin}<1$ we are analysing, the order of the bangs is reversed: $\chi(t)=0$ ($\chi(t)=\chi_{\max}$) in
the time interval $[0,t_{J}]$ ($[t_{J},t_{\fin}]$).  Then we obtain that
\begin{align}
  \calC^{(0)}=& \frac{d}{4}\int_{0}^{t_{\fin}}dt\,
\frac{\dot{T}^{2}}{T^{2}} =\frac{d}{4}\left(\int_{1}^{T_{J}}dT\,
\frac{\dot{T}}{T^{2}}+ \int_{T_{J}}^{T_{\fin}}dt\,
                   \frac{\dot{T}}{T^{2}} \right) \nonumber \\
 =&\frac{d}{4}\left[-\int_{1}^{T_{J}}dT\,
                T^{-1/2}+\int_{T_{J}}^{T_{\fin}}dT\,
\frac{\chi_{\max}-T^{3/2}}{T^{2}}\right] \nonumber \\
  =&\frac{d}{4}\left[\chi_{\max}
                \frac{T_{\fin}-T_{J}}{T_{J}T_{\fin}}-
  2\left(T_{\fin}^{1/2}-1\right)\right].
\end{align}
Taking into account that $\chi_{\max}\gg 1$, the second
term on the rhs is negligible against the first one and thus 
\begin{align}
  \calC^{(0)}\sim \frac{d}{4}\chi_{\max}\frac{T_{\fin}-T_{J}}{T_{J}T_{\fin}}\to\infty, &&
\chi_{\max}\to\infty.
   \label{eq:cost-cooling-G}
\end{align}
In complete analogy to the $T_{\fin}>1$ case,
Eq.~\eqref{eq:cost-cooling-G} continues to give the leading behaviour
of the cost when $\chi_{\max}\gg 1$ but not infinite, and the
connection time and cost are still dominated by the cooling and
heating bangs, respectively. Since
the stochastic thermostat only heats the system in the time interval
$[t_{J},t_{\fin}]$, the energy input is now given by
\begin{align}
  \int_{t_{J}}^{t_{\fin}}\!\!dt\, \chi_{\max}\left(
1+\frac{3}{16}a_{2}^{\st}\right)\!&=\chi_{\max}(t_{\fin}-t_{J})
\left(1+\frac{3}{16}a_{2}^{\st}\right) \! \nonumber \\ & \to
T_{\fin}-T_{J},
\end{align}
which is also finite. For the same reasons as in the case
$T_{\fin}>1$, here the bang-bang protocol also minimises the energy
input from the stochastic thermostat for a fixed connecting time
$t_{\fin}$.

\section{Numerical simulations}\label{sec:numerics}

In order to check our theoretical predictions, we have carried out
numerical simulations of the dynamics of the granular
gas. Specifically, we have carried out Direct Simulation Monte Carlo
(DSMC) for the two-dimensional case and two different values of the
restitution coefficient, $\alpha=0.3$---for which $a_{2}^{\st}$ is
positive---and $\alpha=0.8$---for which $a_{2}^{\st}$ is negative. In
all cases, we start from a high temperature state with a Maxwellian
velocity distribution function and switch on the stochastic thermostat
with a certain intensity $\xi_{\ini}$: the granular gas relaxes
towards the corresponding NESS, in which the temperature $T_{\ini}$
and the noise intensity are related by Eq.~\eqref{eq:Ts}. Recall that
we have employed $T_{\ini}$ to non-dimensionalise the temperature, so
in our units $T_{\ini}=1$. From this initial NESS, we implement the
bang-bang protocols developed in the previous sections.

In the case $T_{\fin}>1$, we proceed as follows in each trajectory of
the simulation. First, the system is instantaneously heated from
$T_{\ini}=1$ to $T_{J}$: we make the velocities of all particles
change as $\bm{v}_{i}\to\bm{v}_{i}+\bm{\eta}_{i}$, where
$\bm{\eta}_{i}$ are independent Gaussian distributed random variables
of a certain variance---the larger the variance, the larger the
temperature increment $T_{J}-1$ and the smaller the excess kurtosis
$a_{2J}$. Second, starting from the previously generated
configuration, we let the system freely cool ($\xi=0$) until $a_{2}$
in the trajectory equals the steady value $a_{2}^{\st}$. This
determines the connecting time $t_{\fin}$, at which the temperature in
the trajectory equals $T_{\fin}$. At this time, we switch on the
stochastic thermostat again but with an intensity $\xi_{\fin}$ such
that the system remains stationary for $t>t_{\fin}$: taking advantage
of the theoretical prediction $\xi\propto T_{\st}^{3/4}$, as given by
Eq.~\eqref{eq:Ts}, we set $\xi_{\fin}=\xi_{\ini}T_{\fin}^{3/4}$.

\begin{figure}
  \centering
  \includegraphics[width=3.375in]{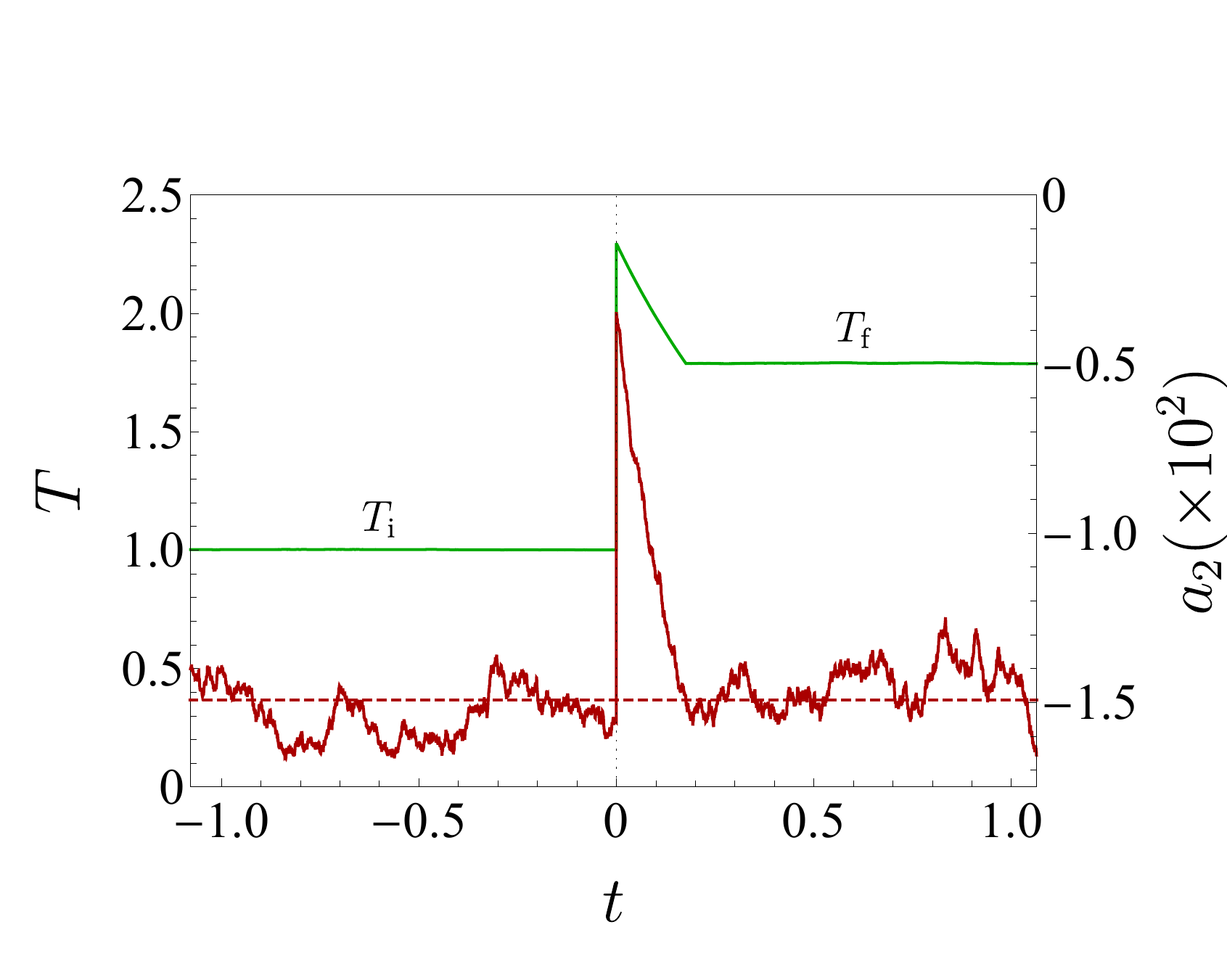}
  \caption{Typical simulation trajectory for the case $T_{\fin}>1$.
    The upper curve corresponds to the temperature (left vertical
    axis) and the lower curve to the excess kurtosis (right vertical
    axis). Negative times correspond to the initial NESS, with
    $T_{\ini}=1$. At $t=0$, the system is instantaneously heated, the
    temperature increases whereas the absolute value of the excess
    kurtosis decreases. Afterwards, the temperature decreases and the
    absolute value of the excess kurtosis increases in the cooling
    stage. The thermostat is switched on with intensity $\xi_{\fin}$
    when $a_{2}$ touches its steady value $a_{2}^{\st}$ (dashed line):
    this determines the connection time $t_{\fin}$. The noise
    intensity $\xi_{\fin}$ corresponds to
    $\chi_{\fin}=T_{\fin}^{3/2}$, see Eq.~\eqref{eq:Ts}, where
    $T_{\fin}$ is the value of the temperature at $t_{\fin}$.}
  \label{fig:traj-heating}
\end{figure}

The quantities $T_{J}$, $a_{2J}$, $t_{\fin}$, $T_{\fin}$, and
$\xi_{\fin}$ fluctuate from one realisation to another. A typical
trajectory of the case $T_{\fin}>1$ is depicted in
Fig.~\ref{fig:traj-heating}. Specifically, the realisation shown
corresponds to $d=2$ and $\alpha=0.8$ in a system with $N=10^{6}$
particles. It is neatly observed how the system remains stationary
after the stochastic forcing is switched on at $t_{\fin}$. Note that
fluctuations in the excess kurtosis are much larger than those of the
temperature---which are basically not seen in the scale of the figure.

\begin{figure}
  \centering
  \includegraphics[width=3.375in]{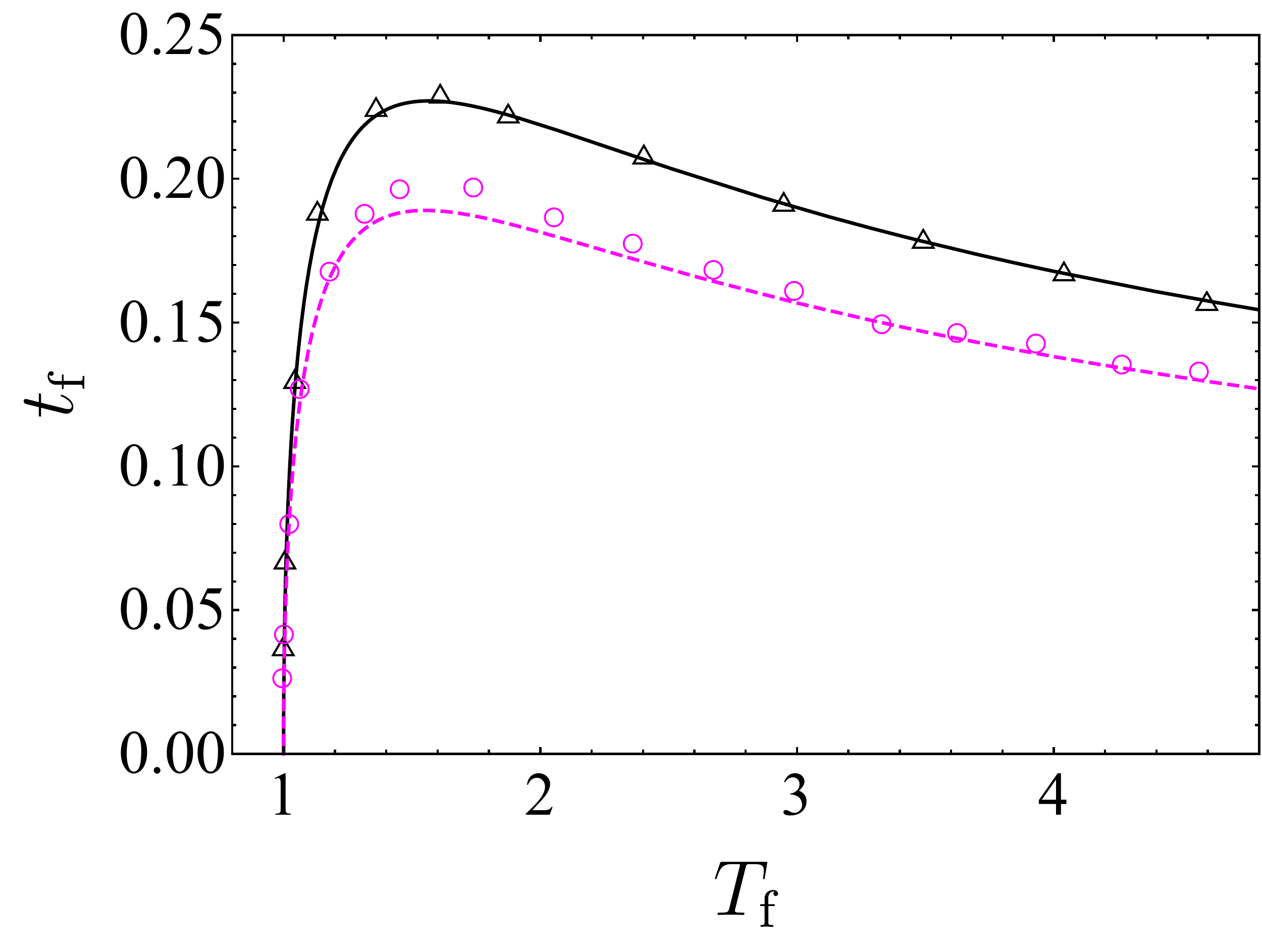}
  \caption{Connection time as a function of the target
    temperature. Simulation results (symbols) are compared with
    Eq.~\eqref{eq:tf0-tJ-hc} (lines) for $d=2$ and two values of
    $\alpha$: $\alpha=0.3$ (open triangles, solid line) and
    $\alpha=0.8$ (open circles, dashed line).  }
  \label{fig:tf-heating-theo-simul}
\end{figure}

Figure~\ref{fig:tf-heating-theo-simul} shows the connecting time
$t_{\fin}$ as a function of the target temperature $T_{\fin}$. Once
more, we consider the two-dimensional case and two different values of
the restitution coefficient, $\alpha=0.3$ and $\alpha=0.8$. The
simulation results are averaged over $100$ trajectories and compared
with the theoretical prediction \eqref{eq:tf0-tJ-hc}, showing a very
good agreement. The simulation curve is smoother for $\alpha=0.3$ than
for $\alpha=0.8$, because $|a_{2}^{\st}|$ is larger for the former.

\begin{figure}
  \centering
  \includegraphics[width=3.375in]{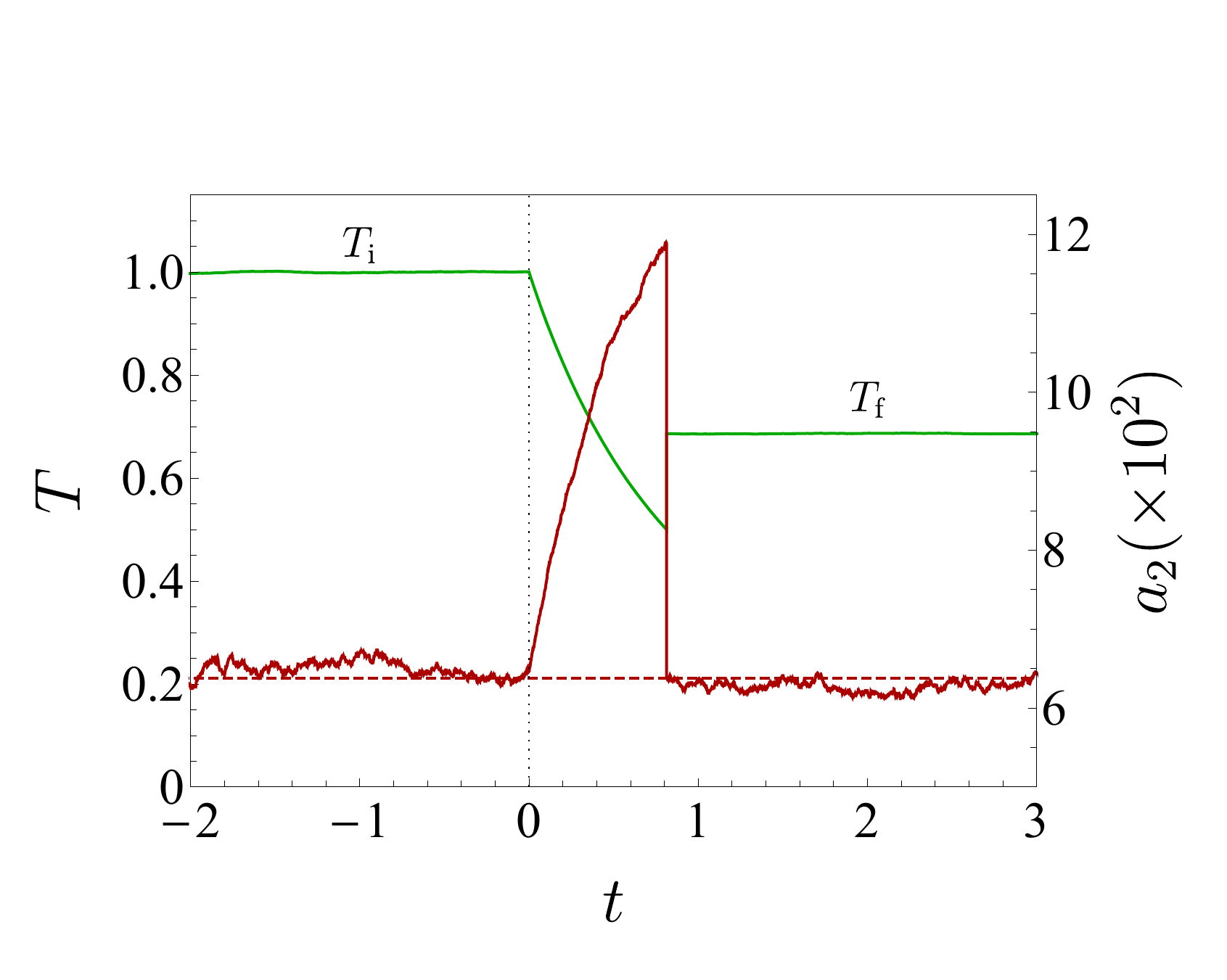}
  \caption{Typical simulation trajectory for the case
    $T_{\fin}<1$. This plot is similar to that in
    Fig.~\ref{fig:traj-heating}, but the order of the bangs is
    reversed: first, the granular gas freely cools ($\xi=0$) in the
    time interval $(0,t_{\fin})$ and second, at $t=t_{\fin}$, the
    system is instantaneously heated. Again, the thermostat is
    switched on with intensity $\chi_{\fin}=T_{\fin}^{3/2}$ at
    $t=t_{\fin}$ and the system remains stationary for $t>t_{\fin}$. }
  \label{fig:traj-cooling}
\end{figure}

In the case $T_{\fin}<1$, the cooling-heating bang-bang trajectory is
generated in the following way. First, the system freely cools from
the initial configuration, with $T_{\ini}=1$, until reaches a certain
configuration with $T_{J}<1$ and a larger---in absolute value---excess
kurtosis $a_{2J}$. Therefrom, we instantaneously heat the system by
changing the velocities as
$\bm{v}_{i}\to\bm{v}_{i}+\sum_{j=1}^{M}\bm{\eta}_{ij}$, where
$\bm{\eta}_{ij}$ are independent Gaussian distributed random variables
of a certain---small---variance. Note that, in contrast to the
heating-cooling case described before, this is not done in one step
but several.  This recurrent procedure stops when the excess
kurtosis---the absolute value of which is decreasing---equals its
steady value $a_{2}^{\st}$: this fixes the number of steps $M$ over
the considered trajectory. At this point, the temperature of the
system is $T_{\fin}$ and, again, the stochastic forcing is switched on
with intensity $\xi_{\fin}=\xi_{\ini}T_{\fin}^{3/4}$---this makes the
system stationary for longer times. A typical trajectory of the case
$T_{\fin}<1$ is depicted in Fig.~\ref{fig:traj-cooling}. Specifically,
the realisation shown corresponds to $d=2$ and $\alpha=0.3$ in a
system with $N=10^{6}$ particles.

\begin{figure}
  \centering
  \includegraphics[width=3.375in]{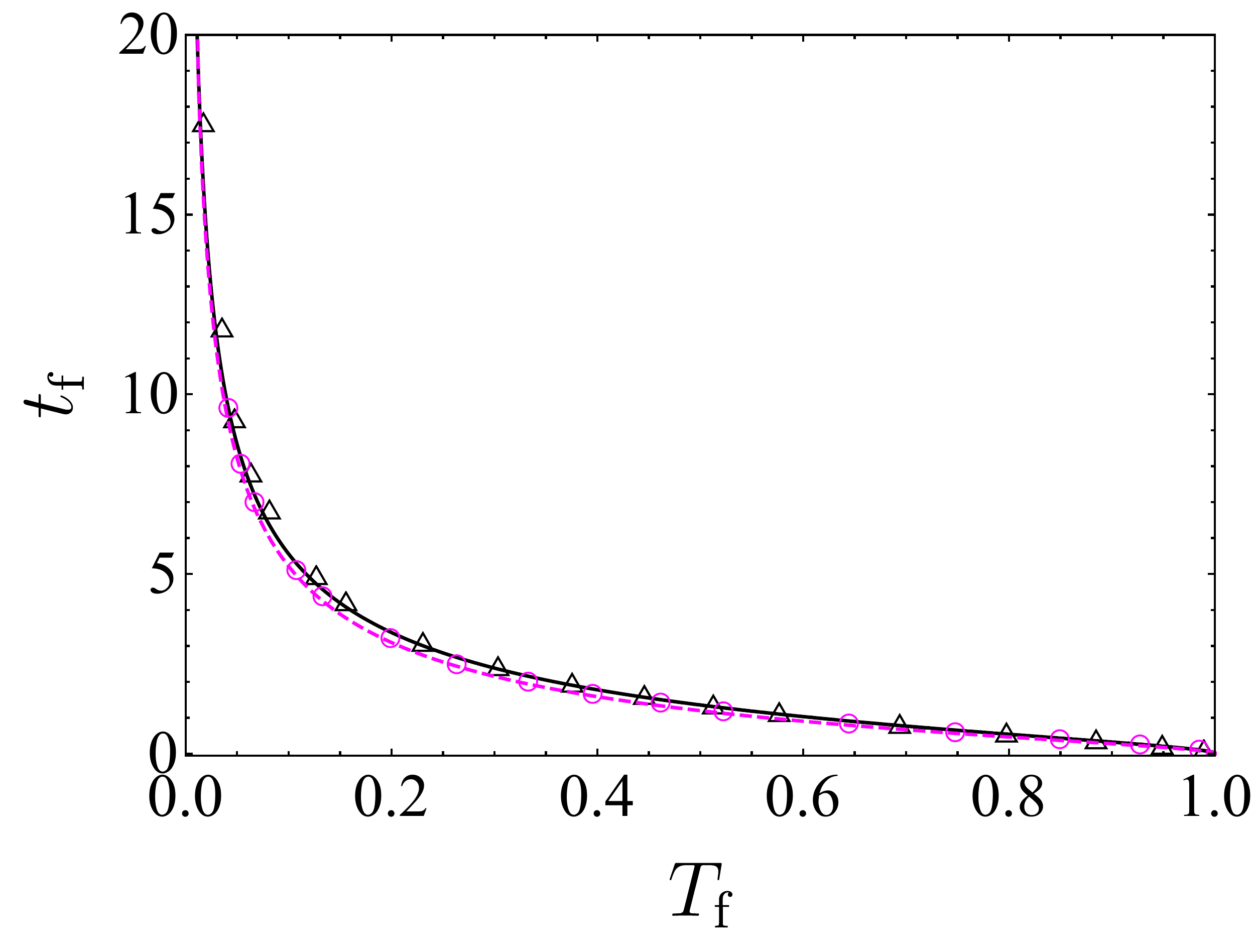}
  \caption{Connecting time as a function of the target
    temperature. Simulation results (symbols) are compared with
    Eq.~\eqref{eq:tf0-tJ-ch} (lines) for $d=2$ and two values of
    $\alpha$: $\alpha=0.3$ (open triangles, solid line) and
    $\alpha=0.8$ (open circles, dashed line).  }
  \label{fig:tf-cooling-theo-simul}
\end{figure}

We compare the numerical results for the connecting time with the
theoretical prediction, as given by Eq.~\eqref{eq:tf0-tJ-hc}, in
Figure~\ref{fig:tf-cooling-theo-simul}. Again, simulations correspond
to $d=2$, and $\alpha=(0.3,0.8)$. The agreement between theory and
simulations is excellent. Relative fluctuations seem to be smaller
than in Fig.~\ref{fig:tf-heating-theo-simul}, but it has to be taken
into account that the connection times here are longer.

\section{Prevalence of optimal bang-bang protocols}
\label{sec:generality-bang-bang}

In this section, we discuss the emergence of similar bang-bang
protocols as the optimal ones---in the sense of minimising the
connection time---in a wide class of physical systems, not only for
the specific case of granular fluids. The evolution equations of the
relevant physical quantities are often linear in the ``control
functions''~\cite{martinez_engineered_2016,chupeau_thermal_2018,kourbane-houssene_exact_2018,manacorda_lattice_2017,martikyan_comparison_2020}. In
addition, the controls (stiffness, temperature, diffusions
coefficient, noise strength) are non-negative in most of these
situations, which gives rise to non-holonomic constraints.

Non-holonomic restrictions make it necessary to resort to Pontryagin's
maximum principle to solve the problem of minimising the connection
time. When the evolution equations involve the controls $\lambda$
linearly, Pontryagin's Hamiltonian $\Pi$ is also linear in
them. Therefore, the the maximum of $\Pi$ is reached at the limit
values of the control, $\lambda_{\min}$ and $\lambda_{\max}$: the
bang-bang protocols arise in this way. Once the bang-bang protocols
are established as the optimal ones, a relevant question is the number
of ``bangs'' that are needed for the connection. The answer to this
question in each specific physical system depends on the number $n$ of
variables that characterise its state. If the evolution equations are
also linear in these physical variables---either exactly or as a
consequence of linearising them around a steady state, a rigorous
mathematical theorem ensures that the optimal bang-bang control has at
most $n-1$ switchings~\footnote{See theorem 10 in Sec.~III.17 of
  Pontryagin's book~\cite{pontryagin_mathematical_1987}, the main
  hypothesis of which is the eigenvalues of the evolution equations
  being real.}.

If the evolution equations are non-linear in the physical quantities,
as is the case of granular fluids, the following argument supports the
arising of a completely similar picture.  A simple bang---with two
possibilities, $\lambda(t)=\lambda_{\min}$ or
$\lambda(t)=\lambda_{\max}$---suffices for $n=1$. Its duration
$t_{\fin}$ makes it possible to adjust the final value of the only one
variable. Two bangs---with two possibilities,
$\lambda_{\max}-\lambda_{\min}$ or
$\lambda_{\min}-\lambda_{\max}$---suffice for $n=2$, with only one
switching at an intermediate time $t_{J}$. The joining time $t_{J}$
together with the connecting time $t_{\fin}$---or the duration of the
bangs $t_{1}=t_{J}$ and $t_{2}=t_{\fin}-t_{J}$---make it possible to
tune the final value of the two variables. This is the case we have
found here when studying the granular fluid in the Sonine
approximation. In general, $n$ bangs---with two possibilities,
starting with either $\lambda_{\max}$ or $\lambda_{\min}$---suffice
for a generic value of $n$, with $n-1$ switchings at intermediate
times $t_{J_{1}}$, $t_{J_{2}}$, \ldots, $t_{J_{n-1}}$. These $n-1$
joining times together with the connecting time $t_{\fin}$---or the
duration of the $n$ bangs $t_{1}=t_{J_{1}}$, $t_{2}=t_{J_{2}}-t_{J_{1}}$,
\ldots, $t_{n}=t_{\fin}-t_{J_{n-1}}$---make it possible to tune the
final value of the $n$ variables.

\subsection{Brownian particle moving in a d-dimensional harmonic potential}

As a proof of concept, let us consider a Brownian particle trapped in
a $d$-dimensional harmonic potential. Our analysis will be carried
out in the overdamped limit, in which the probability distribution
$P(\bm{x},t)$ of the particle's position obeys the Fokker-Planck
equation
\begin{equation}\label{eq:FP-3d-osc}
  \gamma
  \partial_{t}P(\bm{x},t)=\nabla\cdot
  \left[\nabla U_{h}(\bm{x})\,P(\bm{x},t)\right]+
    k_{B}T(t)\, \nabla^{2}\! P(\bm{x},t),
\end{equation}
where $\gamma$ is the friction coefficient and
$U_{h}(\bm{x})=\frac{1}{2}\sum_{j,k=1}^{d}\lambda_{jk}x_{j}x_{k}$. We
want to connect two equilibrium states in the minimum possible time,
by controlling the temperature $T(t)$ of the bath inside which the
colloidal particle is immersed.

It is fitting to describe the particle position in terms of the three
normal coordinates $\xi_{i}$, such that the harmonic potential is
diagonal in them. Therefore, we write
$U_{h}(\bm{x})=U_{h}(\bm{\xi})=\frac{1}{2}\sum_{\beta=1}^{d}\kappa_{\beta}\,
\xi_{\beta}^{2}$, where $\kappa_{\beta}>0$ are the eigenvalues of the
matrix of elements $k_{ij}$. The time evolution of the particle is
uniquely characterised by the value of the variances of the normal
modes, $\sigma_{\beta}^{2}\equiv \langle\xi_{\beta}^{2}\rangle$. In
dimensionless variables, they evolve according to
\begin{equation}\label{eq:sigma-evol}
  \frac{d}{dt}\sigma_{\beta}^{2}=-2\kappa_{\beta}\sigma_{\beta}^{2}+2T(t),
  \quad 1\leq\beta\leq d.
\end{equation}
Similarly to what we did in the analysis of the granular fluid, the
units that non-dimensionalise variables are chosen to simplify our
formulas: here, $T_{\ini}=1$ and $\kappa_{1}=1$. Without loss of
generality, we choose to label the modes such that
$\kappa_{1}=1\leq \ldots \leq \kappa_{d}$. See
Appendix~\ref{sec:3d-harm-pot} for details.

In the initial and final equilibrium states, with respective initial
temperatures $T_{\ini}=1$ and $T_{\fin}$, we have
\begin{equation}\label{eq:sigma-bc}
  \sigma_{\beta,\ini}^{2}\equiv\sigma_{\beta}^{2}(t=0)=
  \frac{1}{\kappa_{\beta}},
  \; \sigma_{\beta,\fin}^{2}\equiv\sigma_{\beta}^{2}(t=t_{\fin})=
  \frac{T_{\fin}}{\kappa_{\beta}}.
\end{equation}
The connection between the initial and final
states is done by controlling the temperature of the bath $T(t)$,
which obeys the non-holonomic constraint $T\geq 0$. Since the
evolution equations \eqref{eq:sigma-evol} are linear in the control
$T(t)$, the optimal connection is of bang-bang type: it comprises
several time windows with either $T(t)=T_{\max}$ or $T(t)=T_{\min}=0$.

An exhaustive analysis of the optimal connection for this system,
investigating in detail the behaviour of the connection time
throughout the whole space of parameters
$(\bm{\kappa},T_{\max},T_{\fin})$, where
$\bm{\kappa}=(\kappa_{1},\ldots,\kappa_{d})$, is outside the scope of this
paper. Here, we focus on the general trends of the optimal connecting
time as a function of the final temperature $T_{\fin}$ for some
specific choices of $\bm{\kappa}$, in order to show the generality of
the arguments we have put forward above. For the sake of simplicity,
we consider the limiting case $T_{\max}\to\infty$--- in analogy with
our study of the granular fluid, in which the maximum noise strength
$\chi_{\max}\to\infty$.

\paragraph*{One-dimensional case.-} The simplest situation is that of
the one-dimensional case, where we have only one equation to control.
Thus, one bang suffices of duration $t_{\fin}^{\tI}$. It is
interesting to remark that the same situation appears in $d=2$ or
$d=3$ for the isotropic or central symmetry situation, in which all
the $\kappa_{\beta}$ are identical: both \eqref{eq:sigma-evol} and
\eqref{eq:sigma-bc} do not depend on the mode and all
$\sigma_{\beta}(t)$ are also identical. From an effective point of
view, we have $n=1$ and the control problem is equivalent to that of
the one-dimensional case.

On the one hand, in the ``heating'' bang, $T(t)=T_{\max}$,
$\sigma_{1}^{2}(t)=T_{\max}-\left(T_{\max}-1\right)e^{-2t}$ and the
connection time is determined by
$\sigma_{1,\fin}^{2}=T_{\max}-\left(T_{\max}-1\right)e^{-2t_{\fin}^{^{\tI}}}
=T_{\fin}$.  In the limit as $T_{\max}\to\infty$, this expression
simplifies to
\begin{equation}
  \sigma_{1,\fin}^{2}=1+\alpha=T_{\fin}, \quad
  \alpha=2T_{\max}t_{\fin}^{\tI} \text{\quad finite.}
\end{equation}
Therefore, we have that
\begin{equation}\label{eq:tfI-h}
  t_{\fin}^{\tI}\sim \frac{1}{2}\frac{T_{\fin}-1}{T_{\max}}\to 0^{+}, \quad
  T_{\fin}>1. 
\end{equation}
Thus, any $T_{\fin}>1$ can be reached by an instantaneous, infinitely
strong, jump to $T_{\max}\to\infty$ that is stopped as soon as
$\sigma_{1}^{2}$ attains its final value. On the other hand, in the
``cooling'' bang, $T(t)=0$, $\sigma_{1}^{2}=e^{-2t}$ and the
connection time is determined by
\begin{equation}\label{eq:tfI-c}
  \sigma_{1,\fin}^{2}=e^{-2t_{\fin}^{\tI}}=T_{\fin} \implies
  t_{\fin}^{\tI}=-\frac{1}{2}\ln T_{\fin}, \quad T_{\fin}<1. 
\end{equation}
As a consequence of the non-holonomic constraint $T(t)\geq 0$, we have
that the optimal time is finite for $T_{\fin}<1$. 

\paragraph*{Two-dimensional case.-} Next we look into a the
two-dimensional case, with $\kappa_{2}\ne\kappa_{1}$. We have two
equations to control and thus $n=2$.  It is clear that only one bang
is not enough: in a ``cooling'' bang of duration $t_{\fin}^{\tI}$ we
would have
$\sigma_{2,\fin}^{2}=e^{-2\kappa_{2}t_{\fin}^{\tI}}/\kappa_{2}\ne
T_{\fin}/\kappa_{2}$. Therefore, we need two bangs: either
``heating-cooling'' or ``cooling-heating'', consistently with our
general discussion for $n=2$. Note that for $d=3$ but with axial
symmetry, for example $\sigma_{2}(t)=\sigma_{3}(t)$, the control
problem is equivalent to the two-dimensional situation considered here.

First, we analyse the ``heating-cooling'' bang-bang. In the first
step, we have an instantaneous heating with $T(t)=T_{\max}\to\infty$
that leads to
\begin{equation}\label{eq:sigmabJ-two-var}
  \sigma_{\beta,J}^{2}=\frac{1}{\kappa_{\beta}}+\alpha, \quad
  \alpha=2T_{\max}t_{J}, \qquad \beta=1,2.
\end{equation}
The duration $t_{1}=t_{J}$ of the heating bang vanishes in the limit as
$T_{\max}\to\infty$. In the second step, we put $T(t)=0$ and thus
$\sigma_{\beta}^{2}(t)=\sigma_{\beta,J}^{2}e^{-2\kappa_{\beta}(t-t_{J})}$. Bringing
to bear \eqref{eq:sigmabJ-two-var}, we have for the final time $t=t_{\fin}^{\tII}$
\begin{equation}\label{eq:sigmabf-two-var}
  \sigma_{\beta,\fin}^{2}=\left(\frac{1}{\kappa_{\beta}}+\alpha\right)
  e^{-2\kappa_{\beta}t_{\fin}^{\tII}}=
  \frac{T_{\fin}}{\kappa_{\beta}}, \qquad \beta=1,2.
\end{equation}
This equation provides us with both $\alpha$ and $t_{\fin}^{\tII}$ as
functions of $T_{\fin}$. Specifically, we can eliminate $\alpha$ to
get,
\begin{equation}\label{eq:tfII-h}
  T_{\fin}=\frac{\kappa_{2}-1}{\kappa_{2}\,e^{2t_{\fin}^{\tII}}-
e^{2\kappa_{2}t_{\fin}^{\tII}}}, \quad T_{\fin}>1,
\end{equation}
which gives the optimal connection time $t_{\fin}^{\tII}$ as an
implicit function of $T_{\fin}$, but only for $T_{\fin}>1$ as
marked. For $T_{\fin}<1$, it does not have positive solutions for the
connection time. Again in analogy with the granular case, the
``heating-cooling'' bang-bang allows us to connect states in which the
final temperature is larger than the initial one.

Second, we study the ``cooling-heating'' bang-bang. In this case, the
first step has a duration $t_{1}=t_{J}$ whereas the second one is
instantaneous, $t_{2}\to 0$. Therefore, $t_{\fin}^{\tII}=t_{J}$ and we can
directly write
\begin{equation}
  \sigma_{\beta,J}^{2}=\frac{1}{\kappa_{\beta}}e^{-2\kappa_{\beta}t_{\fin}^{\tII}},
  \quad \sigma_{\beta,\fin}^{2}=\sigma_{\beta,J}^{2}+\alpha, \qquad \beta=1,2,
\end{equation}
with $\alpha=2T_{\max}t_{2}$. Again, eliminating $\alpha$ we obtain
\begin{equation}\label{eq:tfII-c}
  T_{\fin}=\frac{\kappa_{2}\,e^{-2t_{\fin}^{\tII}}-e^{-2\kappa_{2}t_{\fin}^{\tII}}}
  {\kappa_{2}-1},  \quad T_{\fin}<1.
\end{equation}
Now, the above equation does not have physical solutions---positive
connection time---for $T_{\fin}>1$. The ``cooling-heating'' bang-bang
connects equilibrium states with smaller final temperature.

\paragraph*{Three-dimensional case.-} Finally, we investigate the case
in which $d=3$ and all the $\kappa_{\beta}$ are different.  Then, we
have three equations to control and $n=3$~\footnote{This corresponds
  to an anisotropic potential, recall that the axial symmetry and
  spherical symmetry situations correspond to $n=2$ and $n=1$,
  respectively.}. None of the previous one-bang or two-bangs protocols
can accomodate the final value of the three variances
$\sigma_{\beta}^{2}$: we need two intermediate jumps at times
$t_{J_{1}}$ and $t_{J_{2}}$. Then, a three-bang protocol of duration
$t_{\fin}^{\tIII}$ emerges with two possibilities:
``heating-cooling-heating'' (HCH) or ``cooling-heating-cooling'' (CHC). From
our previous discussions of the cases $n=1$ and $n=2$, physical
intuition tells us that HCH corresponds to $T_{\fin}>1$ and CHC to
$T_{\fin}<1$ for $n=3$. In the following, we rigorously show that this
is indeed the case.

We start by analysing the HCH bang-bang. In the limit as
$T_{\max}\to\infty$, we know that the first and third heating steps
are instantaneous: their durations are $t_{1}=t_{J_{1}}\to 0$,
$t_{3}=t_{\fin}^{\tIII}-t_{J_{2}}\to 0$. Therefore, the second cooling
step has a duration $t_{2}=t_{J_{2}}-t_{J_{_{1}}}\to t_{\fin}$. In
addition, we introduce the notation $\alpha_{1}=2T_{\max}t_{J_{1}}$
and $\alpha_{2}=2T_{\max}(t_{\fin}^{\tIII}-t_{J_{2}})$.  We obtain
\begin{widetext}
\begin{equation}                            \sigma_{\beta,J_{1}}^{2}=\frac{1}{\kappa_{\beta}}+\alpha_{1}, \quad
\sigma_{\beta,J_{2}}^{2}=\sigma_{\beta,J_{1}}^{2} 
e^{-2\kappa_{\beta}t_{\fin}^{\tIII}}, \quad
\sigma_{\beta,\fin}^{2}=\sigma_{\beta,J_{2}}^{2}+\alpha_{2},
\end{equation}
which entails
\begin{equation}\label{eq:sigmabf-hch}
  \frac{T_{\fin}}{\kappa_{\beta}}=\left( \frac{1}{\kappa_{\beta}}+
    \alpha_{1}\right)e^{-2\kappa_{\beta}t_{\fin}^{\tIII}} +\alpha_{2}, \qquad \beta=1,2,3.  
\end{equation}
Equation \eqref{eq:sigmabf-hch} gives us
$(\alpha_{1},\alpha_{2},t_{\fin}^{\tIII})$ as a function of
$T_{\fin}$. Eliminating $\alpha_{1}$ and $\alpha_{2}$, we get
\begin{equation}\label{eq:tfIII-h}
  T_{\fin}=\frac{e^{2t_{\fin}^{\tIII}}(\kappa_{3}-\kappa_{2})-
    e^{2\kappa_{2}t_{\fin}^{\tIII}}(\kappa_{3}-1)+e^{2\kappa_{3}t_{\fin}^{\tIII}}(\kappa_{2}-1)}
  {e^{2(1+\kappa_{3})t_{\fin}^{\tIII}}\kappa_{2}(\kappa_{3}-1)-
    e^{2(1+\kappa_{2})t_{\fin}^{\tIII}}\kappa_{3}(\kappa_{2}-1)
    -e^{2(\kappa_{2}+\kappa_{3})t_{\fin}^{\tIII}}(\kappa_{3}-\kappa_{2})}.
\end{equation}
Plotting the rhs as a function of $t_{\fin}^{\tIII}$ shows that in fact
$T_{\fin}>1$, in agreement with our physical intuition.

The analysis of the CHC bang-bang follows the same lines as above. The
first cooling step has duration $t_{1}=t_{J_{1}}$, the second heating
step is instantaneous $t_{2}=t_{J_{2}}-t_{J_{1}}\to 0$, and the third
cooling step lasts for $t_{3}=t_{\fin}^{\tIII}-t_{J_{2}}\to
t_{\fin}^{\tIII}-t_{1}$. The composition of the three bangs gives at the final
time
\begin{equation}
  \frac{T_{\fin}}{\kappa_{\beta}}=
  \left(\frac{1}{\kappa_{\beta}}e^{-2\kappa_{\beta}t_{1}}+
    \alpha\right)e^{-2\kappa_{\beta}(t_{\fin}^{\tIII}-t_{1})}, \qquad \beta=1,2,3,
\end{equation}
where $\alpha=2T_{\max}t_{2}$ is the strength of the heating
bang. Solving for $t_{\fin}^{\tIII}$ and $T_{\fin}$, we get
\begin{equation}\label{eq:tfIII-c}
  \left(T_{\fin}-e^{-2t_{\fin}^{\tIII}}\right)^{\kappa_{3}-\kappa_{2}}
  \left(T_{\fin}-e^{-2\kappa_{2}t_{\fin}^{\tIII}}\right)^{1-\kappa_{3}}
   \left(T_{\fin}-e^{-2\kappa_{3}t_{\fin}^{\tIII}}\right)^{\kappa_{2}-1}=
   \kappa_{3}^{\kappa_{2}-1}\kappa_{2}^{1-\kappa_{3}}. 
 \end{equation}
 \end{widetext}
 The condition $\alpha\geq 0$ entails that all the terms inside the
 parentheses must be non-negative. Since $\kappa_{3}>\kappa_{2}>1$, this
 means that $T_{\fin}>e^{-2t_{\fin}^{\tIII}}$ or
 $t_{\fin}^{\tIII}>-\ln T_{\fin}/2=t_{\fin}^{\tI}$. In fact,
 we would have $t_{\fin}^{\tIII}>t_{\fin}^{\tII}>t_{\fin}^{\tI}$ if,
 from going from I to III, one incorporates a new value of $\kappa$
 while keeping the previous ones---i.e. if $\kappa_{1}$ and $\kappa_{2}$
 in III are the same as in II. This is logical, the incorporation
 of additional variables reduces the size of the set of ``admissible''
 controls---i.e. those connecting the initial and final states---and
 thus increases the minimum connection time.

\begin{figure}
  \centering
  \includegraphics[width=3.375in]{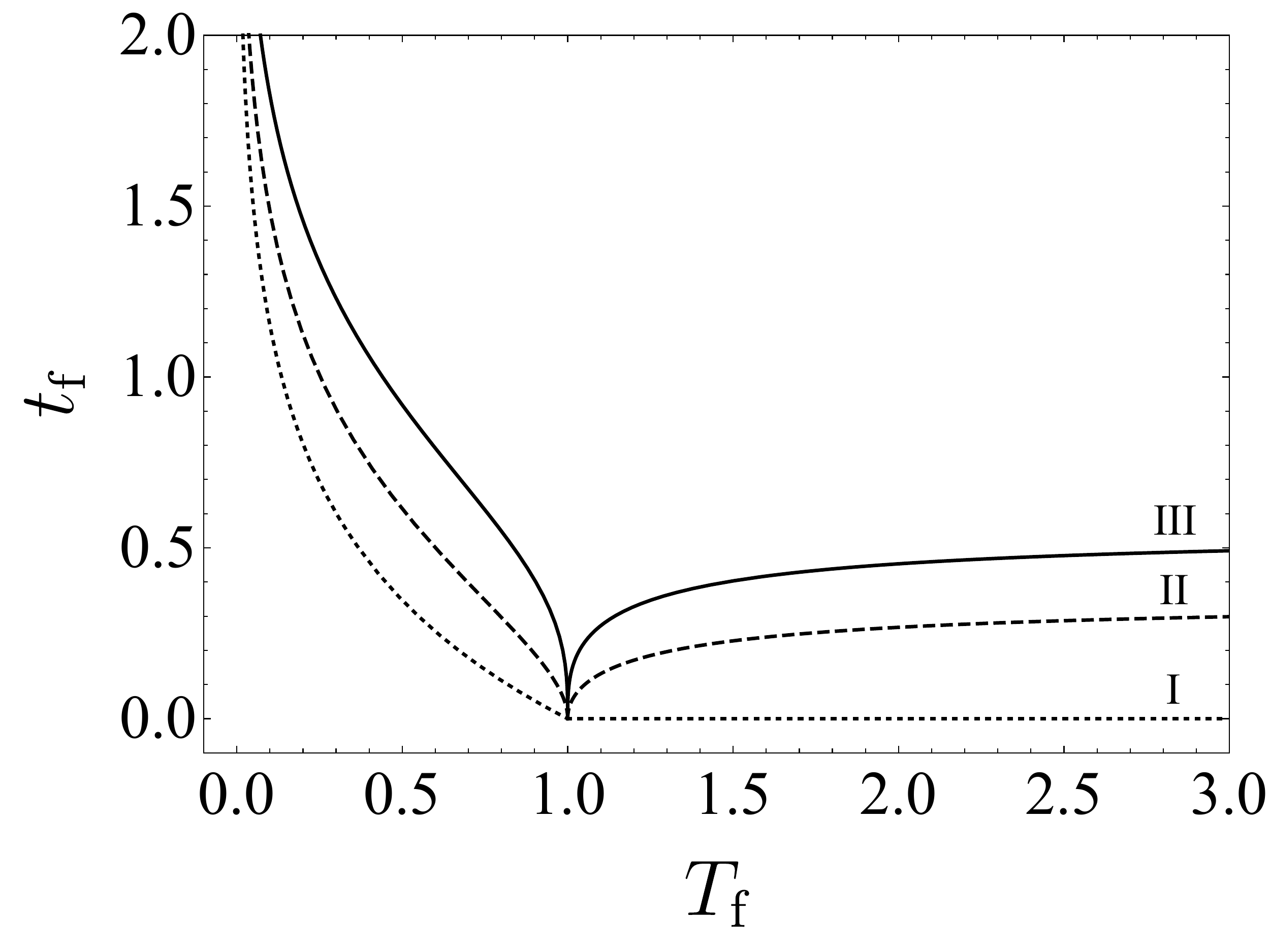}
  \caption{Connecting time as a function of the target temperature for
    the $d$-dimensional oscillator. Specifically, we have plotted
    three cases: (a) $t_{\fin}^{\tI}$ given by \eqref{eq:tfI-h} and
    \eqref{eq:tfI-c} for the one-dimensional oscillator,
    $\kappa_{1}=1$ (or $d$-dimensional one with central simmetry,
    $\kappa_{\beta}=1$ $\forall\beta$); (b) $t_{\fin}^{\tII}$ given by
    \eqref{eq:tfII-h} and \eqref{eq:tfII-c} for the two-dimensional
    oscillator (or three-dimensional with axial symmetry), with
    $\bm{\kappa}=(1,2)$; and (c) $t_{\fin}^{\tIII}$ given by
    \eqref{eq:tfIII-h} and \eqref{eq:tfIII-c} for the
    three-dimensional case, with $\bm{\kappa}=(1,2,3)$.  }
  \label{fig:tf-vs-Tf-three-modes}
\end{figure}

 Figure~\ref{fig:tf-vs-Tf-three-modes} shows the connection time as a
 function of the final temperature, for three sets of the model
 parameters $(\kappa_{1}=1,\kappa_{2},\kappa_{3})$. In accordance with
 the physical picture of the previous paragraph, it is neatly observed
 that $t_{\fin}^{\tIII}>t_{\fin}^{\tII}>t_{\fin}^{\tI}$. In case I,
 the minimum connecting time bahaves similarly to the Gaussian
 approximation of the granular system, $t_{\fin}^{\tI}=0$ for
 $T_{\fin}>1$ whereas it increases as the final temperature decreases
 for $T_{\fin}<1$, diverging in the limit as $T_{\fin}\to 0$. In case
 II, the observed behaviour is also qualitatively similar with that of
 the Sonine description of the granular system, the main difference
 being that $t_{\fin}^{\tII}$ is no longer zero for $T_{\fin}>1$. The
 existence of two variables makes it impossible to connect the two
 states instantaneously, because the cooling part also involves a
 finite time due to the non-holonomic constraint $T\geq 0$---analogous
 to $\chi\geq 0$ in the granular case. The main difference is the
 finite value of $t_{\fin}^{\tII}$ for the two-dimensional
 oscillator---or a three-dimensional oscillator with axial
 symmetry---in the limit as $T_{\fin}\to\infty$, in contrast with the
 vanishing of the connecting time for the granular case. The latter
 behaviour stems from the collision rate being proportional to
 $\sqrt{T}$, which also accelerates the cooling step of the bang-bang
 in the granular case, whereas the relaxation rates of the harmonic
 modes $\kappa_{\beta}$ are independent of the temperature. In case
 III, the incorporation of the third mode increases the connection
 time, $t_{\fin}^{\tIII}>t_{\fin}^{\tII}$, but keeps the qualitative
 picture unchanged.

\section{Discussion}\label{sec:discussion}

In this work, we have applied information geometry and control theory
ideas to a system described at the kinetic level. The resulting
framework is physically appealing. On the one hand, information
geometry put lower bounds on the operating time: the classical speed
limits in
Refs.~\cite{ito_stochastic_2020,nicholson_timeinformation_2020} apply
although the dynamics is not Markovian. On the other hand, control
theory makes it possible to build protocols that entail large
accelerations of the system dynamics, by minimising the connection
time.

It is known that inverse engineering techniques---such as engineered
swift
equilibration~\cite{martinez_engineered_2016,
  muratore-ginanneschi_application_2017,li_shortcuts_2017,
  chupeau_engineered_2018,baldassarri_engineered_2020}---can
connect states in times that are shorter than the empirical relaxation
time. Here, we show that optimal control protocols not only are able
to beat the empirical relaxation time for relaxation---by more than an
order of magnitude---but also the recently derived classical speed
limits---which are considerably shorter than the empirical time. The
latter 
play in classical systems a role
similar to that of the quantum speed limits---associated with the
time-energy uncertainty relation---in quantum systems. This beating of
the classical speed limit for relaxation does not represent a
contradiction, since the optimal control protocols involve a
time-dependent driving.

There appears a clear asymmetry between the cases $T_{\fin}>1$ and
$T_{\fin}<1$---recall that in our dimensionless units the initial
temperature equals unity. For the case $T_{\fin}>1$, the optimal
connecting times are rather small, vanishing in the limits
$T_{\fin}\to 1$ and $T_{\fin}\to\infty$. The smallness of the minimum
connecting times can be understood in a physical way: in the Gaussian
approximation, the minimum connecting time vanishes because the
optimal protocol is clearly a pulse of very high noise intensity such
that the granular temperature instantaneously changes from $1$ to
$T_{\fin}$. Therefore, it is non-Gaussianities---specifically, the
excess kurtosis $a_{2}$ that is small---that make impossible to
instantaneously connect the two NESS for $T_{\fin}>1$. The excess
kurtosis decreases and therefore the state
after the instantaneous heating pulse is not stationary.

For the case $T_{\fin}<1$, the minimum connecting times are longer
than those for heating. Again, this can be understood from the
Gaussian approximation: therein, the optimal protocol is letting the
system freely cool, i.e. with driving intensity $\chi=0$. At
difference with the case $T_{\fin}>1$, the minimum connection time for
$T_{\fin}<1$ does not vanish because free cooling involves a finite
time.  Interestingly, both for $T_{\fin}>1$ and $T_{\fin}<1$
non-Gaussianities make the connecting times longer: this is physically
understood by taking into account that non-Gaussianities stem from the
inelasticity of collisions.

One of the main results of our paper is the emergence of optimal
bang-bang protocols for minimising the connecting time.  In our case,
the bang-bang processes comprise two steps (i.e. one switching): (i)
instantaneous heating with a very high driving intensity
$\chi_{\max}\to\infty$ and (ii) free cooling, i.e. no driving,
$\chi=0$. The order of the bangs is different for $T_{\fin}>1$ and
$T_{\fin}<1$: heating-cooling for $T_{\fin}>1$, but cooling-heating
for $T_{\fin}<1$.  Qualitatively, this can be understood as follows:
in both cases, the first step corresponds to what would be done in the
Gaussian approximation. However, the existence of non-Gaussianities
entail that the excess kurtosis does not have the stationary value at
the end of the first step. This imbalance is somehow mended by the
second step of the bang-bang.


Indeed, bang-bang processes are expected to emerge as the optimal
protocols---in the sense of minimising the connection time---in a wide
variety of physical situations, not only for the specific case of
granular fluids. The evolution equations of the relevant physical
properties typically include ``control functions'': other quantities,
the time-dependence of which can be externally controlled. Often, the
control function $\lambda$---stiffness of a harmonic trap, temperature
of the bath, diffusion coefficient, noise strength, density,
etc. depending on the physical context---verify that (i) the evolution
equations of the physical properties are linear in them, and (ii) a
non-holonomic constraint limits their physically acceptable values,
$\lambda_{\min}\leq\lambda\leq\lambda_{\max}$. Examples abound, from
the trapped Brownian
particle~\cite{martinez_engineered_2016,chupeau_thermal_2018} or
active
systems~\cite{kourbane-houssene_exact_2018,manacorda_lattice_2017} to
a particle moving in an electric
field~\cite{martikyan_comparison_2020}). Therein, the mathematical
structure of Pontryagin's principle ensures that the optimal controls
minimising the connecting time are of bang-bang type.

Let us consider thus a physical system such that, at the macroscopic
(or hydrodynamic, or thermodynamic,\ldots) level of description is
described by $n$ physical quantities~\footnote{Also at the mesoscopic
  level of description (fluctuating hydrodynamics, stochastic
  thermodynamics,\ldots), which incorporates fluctuations of these
  quantities to the picture.}. The number $n$ is thus small, in the
examples above
$n=1$~\cite{martinez_engineered_2016,chupeau_thermal_2018,martikyan_comparison_2020}
or
$n=2$~\cite{kourbane-houssene_exact_2018,manacorda_lattice_2017}---as
is the case of the granular fluid in the Sonine approximation. A
relevant question is: how many bangs, i.e. how many time windows
inside which either $\lambda(t)=\lambda_{\min}$ or
$\lambda(t)=\lambda_{\max}$, are necessary to make the optimal
connection? For $n=1$, a simple bang---either
$\lambda(t)=\lambda_{\min}$ or $\lambda(t)=\lambda_{\max}$---of
duration $t_{\fin}$ makes it possible to adjust the final value of the
single variable. 
For $n=2$, there will be a mismatch between the target value of the
second variable and the one obtained with a simple bang that tunes the
final value of the first variable. This makes it necessary to
introduce a second step with the control being switched to the
opposite limit: two bangs---either $\lambda(t)=\lambda_{\max}$
followed by $\lambda(t)=\lambda_{\min}$, or viceversa---of durations
$t_{1}$ and $t_{2}$, with $t_{\fin}=t_{1}+t_{2}$, allow for matching
the final values of two variables, and thus suffices for $n=2$. In
general, we have to introduce $n-1$ jumps at intermediate times to
allow for matching all $n$ variables: the number of bangs equals the
number of variables.

Specifically, we have shown that the above picture is indeed the
correct one by solving a simple but relevant physical situation: the
compression/decompression of a Brownian particle trapped in a
$d$-dimensional harmonic potential, by controlling the temperature
of the bath~\footnote{Optical confinement makes it possible to control
  the time dependence of the effective temperature seen by the
  Brownian particle by randomly shaking the confining trap or by using
  Brownian particles with an inherent charge submitted to a random
  electric field~\cite{martinez_effective_2013}.}. 
For $n=2$ (axial symmetry for $d=3$ or $d=2$), a
two-step bang-bang process, qualitatively similar to that found in the
granular fluid, carries out the optimal connection---heating-cooling
or cooling-heating, depending on the final value of the temperature
being larger or smaller than the initial one. For $d=3$, an optimal three-step bang-bang process arises:
heating-cooling-heating or cooling-heating-cooling, again depending on
the final value of the temperature.


Our work has been focused on the minimisation of the connecting time
$t_{\fin}$ between two NESS of the granular fluid. Not only is the
optimisation of the connection time between two given states a
relevant problem from a fundamental point of view, but also has
potential applications in different contexts. For example, minimising
the connection time in the adiabatic---in the sense of zero
heat---branches is essential for building a finite-time version of
Carnot's heat engine that maximises the delivered
power~\cite{plata_building_2020}. Also, the optimisation of the
relaxation route to equilibrium or to a NESS is of interest in
connection with behaviours such as the Mpemba effect, which is
currently a very active field of
research~\cite{lu_nonequilibrium_2017,lasanta_when_2017,
  baity-jesi_mpemba_2019,gal_precooling_2020,
  kumar_exponentially_2020,lapolla_faster_2020}.

In addition, we have considered the associated statistical length
$\calL$ and cost $\calC$ over the optimal processes in the granular
fluid. The length of the optimal bang-bang protocols is always longer
that that of the relaxation process, which contributes to increase the
bound for the connecting time---recall that
$t_{\fin}\geq \calL^{2}/(2\calC)$. However, this is compensated by the
cost, which is dominated by a term proportional to the maximum value
of the noise intensity $\chi_{\max}\to\infty$---i.e. the cost
diverges. Therefore, the bound goes to zero and the connecting time
may---and we have shown that this is indeed the case here---beat the
speed limit for relaxation.

Had we minimised the cost, we would have obtained an infinite
operation time. Therein, the system would be for all times in the NESS
corresponding to the instantaneous value of the noise intensity and
thus the cost would vanish. The divergence of the operation
time---when minimising the cost---is the counterpart of the divergence
of the cost---when minimising the connection time. But the analogy
ends there. We have shown in the granular gas that the minimum
connecting time does not vanish despite the diverging cost: the
cooling part of the bang-bang protocol involves a finite time. Our
general arguments above, and the specific example of the
$d$-dimensional oscillator, show that this will be also the case in
many physical situations where a non-holonomic constraint is present,
e.g. when controls are non-negative.

Our approach opens interesting perspectives for further research. In
the context of granular systems, it is far from trivial to rigorously
prove the global stability of the long-time NESS. Indeed, there are
strong signs, but not a formal proof, that it is the relative
Kullback-Leibler divergence with respect to the stationary
distribution---and not Shannon's entropy---that acts as a Lyapunov
functional~\cite{marconi_about_2013,garcia_de_soria_towards_2015,
  plata_global_2017,megias_kullbackleibler_2020}. In
this sense, the role of the Fisher information for rigorously
establishing the H-theorem for granular gases is worth
investigating.

For Markovian dynamics, the cost $\calC$ has been related in general
to entropy
production~\cite{nicholson_nonequilibrium_2018,ito_stochastic_2018,
  hasegawa_uncertainty_2019,ito_stochastic_2020}. In
the realm of kinetic theory and, more specifically, for the granular
case, the situation is far more complicated. Even admitting Shannon's
as the good definition of entropy for the granular case, there is not
a clear-cut way of splitting entropy production into ``irreversible''
and ``flux'' contributions, as discussed in
Ref.~\cite{bena_stationary_2006}. Therefore, elucidating the physical
meaning of information geometry's cost---beyond stating that it is the
physical quantity appearing in the thermodynamic uncertainty relation
for the connecting time---in granular fluids is a
relevant problem that remains to be solved.

Kinetic theory tools are not restricted to low density---or moderate
density if using Enskog's equation instead of Boltzmann's---gases,
either molecular or granular. They have also been successfully applied
to other intrinsically non-equilibrium systems such as active
matter~\cite{baskaran_enhanced_2008,baskaran_nonequilibrium_2010,
  ihle_kinetic_2011,marchetti_hydrodynamics_2013,
  ihle_chapmanenskog_2016,bonilla_contrarian_2019}. Also, the
classical kinetic approach holds for dilute ultracold gases: despite
the very low temperatures, they are still far from the threshold at
which quantum effects cease to be
negligible~\cite{lobser_observation_2015,guery-odelin_ultracold_2015,
  hohmann_individual_2017}. Therefore, it is worth looking into the
application of information-geometry concepts and, in general, the
extension of our results to these physical contexts.

\appendix
\section{Classical speed limits and geodesic properties in the
  Gaussian approximation }\label{sec:inform-geom}

In this appendix, we derive explicit expressions for the speed limits
for the relaxation process and analyse some properties of the geodesic
in probability space. The analysis is carried out in the Gaussian
approximation, where the granular gas is completely described by the
granular temperature $T(t)$. The velocity distribution function is
assumed to be the $d$-dimensional Maxwellian
\begin{equation}
  \label{eq:Maxwell-dim-d}
  P_{G}(\bm{v};T(t))=
  \left(2\pi T\right)^{-d/2}
  \exp\left(-\frac{v^{2}}{2T}\right).
\end{equation}
The temperature obeys the evolution equation
\begin{equation}\label{eq:T-evol-Gaussian}
  \dot{T}=\chi-T^{3/2},
\end{equation}
which stems from Eqs.~\eqref{eq:evol-non-dim-T} and
\eqref{eq:f1-def} with $a_{2}^{\st}=0$. Throughout, we employ the
subindex $G$ to denote those quantities calculated within the Gaussian
approximation.

\subsection{Classical speed limits for the relaxation process}

First, we calculate the Fisher information. Its definition
\eqref{eq:Fisher-inf-def} entails that
$I_{G}(t)=\langle\left[\partial_{t}\ln
  P_{G}(\bm{v},T(t))\right]^{2}\rangle_{G}$, where
$\langle\cdots\rangle_{G}$ means average with the Gaussian
distribution in Eq.~\eqref{eq:Maxwell-dim-d}. Making use of
\begin{equation}
  \label{eq:dt-log-PG}
  \partial_{t}\ln
  P_{G}(\bm{v};T(t))=
  \frac{\dot{T}(t)}{2T(t)}\left(-d+\frac{v^{2}}{T}\right),
\end{equation}
and the fact that
$\langle v^{4}\rangle=(d+2)\langle v^{2}\rangle^{2}/d=d(d+2)
T^{2}$ for a Gaussian distribution, it is readily shown that
\begin{align}
  \label{eq:I-Gaussian}
  I_{G}(t)&=
            \frac{d}{2}\left( \frac{\dot{T}(t)}{T(t)}\right)_{\!\!G}^{\!\!2}.
\end{align}
The subindex $G$ on the rhs means that we have to consistently
evaluate $T(t)$ within the Gaussian approximation, i.e. over the
solution of Eq.~\eqref{eq:T-evol-Gaussian}.

Second, the
Bhattacharyya angle is also obtained from its definition,
Eq.~\eqref{eq:bhatta-angle}. Specifically, we calculate the angle
between the Gaussian distributions corresponding to the initial
temperature---recall that $T_{\ini}=1$ with our choice of units---and
the final one $T_{\fin}$. Taking into account that (i) the integrand
is Gaussian and (ii) the $d$-dimensional integral factorises into the
product of $d$ identical integrals, we have that
\begin{equation}
  \int
  d\bm{v}\sqrt{P_{G}(\bm{v},T_{i}=1)P_{G}(\bm{v},T_{\fin})}=
  \left(\frac{2\sqrt{T_{\fin}}}{1+T_{\fin}}\right)^{d/2},
\end{equation}
and
\begin{equation}
  \label{eq:Bhatta-angle-Gaussian}
  \Lambda_{G}=2\arccos\left[\left(
      \frac{2\sqrt{T_{\fin}}}{1+T_{\fin}}\right)^{d/2}\right].
\end{equation}

We can also derive analytical expressions for the statistical length
and the cost in the Gaussian case. In particular, we are interested
here in the relaxation process between the initial and final NESS,
with time-independent driving $\chi=\chi_{\fin}=T_{\fin}^{3/2}$. In
the Gaussian approximation, the evolution of the temperature is
monotonic and therefore we can integrate over the temperature instead
of over time. For the statistical length, we get
\begin{align}
  \calL_{G}^{\rel}=&\sqrt{\frac{d}{2}}
\int_{0}^{t_{\fin}}dt\left|\frac{\dot{T}(t)}{T(t)}\right|=
\sqrt{\frac{d}{2}}\left|\int_{0}^{t_{\fin}}dt\frac{\dot{T}(t)}{T(t)}\right|\nonumber
\\ &=\sqrt{\frac{d}{2}}\left|\int_{1}^{T_{\fin}}\frac{dT}{T}\right|=
\sqrt{\frac{d}{2}}\,|\ln T_{\fin}|,
\end{align}
whereas for the cost we have that
\begin{align}
  \calC_{G}^{\rel}=&\frac{d}{4}
          \int_{0}^{t_{\fin}}dt\left(\frac{\dot{T}(t)}{T(t)}\right)^{2}=
\frac{d}{4}\int_{1}^{T_{\fin}}dT\frac{\dot{T}}{T^{2}}\nonumber \\
&=\frac{d}{4}\int_{1}^{T_{\fin}}dT\frac{T_{\fin}^{3/2}-T^{3/2}}{T^{2}}=
\frac{d}{4}\left(T_{\fin}^{3/2}-3T_{\fin}^{1/2}+2\right).
\end{align}

Let $t_{\fin}^{\rel}$ be the time for connecting the initial and final
NESS in the relaxation process. The classical speed limits derived in
Ref.~\cite{ito_stochastic_2020} ensure that, within the Gaussian
approximation we are employing
\begin{equation}
  \label{eq:deltat-general}
  t_{\fin}^{\rel}\geq \frac{\calL_{G}^{2}}{2\calC_{G}}\geq \frac{\Lambda_{G}^{2}}{2\calC_{G}},
\end{equation}
i.e.
\begin{equation}
  \label{eq:deltat-general-bis}
  t_{\fin} \geq \frac{\left|\ln T_{\fin}\right|^{2}}{2T_{\fin}^{3/2}-3T_{\fin}^{1/2}+2}\geq \frac{8\arccos^{2}\left[\left(
        \dfrac{2\sqrt{T_{\fin}}}{1+T_{\fin}}\right)^{d/2}\right]}
  {d\,\left(T_{\fin}^{3/2}-3T_{\fin}^{1/2}+2\right)}.
\end{equation}
The above inequalities are equivalent to those in
Eqs.~\eqref{eq:phi-def}--\eqref{eq:tR1-tR2} of the main text.

\subsection{Geodesic in the Gaussian approximation}\label{sec:geodesic-Gaussian}

The geodesic in probability space can be further characterised. To do
so, it is useful to introduce a rescaled time $\tau=t/t_{\fin}$ for a
process of duration $t_{\fin}$, with $0\leq\tau\leq 1$. The
probability distribution over the geodesic $P^{*}(\bm{v},\tau)$ is
obtained by minimising the statistical length $\calL$ with the
constraint $\int d\bm{v}\, P(\bm{v},\tau)=1$, $\forall\tau$. A
straightforward but rather lengthy calculation leads to the
result~\cite{ito_stochastic_2020}
\begin{equation}\label{eq:P(v)-geodesic}
  \sqrt{P^{*}(\bm{v},\tau)}=\frac{\sqrt{P_{\ini}(\bm{v})}
    \sin\left(\frac{\Lambda}{2}(1-\tau)\right)+
    \sqrt{P_{\fin}(\bm{v})}
    \sin\left(\frac{\Lambda}{2}\tau\right)}
  {\sin\left(\frac{\Lambda}{2}\right)}.
\end{equation}
Over the geodesic, the cost is also minimised but it depends on the
connecting time $t_{\fin}$---i.e. on the parametrisation of the
geodesic~\cite{crooks_measuring_2007}. Specifically, it is obtained
that $\calC^{*}=\Lambda^{2}/(2t_{\fin})$. It is worth stressing that
Eq.~\eqref{eq:P(v)-geodesic} is exact and general, valid for any
dynamics, as shown in Ref.~\cite{ito_stochastic_2020}.

Let us analyse the geodesic in more detail, for the specific case of
the granular fluid. The granular temperature over the geodesic is
directly obtained by taking the second moment of the probability
distribution in Eq.~\eqref{eq:P(v)-geodesic},
\begin{align}\label{eq:temp-geodesic}
  T^{*}(\tau)=\frac{1} {\sin^{2}\left(\frac{\Lambda_{G}}{2}\right)}
\bigg\{&\sin^{2}\left[\frac{\Lambda_{G}}{2}(1-\tau)
\right]\!\!+T_{\fin}\,\sin^{2}\left(\frac{\Lambda_{G}}{2}\tau \right)
\nonumber\\
&+\frac{4T_{\fin}}{1+T_{\fin}}\cos\left(\frac{\Lambda_{G}}{2}\right)\sin\left(\frac{\Lambda_{G}}{2}\tau\right)\nonumber
\\ &\quad\times \sin\left[\frac{\Lambda_{G}}{2}(1-\tau)\right]
\bigg\}.
\end{align}
The first and second terms on the rhs come from $P_{\ini}(\bm{v})$ and
$P_{\fin}(\bm{v})$, respectively, and are exact. The third term stems
from the product $\sqrt{P_{\ini}(\bm{v})P_{\fin}(\bm{v})}$ and has
been written in the Gaussian approximation. Consistently, we have
substituted $\Lambda$ with $\Lambda_{G}$, which is given as a function
of $T_{\fin}$ by Eq.~\eqref{eq:Bhatta-angle-Gaussian}, in the three
terms. Of course, $T^{*}(\tau)>0$ because all terms on the rhs of
Eq.~\eqref{eq:temp-geodesic} are non-negative.

A relevant issue is whether it is possible for the granular gas to
move over the geodesic or not. We can answer this question in the
Gaussian approximation we are employing. The temperature program over
the geodesic follows from the time-dependent protocol
for the driving
\begin{equation}
  \chi_{G}^{*}(t)=\dot{T}^{*}(t)+\left[T^{*}(t)\right]^{3/2},
\end{equation}
making use of Eq.~\eqref{eq:T-evol-Gaussian}. In the scaled time
$\tau$, the driving is thus
\begin{equation}
  \chi_{G}^{*}(\tau)=\frac{1}{t_{\fin}}\frac{dT^{*}(\tau)}{d\tau}
  +\left[T^{*}(\tau)\right]^{3/2}.
\end{equation}
The second term corresponds to the ``quasi-static'' driving: in the
NESS, $\chi=T^{3/2}$. The first term is the finite-time contribution,
which vanishes in the limit as $t_{\fin}\to\infty$.

Taking the derivative of Eq.~\eqref{eq:temp-geodesic}, after a little bit
of algebra one gets
\begin{equation}\label{eq:temp-deriv-geodesic}
  \frac{dT^{*}}{d\tau}=\frac{\Lambda_{G}}
  {2\sin^{2}\left(\frac{\Lambda_{G}}{2}\right)}  
  \frac{T_{\fin}-1}{T_{\fin}+1}
  \left[ T_{\fin}\sin(\Lambda_{G}\tau)+\sin(\Lambda_{G}(1-\tau))\right].
\end{equation}
The temperature evolution is monotonic over the geodesic: the sign of
the derivative is completely encoded in $T_{\fin}-1$, because all the
other terms are strictly positive. This introduces an asymmetry
between the cases $T_{\fin}>1$ and $T_{\fin}<1$.  For $T_{\fin}>1$,
the finite-time contribution $t_{\fin}^{-1}dT^{*}(\tau)/d\tau$ is
always positive and there is a well-defined driving that makes the
temperature sweep the geodesic curve, even in the limit
$t_{\fin}\to 0^{+}$. For $T_{\fin}<1$, the finite-time contribution
$t_{\fin}^{-1}dT^{*}(\tau)/d\tau$ is negative: for short enough
connecting time $t_{\fin}$, it will become larger---in absolute
value---than the quasi-static driving and make
$\chi_{G}<0$.

The above discussion implies that the geodesic cannot be swept for too
short connecting times for $T_{\fin}<1$. It is illustrative
to consider the particularisation of
Eq.~\eqref{eq:temp-deriv-geodesic} for $\tau=1/2$  to give an estimate
for the connecting time such that $\chi_{G}$ becomes negative,
\begin{equation}
  \left.\frac{dT^{*}}{d\tau}\right|_{\tau=1/2}=\frac{\Lambda_{G}}
  {2\sin\left(\frac{\Lambda_{G}}{2}\right)}  
  (T_{\fin}-1).
\end{equation}
The condition for having $\chi_{G}(\tau=1/2)<0$ is
\begin{equation}
  t_{\fin}<\frac{\Lambda_{G}}
  {2\sin\left(\frac{\Lambda_{G}}{2}\right)}  
  \frac{1-T_{\fin}}{[T^{*}(\tau=1/2)]^{3/2}},
\end{equation}
which can be ensured if
\begin{equation}\label{eq:tf-chi-neg-geod}
  t_{\fin}<\frac{\Lambda_{G}}
  {2\sin\left(\frac{\Lambda_{G}}{2}\right)}  
  (1-T_{\fin}).
\end{equation}
Note the time for which $\chi_{G}$ first becomes negative is longer
than the rhs of Eq.~\eqref{eq:tf-chi-neg-geod}.

\section{Simple ESR polynomial connection}\label{sec:simple-ESR}

Here we discuss how the ESR protocol is built from a simple
polynomial. We need at least a fourth-order polynomial with five
coefficients: four to adjust the boundary conditions
\eqref{eq:Tp-conditions} for the temperature, and one extra parameter
to impose that $A_{2p}(t_{\fin})=1$.

To start with, it is adequate to employ the scaled time
$\tau=t/t_{\fin}$ introduced in Appendix~\ref{sec:geodesic-Gaussian} and
to work with the thermal velocity
$v_{\ther}\equiv\sqrt{T}$. Consistently,
$v_{\ther,p}(\tau)=\sqrt{T}_{p}(\tau)$, and we rewrite
Eq.~\eqref{eq:chi-p} as
\begin{equation}
  \label{eq:chi-p-tau}
  \chi_{p}(\tau)= \frac{v_{\ther,p}(\tau)\left[\frac{2}{t_{\fin}}
\frac{dv_{\ther,p}(\tau)}{d\tau}+
v_{\ther,p}^{2}(\tau)\left(1+\frac{3}{16}a_{2}^{\st}A_{2p}(\tau)\right)\right]}
{1+\frac{3}{16}a_{2}^{\st}}.
\end{equation}
Insertion of this expression for the noise intensity into the
evolution equation of the excess kurtosis gives, after some algebra,
\begin{align}  \frac{dA_{2p}(\tau)}{d\tau}=&-\frac{4}{1+\frac{3}{16}a_{2}^{\st}}\frac{d\ln
v_{\ther,p}(\tau)}{d\tau}A_{2p}(\tau) \nonumber \\ & -2
t_{\fin}\left(B+\frac{3a_{2}^{\st}A_{2p}(\tau)}{16+32a_{2}^{\st}}\right)v_{\ther,p}(\tau)\left(A_{2p}(\tau)-1\right).
                                \label{eq:a2p-evol-tau}
\end{align}
We solve this equation---with the initial condition
$A_{2p}(0)=1$---with the following 4-th order polynomial for
the thermal velocity,
\begin{equation}
  \label{eq:vth-4th-order-pol}
  v_{\ther,p}(\tau)=1+c\tau^{2}+(4\Delta v_{\ther}-2c)\tau^{3}+(c-3\Delta
  v_{\ther})\tau^{4} .
\end{equation}
where $\Delta v_{\ther}\equiv\Delta\sqrt{T}=\sqrt{T_{\fin}}-1$. The
parameter $c$ is tuned to meet the boundary condition
$A_{2p}(t_{\fin})=1$: there is only one 4-th order
polynomial making the connection.
\begin{figure}
  \centering
  \includegraphics[width=1.68in]{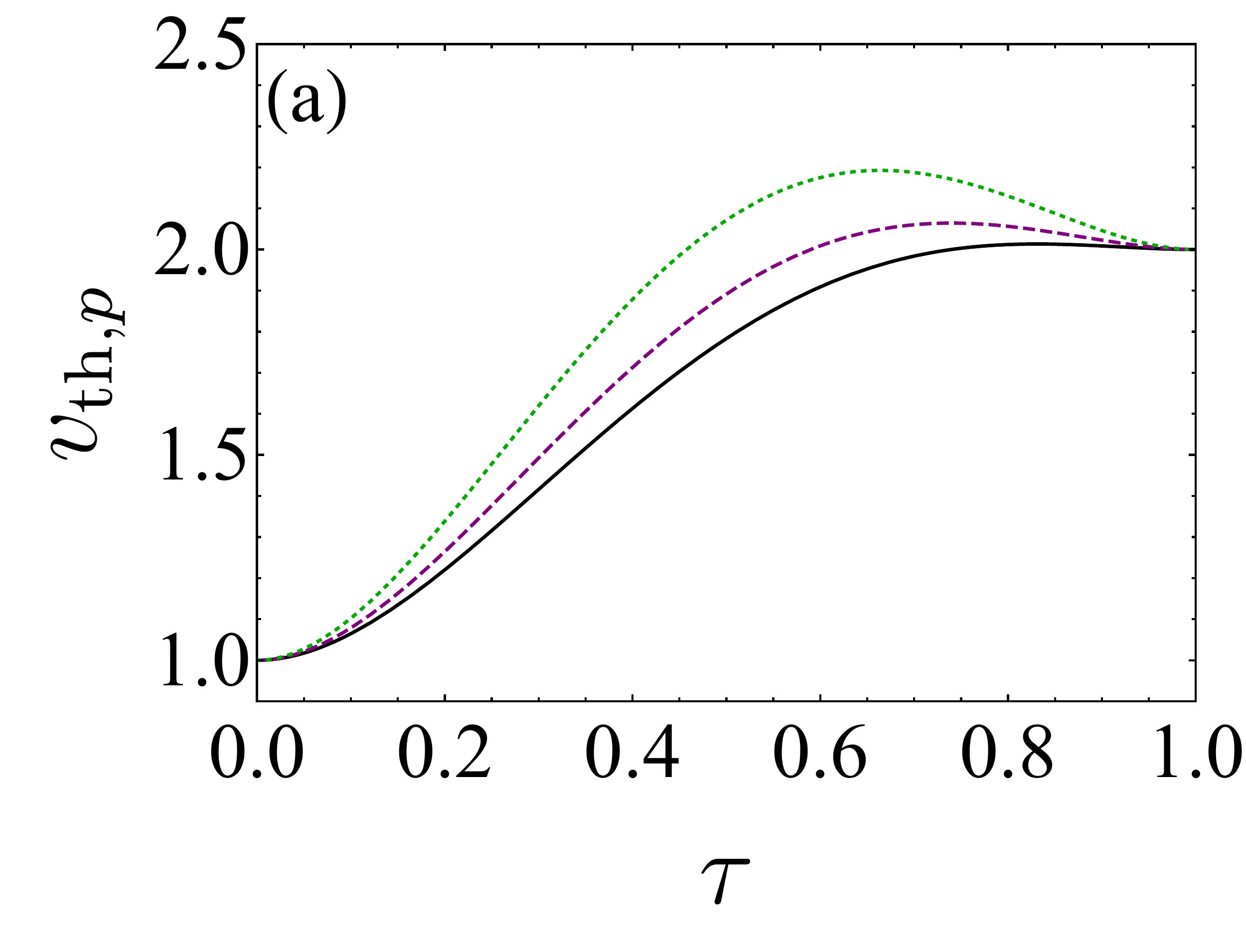}
  \includegraphics[width=1.68in]{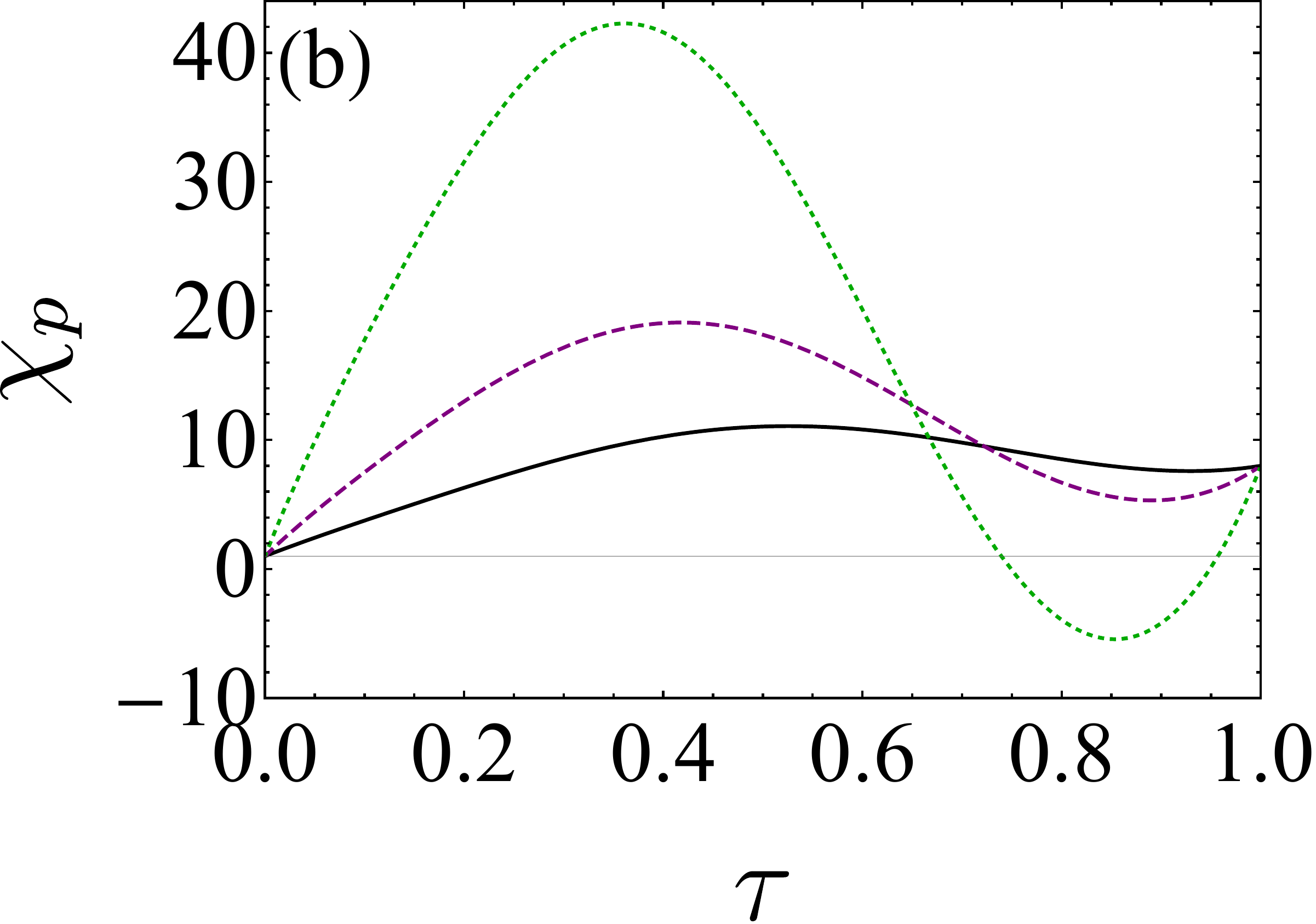}
    \includegraphics[width=1.68in]{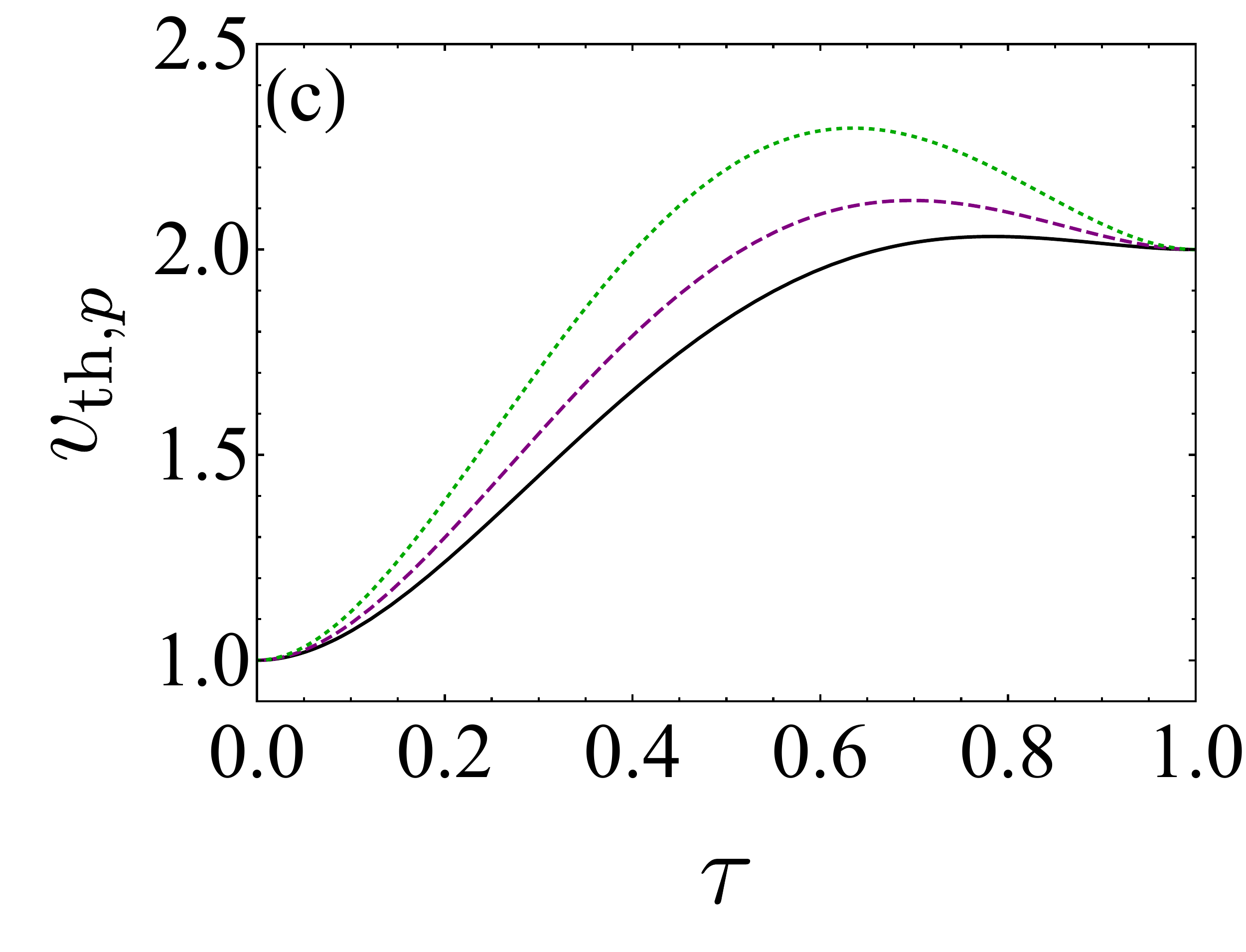}
  \includegraphics[width=1.68in]{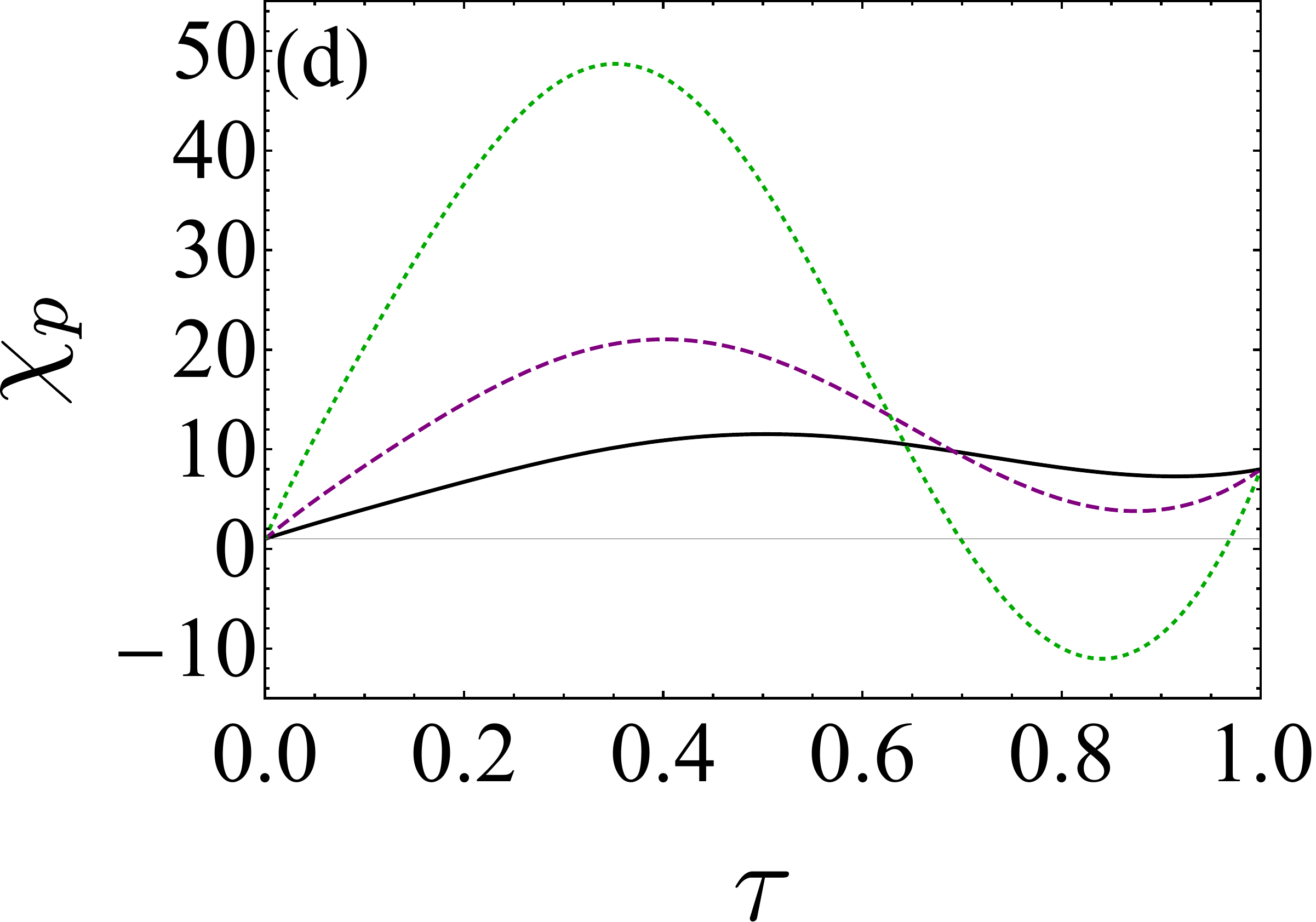}
  \caption{%
    Thermal velocity $v_{\ther,p}$ and noise intensity
    $\chi_{p}$ for the third-order polynomial connection, as a
    function of the normalised time $\tau=t/t_{\fin}$. All panels are
    for the two-dimensional case, panels (a) and (b) correspond to
    $\alpha=0.8$ and panels (c) and (d) correspond to $\alpha=0.3$. In
    each panel, three curves are plotted for different connection
    times: from bottom to top, $t_{\fin}=1$ (solid black),
    $t_{\fin}=0.5$ (dashed purple) and $t_{\fin}$=0.25 (dotted
    green). For the shortest connection time, $\chi_{p}(t)$ becomes
    negative inside a certain time window.}
  \label{fig:ESR-polynom}
\end{figure}

We have carried out the above procedure by numerically solving
Eq.~\eqref{eq:a2p-evol-tau} for the two-dimensional case---i.e. hard
disks. We show the numerical results for $T_{\fin}>1$ in
Fig.~\ref{fig:ESR-polynom}, specifically for $T_{\fin}=4$
($\Delta v_{\ther}=1$). The following qualitative behaviour is
observed: as the connecting time $t_{\fin}$ in decreased, the driving
$\chi_{p}(t)$ goes to very high values before decreasing to lower,
even negative, values. Evidently, the noise intensity $\chi_{p}(t)$
cannot become negative, so this means that the ESR connection cannot
be done with a 4-th order polynomial for too short times~\footnote{A
  similar behaviour is found for $T_{\fin}<1$, but the driving first
  decreases---taking also negative values for short enough
  $t_{\fin}$---and afterwards increases to overshoot its final value.}.

The observed behaviour hints at the emergence of a minimum,
non-vanishing, value of the connecting time for the ESR protocol, both
for $T_{\fin}>1$ and $T_{\fin}<1$. This feeling is reinforced by
employing higher order polynomials. For example, in the 5-th order
case, there is a mono-parametric family of polynomials connecting the
initial and final NESS. Nevertheless, $\chi_{p}(t)$ becomes negative
for $t_{\fin}$ below a certain value, over the whole family of
polynomials making the connection.

\section{Fisher information in the Sonine approximation}\label{sec:Fisher-info-Sonine}

Here, we look into the Fisher information $I(t)$ within the Sonine
approximation we have considered throughout.  The velocity
distribution function is expanded as
\begin{equation}
  \label{eq:Sonine-approx}
  P(\bm{v},t)=P_{G}(\bm{v},T(t))\left[1+\sum_{k=2}^{\infty}a_{k}(t)
    S_{k}\!\left(\frac{v^{2}}{2T}\right)\right],
\end{equation}
where $P_{G}(\bm{v},v_{\ther}(t))$ is the Maxwellian distribution of
Eq.~\eqref{eq:Maxwell-dim-d}, $a_{k}(t)$ are the coefficients of the
expansion---which are related to the cumulants, and finally
$S_{k}(x)\equiv L_{k}^{(\frac{d-2}{2})}(x)$ are the associated
Laguerre (or Sonine) polynomials~\cite{abramowitz_handbook_1988}. The
explicit expression for the first Sonine polynomials
are~\cite{van_noije_velocity_1998}
\begin{subequations}\label{eq:S(x)}
  \begin{align}
    S_{0}(x)&=1, \quad
S_{1}(x)=-x+\frac{d}{2} ,\\
S_{2}(x)&=
\frac{1}{2}x^{2}-\frac{d+2}{2}x+\frac{d(d+2)}{8}.
  \end{align}
\end{subequations}

The first Sonine approximation consists in keeping only the first term
in the expansion, with coefficient $a_{2}$ that equals the excess
kurtosis, and neglecting non-linear contributions in $a_{2}$ in all
the expressions derived from
Eq.~\eqref{eq:Sonine-approx}~\footnote{The first polynomial $S_{1}$
  does not appear in the expansion because the Gaussian distribution
  gives the correct value for the temperature $T(t)$, i.e. the
  corresponding coefficient $a_{1}(t)$ vanishes identically.}. For our
purposes, it is convenient to rewrite Eq.~\eqref{eq:Sonine-approx} as
follows: we introduce a dimensionless velocity
\begin{equation}
  \bm{c}(\bm{v},T(t))=\bm{v}/\sqrt{2T(t)},
\end{equation}
and the order of unity quantity $A_{2}(t)$ defined in
Eq.~\eqref{eq:A2-def}, so that
  \begin{align}\label{eq:Sonine-approx-A2}
    P(\bm{v},t)&=\frac{e^{-c^{2}}}{[2\pi T(t)]^{d/2}}
                 \left[1+a_{2}^{\st}A_{2}(t)
                 S_{2}(c^{2})\right].
    \end{align}

Now we proceed to calculate the Fisher information. For that, we take
into account that
\begin{equation}
  \partial_{t}f(c^{2})=\frac{df(c^{2})}{d(c^{2})}\partial_{t}c^{2}=
  -\frac{\dot{T}(t)}{T(t)} c^{2} \frac{df(c^{2})}{d(c^{2})}
\end{equation}
to write
\begin{align}
  \partial_{t}\ln
  P&(\bm{v},t)=-\frac{\dot{T}}{T}
  S_{1}(c^{2}) \nonumber \\
  &+a_{2}^{\st}\left[\dot{A}_{2}
    S_{2}(c^{2})-A_{2}\frac{\dot{T}}{T}c^{2}
    \frac{dS_{2}(c^{2})}{d(c^{2})}\right]\!. \label{eq:dt-lnP-Sonine}
\end{align}
In order to obtain $I(t)$, Eq.~\eqref{eq:dt-lnP-Sonine} is squared and
averaged with the probability distribution
\eqref{eq:Sonine-approx-A2}---neglecting non-linear terms in
$a_{2}^{\st}$. After a little algebra, one gets
\begin{align} I(t)=I_{G}(t)+&a_{2}^{\st}A_{2}\left(\frac{\dot{T}}{T}\right)^{\!\!2}
  \nonumber \\
  & \times\!\left[
    \overline{S_{1}^{2}(c^{2})
      S_{2}(c^{2})} +
    2\,\overline{c^{2}S_{1}(c^{2})
  \frac{dS_{2}(c^{2})}{d(c^{2})}}\,
    \right]\!,
    \label{eq:I=IG+a2}
\end{align}
where we have omitted the time dependence of $T(t)$ and $A_{2}(t)$ to
simplify the notation, and defined
\begin{equation}
  \overline{f(\bm{c})}\equiv \int d\bm{c}\,f(\bm{c})\,\phi(\bm{c}),
  \qquad \phi(\bm{c})=\pi^{-d/2}e^{-c^{2}}
\end{equation}
as the average of $f(\bm{c})$ with the dimensionless Gaussian
distribution $\phi(\bm{c})$. The averages in Eq.~\eqref{eq:I=IG+a2}
are thus $d$-dimensionless integrals of polynomials with the Gaussian
distribution, which result in
\begin{align}\label{eq:I(t)-Sonine-result}
  I(t)& =I^{(0)}(t)\!\left[1-
    \frac{d+2}{2}a_{2}^{\st}A_{2}(t)\right]\!, \;\;
        I^{(0)}(t)=\frac{d}{2}\!
        \left(\frac{\dot{T}(t)}{T(t)}\right)^{\!\!2}
        \!\! . 
\end{align}
There is no contribution coming from the term proportional to
$\dot{A}_{2}$ in Eq.~\eqref{eq:dt-lnP-Sonine} because of the
orthogonality of Sonine polynomials,
$\overline{S_{j}(c^{2})S_{k}(c^{2})}=0$
for $j\ne k$. Also, note that $I^{(0)}(t)\ne I_{G}(t)$ because we no
longer set the excess kurtosis to zero in the first Sonine
approximation. Notwithstanding, the smallness of $a_{2}^{\st}$ implies
that the main contribution to the Fisher information comes from $I^{(0)}(t)$.

\section{Normal modes for the $d$-dimensional harmonic potential}\label{sec:3d-harm-pot}

Our starting point is the Fokker-Planck equation \eqref{eq:FP-3d-osc},
for the harmonic potential case. The transformation to normal
modes is orthogonal, i.e. there exists
an orthogonal matrix of elements $C_{jk}$ such that
\begin{equation}
  x_{j}=\sum_{\beta=1}^{d}C_{j\beta}\xi_{\beta}, \quad
  \sum_{\beta=1}^{d}C_{j\beta}C_{k\beta}=\delta_{jk},
\end{equation}
which diagonalises the symmetric matrix---with elements
$\lambda_{jk}$---of the harmonic well $U_{h}(\bm{x})$,
\begin{align}
  &\sum_{\beta=1}^{d}\sum_{\beta'=1}^{d}
    C_{j\beta}\lambda_{jk}C_{k\beta'}=\kappa_{\beta}\delta_{\beta,\beta'}, \\ 
  & U_{h}(\bm{\xi})=\frac{1}{2}\sum_{\beta=1}^{d}\kappa_{\beta}\xi_{\beta}^{2}.
\end{align}

Therefore, the Fokker-Planck equation can be rewritten in terms of the
probability $P(\bm{\xi},t)=P(\bm{x},t)$ as
\begin{equation}\label{eq:FP-3d-osc-xi}
  \gamma\,
  \partial_{t}P(\bm{\xi},t)=\nabla_{\bm{\xi}}\cdot
  \left[\nabla_{\bm{\xi}} U_{h}(\bm{\xi})\,P(\bm{\xi},t)\right]+
    k_{B}T(t)\, \nabla_{\bm{\xi}}^{2}\! P(\bm{\xi},t),
\end{equation}
At equilibrium, the initial distribution $P(\bm{\xi},t=0)$ factorises
into the product of $d$ Gaussian distributions with zero mean and
variances $T_{\ini}/\kappa_{\beta}$, one for each normal mode. Since
$\partial_{\xi_{\beta}}U_{h}(\bm{\xi})=\kappa_{\beta}\xi_{\beta}$, the
joint distribution $P(\bm{\xi},t)$ still factorises into $d$
Gaussian distributions with zero mean for all times. Therefore, it is
completely characterised by the variances of the modes $\langle
\xi_{\beta}^{2}\rangle$, which obey the uncoupled equations
\begin{equation}\label{eqs:xi-evo-dimensions}
  \gamma\,
  \frac{d}{dt}\langle\xi_{\beta}^{2}\rangle=
  -2\kappa_{\beta}\langle\xi_{\beta}^{2}\rangle+ 2k_{B}T(t).
\end{equation}

It is convenient to go to dimensionless variables, by introducing
suitable units for time, length and temperature. We label the modes in
such a way that $\kappa_{1}\leq\cdots\leq\kappa_{d}$. We define 
\begin{equation}
  t^{*}=\frac{\kappa_{1}}{\gamma}t, \quad
  \xi_{\beta}^{*}=\frac{\xi_{\beta}}{\sqrt{k_{B}T_{\ini}/\kappa_{1}}},
  \quad T^{*}(t)=\frac{T(t)}{T_{\ini}}.
\end{equation}
With our choice of units, $\langle(\xi_{1}^{*})^{2}\rangle_{\ini}=1$
and $T^{*}_{\ini}=1$. We can rewrite~\eqref{eqs:xi-evo-dimensions} as
\begin{equation}\label{eqs:xi-evo-dimensionless}
  \frac{d}{dt^{*}}\langle(\xi_{\beta}^{*})^{2}\rangle=-2\kappa_{\beta}^{*}
  \langle(\xi_{\beta}^{*})^{2}\rangle +2 T^{*}(t),
\end{equation}
where
\begin{equation}
  \kappa_{1}^{*}=1\leq \cdots\leq \kappa_{d}^{*}.
\end{equation}
Equation~\eqref{eqs:xi-evo-dimensionless} is equivalent
to~\eqref{eq:sigma-evol} of the main text. Therein, we have dropped
the asterisks in order not to clutter our formulas.

\begin{acknowledgments}
  I thank Carlos A. Plata, Andr\'es Santos, and Emmanuel Trizac for
  discussions and their critical reading of the manuscript. Financial
  support from the Spanish Agencia Estatal de Investigaci\'on through
  Grant No.\ PGC2018-093998-B-I00, partially financed by the European
  Regional Development Fund, is also acknowledged.
\end{acknowledgments}

\bibliography{Mi-biblioteca-23-mar-2021}

\end{document}